\documentclass[12pt]{article}
\pdfoutput=1
\usepackage{jheppub}
\usepackage{amsmath,bbm,array,amsfonts,graphicx,wrapfig,lscape,float,mathtools,multirow,longtable,rotating,subfigure,ulem,cancel}
\usepackage[section]{placeins}
\usepackage[all]{xy}

\newcommand{\be}{\begin{equation}}
\newcommand{\ee}{\end{equation}}
\newcommand{\beq}{\begin{equation}}
\newcommand{\beql}[1]{\begin{equation}\label{#1}}
\newcommand{\eeq}{\end{equation}}
\newcommand{\ba}{\begin{array}}
\newcommand{\ea}{\end{array}}
\newcommand{\bea}{\begin{eqnarray}}
\newcommand{\beal}[1]{\begin{eqnarray}\label{#1}}
\newcommand{\eea}{\end{eqnarray}}
\newcommand{\ben}{\begin{enumerate}}
\newcommand{\een}{\end{enumerate}}
\newcommand{\bean}{\begin{eqnarray*}}
\newcommand{\eean}{\end{eqnarray*}}

\newcommand{\btab}[1]{\begin{tabular}{#1}}
\newcommand{\etab}{\end{tabular}}

\newcommand{\comment}[1]{}

\newenvironment{definition}[1][Definition]{\begin{trivlist}
\item[\hskip \labelsep {\bfseries #1}]}{\end{trivlist}}

     \let\P=\Pi

\newcommand{\qed}{\nobreak \ifvmode \relax \else
      \ifdim\lastskip<1.5em \hskip-\lastskip
      \hskip1.5em plus0em minus0.5em \fi \nobreak
      \vrule height0.75em width0.5em depth0.25em\fi}

\numberwithin{equation}{section}

\newcommand{\slice}{$\mathcal{S}_{\mathcal N,\rho}$ }
\newcommand{\rbv}{d_{BV}(\rho)}

\usepackage{mathabx}

\newcommand{\Pija}{\mathcal{P}_{ij}(a)}

\def\node#1#2{\overset{#2}{\underset{#1}{{\displaystyle    \color {blue}   \newmoon} }   }    }
\def\wver#1#2{\underset{{\llap{$\scriptstyle#1$}{ \displaystyle   {\color {red} \blacksquare}}  {\rlap{$\scriptstyle#2$}}}}{\scriptstyle \vert}}
\def\fu#1#2{\underset{{\llap{$\scriptstyle#1$}\displaystyle{ \color{red} \blacksquare}{\rlap{$\scriptstyle#2$}}}}{\scriptstyle\downarrow}}
\def\fd#1#2{\underset{{\llap{$\scriptstyle#1$}\displaystyle{ \color{red} \blacksquare}{\rlap{$\scriptstyle#2$}}}}{\scriptstyle\uparrow}}
\def\fdu#1#2{\underset{{\llap{$\scriptstyle#1$}\displaystyle{\color{red} \blacksquare}{\rlap{$\scriptstyle#2$}}}}{\scriptstyle\updownarrows}}

\def\rarr{{\scriptstyle\rightarrow}}
\def\larr{{\scriptstyle\leftarrow}}
\def\rlarr{{\scriptstyle\leftrightarrows}}
\def\lrarr{{\scriptstyle\rightleftarrows}}
\def\CS{\mathcal S}
\def\lb{\left(}
\def\rb{\right)}

\def\slod#1{\mathcal{S}_{\mathcal N,(#1)}}
\def\ba{\begin{equation} \begin{aligned}}
\def\ea{\end{aligned}\end{equation}}

\DeclareMathOperator{\tr}{tr}

\title{Quiver Theories and Formulae for Slodowy Slices of Classical Algebras}

\author{Santiago Cabrera,}
\author{Amihay Hanany,}
\author{Rudolph Kalveks}

\affiliation{
Theoretical Physics Group, The Blackett Laboratory,
Imperial College London, \\
Prince Consort Road, London SW7 2AZ, United Kingdom
}

\emailAdd{santiago.cabrera13@imperial.ac.uk, a.hanany@imperial.ac.uk, rudolph.kalveks09@imperial.ac.uk }

\preprint{Imperial/TP/18/AH/05}

\abstract{ 
We utilise SUSY quiver gauge theories to compute properties of Slodowy slices; these are spaces transverse to the nilpotent orbits of a Lie algebra $\mathfrak g$. We analyse classes of quiver theories, with Classical gauge and flavour groups, whose Higgs branch Hilbert series are the intersections between Slodowy slices and the nilpotent cone $\mathcal S\cap \mathcal N$ of $\mathfrak{g}$. We calculate refined Hilbert series for Classical algebras up to rank $4$ (and $A_5$), and find descriptions of their representation matrix generators as algebraic varieties encoding the relations of the chiral ring. We also analyse a class of dual quiver theories, whose Coulomb branches are intersections $\mathcal S\cap \mathcal N$; such dual quiver theories exist for the Slodowy slices of $A$ algebras, but are limited to a subset of the {Slodowy slices} of $BCD$ algebras. The analysis opens new questions about the extent of $3d$ mirror symmetry within the class of SCFTs known as $T_\sigma^\rho(G)$ theories. We also give simple group theoretic formulae for the Hilbert series of Slodowy slices; these draw directly on the $SU(2)$ embedding into $G$ of the associated nilpotent orbit, and the Hilbert series of the nilpotent cone.\\

~\today}

\begin{document}

\maketitle

\listoftables

\listoffigures

\section{Introduction}
\label{sec:intro}

The relationships between supersymmetric (``SUSY") quiver gauge theories, the Hilbert series (``HS") of their Higgs and Coulomb branches, and the nilpotent orbits (``NO") of simple Lie algebras $\mathfrak{g}$ were analysed in two recent papers \cite{Hanany:2016gbz, Hanany:2017ooe}. Closures of classical nilpotent orbits appear as Higgs branches on ${\cal N}=2$ quiver theories in $4d$, and also as Coulomb branches on ${\cal N}=4$ quiver theories in $2+1$ dimensions. Both these types of theory have 8 supercharges.

The aim herein is to examine systematically the relationships between these SUSY quiver gauge theories and the spaces transverse to nilpotent orbits, known as Slodowy slices. The focus herein is the Slodowy slices of the nilpotent orbits of Classical algebras, which are associated with a rich array of $3d {~} \mathcal {N}=4$ quiver theories and dualities. The relationships between SUSY quiver gauge theories and the Slodowy slices of nilpotent orbits of Exceptional algebras remain to be treated.

The mathematical study of Slodowy slices has its roots in \cite{slodowy_1980}, which built on earlier work by Brieskorn \cite{brieskorn1970singular}, Grothendieck and Dynkin \cite{Dynkin:1957um}. This showed that each nilpotent orbit $\cal O_\rho$ of a Lie algebra $\mathfrak g$ of a Classical group $G$ has a transverse slice, or Slodowy slice $\cal S_\rho$, lying within the algebra $\mathfrak g$.\footnote{$\rho$ identifies the embedding of $\mathfrak{su(2)}$ into $\mathfrak g$ that defines the nilpotent orbit.} There is a variety defined by the intersection between the Slodowy slice and the nilpotent cone ${\cal N}$ of the algebra: ${\cal S}_{{\cal N},\rho} \equiv {\cal N} \cap \cal S_{\rho}$. In this paper, we deal almost entirely with these intersections $S_{\cal N,\rho}$ and refer to them loosely as Slodowy slices (except where the context requires otherwise). Each such slice is a singularity that can be characterised by a sub-algebra $\mathfrak f$ of $\mathfrak g$ that commutes with (or stabilises) the $\mathfrak{su(2)}$. In the case of the sub-regular nilpotent orbit \slice is a Kleinian singularity of type $ADE$.\footnote{For general background on nilpotent orbits the reader is referred to \cite{Collingwood:1993fk}.}

The connection between nilpotent orbits and their Slodowy slices, and instanton moduli spaces, i.e. the solutions of self dual Yang-Mills equations, was made in \cite{Kronheimer:1990ay}. The use of Dynkin diagrams and quiver varieties to define instantons on ALE spaces was discussed in \cite{nakajima_1994}. The relevance of nilpotent orbits and Slodowy slices to  $3d$ ${\cal N} = 4$ quiver theories was later explored in detail in \cite{Gaiotto:2008sa} and \cite{Gaiotto:2008ak}. In this context, they appear as effective gauge theories describing the brane dynamics of a system in Type IIB string theory. Brane systems of the type of \cite{Hanany:1996ie} are relevant for the $A$ series and systems with three dimensional orientifold planes \cite{Feng:2000eq} for the $BCD$ series\footnote{Note that these brane systems can explicitly realize the transverse slices developed by Brieskorn and Slodowy \cite{brieskorn1970singular,slodowy_1980}. A systematic analysis of transverse slices was carried out by Kraft and Procesi \cite{Kraft:1982fk} and the physics realization was studied in \cite{Cabrera:2016vvv, Cabrera:2017njm}. The concept of transverse slices can be further extended as an operation of subtractions between two quivers \cite{Cabrera:2018ann}.}.

In the course of these latter papers, a class of superconformal field theories (``SCFT") was proposed, with moduli spaces defined by the intersections between Slodowy slices and nilpotent orbits. These are termed $T_{\sigma}^{\rho} (G)$ theories, where $G$ is a Lie group. Several types of Classical quiver theories were identified, along with associated brane configurations, including theories whose Higgs or Coulomb branches correspond to certain varieties ${\cal S}_{{\cal N},\rho}$, and a relationship between S-duality and dualities arising from the $3d$ mirror symmetry \cite{Intriligator:1996ex} of Classical quiver theories was conjectured\footnote{In the case of nilpotent orbits of C and D type, the precise match between quivers and orbits was given in \cite{Benini:2010uu}. Subsequently, \cite{Chacaltana:2012zy} described the relation for B type and unified all classical cases via the introduction of the Barbasch-Vogan map \cite{barbasch_vogan_1985}.}. 

For example, in the case of an $A$ series nilpotent orbit $\cal O_{\rho}$, where $\rho$ describes the embedding of $\mathfrak{su(2)}$ into $\mathfrak{su(n)}$ that defines the nilpotent orbit, and $\rho = (1^N)$ corresponds to the trivial nilpotent orbit, these dualities entail that the \textit{Higgs} branch of a linear quiver based on a partition $\rho^T$, yields the closure of the nilpotent orbit $\cal {\bar O}_{\rho}$, while the \textit{Coulomb} branch of a linear quiver based on the partition $\rho$ gives its Slodowy slice ${\cal S}_{{\cal N},\rho}$. The application of $3d$ mirror symmetry to this pair of linear quivers yields a pair of ``balanced" quivers, with the \textit{Coulomb} branch of the former yielding $\cal {\bar O}_{\rho}$ and the \textit{Higgs} branch of the latter yielding ${\cal S}_{{\cal N},\rho}$.\footnote{The notation in the Literature regarding partitions and their dual maps has changed a great deal; see \cite[sec.~4]{Cabrera:2017njm} for a summary of the different maps that are relevant to our study and an explicit review of the different conventions used in mathematics and physics.} 

More recently, in \cite{Chacaltana:2012zy} and  \cite{Mekareeya:2012tn}, nilpotent orbits and Slodowy slices have been used in the study of $6d$ ${\cal N} =(2, 0)$ theories on Riemann surfaces. Relationships between diagram automorphisms of quiver varieties and Slodowy slices are explored in \cite{henderson_licata_2014}. In \cite{sole_kac_valeri_2016} the algebras of polynomial functions on Slodowy slices were shown to be related to classical (finite and affine) W-algebras. 

Each Slodowy slice of a sub-algebra ${\mathfrak f}$ of ${\mathfrak g}$ has a ring of holomorphic functions transforming in irreps of the sub-group $F$ of $G$. Our approach is to compute the Hilbert series of these rings. Presented in refined form, such Hilbert series faithfully encode the class function content of Slodowy slices, and can be subjected to further analysis using the tools of the Plethystics Program \cite{Benvenuti:2006qr, Feng:2007ur, Hanany:2014dia}.

Importantly, following a result in \cite{slodowy_1980}, the Hilbert series of Slodowy slices $\mathcal{S}_{\mathcal{N},\rho}$ are always complete intersections, i.e. quotients of geometric series. It was shown in \cite{Cremonesi:2014kwa} how the HS of the Slodowy slices of $A$ series and certain $BCD$ series algebras can be calculated from the Coulomb branches of linear quivers (or from the Higgs branches of their $3d$ mirror duals). \cite{Cremonesi:2014kwa} also identified a relationship between Slodowy slices and the (modified) Hall Littlewood polynomials of $\mathfrak g$, under the mapping $\mathfrak g \to \mathfrak{su(2)} \otimes \mathfrak{f}$.

Methods of calculating Hilbert series for $T_{\sigma}^{\rho} (G)$ theories with multi-flavoured quivers of unitary or alternating $O/USp$ type were developed in \cite{Cremonesi:2014uva}, using both Coulomb branch and Higgs branch methods. As elaborated in \cite{Cabrera:2017ucb}, the calculation of Coulomb branches of quivers of $O/USp$ type requires close attention to the distinction between $SO$ and $O$ groups.

This paper builds systematically on such methods to calculate the Hilbert series of Slodowy slices of closures of nilpotent orbits of low rank Classical Lie algebras and to identify relevant generalisations to arbitrary rank.

In Section \ref{sec:Slodowy} we summarise some facts about nilpotent orbits and review the relationship between a Slodowy slice ${\cal S}_{{\cal N},\rho}$ and the homomorphism $\rho$ defining the embedding of $\mathfrak {su(2)}$ into ${\mathfrak g}$ (and thus of the mapping of irreps of $G$ into the irreps of $SU(2)$) associated with a nilpotent orbit $\cal O_{\rho}$. We give some simple representation theoretic formulae for calculating the dimensions and Hilbert series of a Slodowy slice, given a homomorphism $\rho$.

In Section \ref{sec:ASeries} we treat $A$ series Slodowy slices, summarising the relevant Higgs branch and Coulomb branch formulae, describing the quivers upon which they act, and tabulating the commutant global symmetry group and the Hilbert series of Slodowy slices for all nilpotent orbits up to rank 5. We also build upon the language of $T_{\sigma}^{\rho}(SU(N))$ theories to summarise the known exact $A$ series dualities between quiver theories for Slodowy slices and nilpotent orbits. We find matrix formulations for the generators of $A$ series Slodowy slices and their relations, which explicate the mechanism of symmetry breaking and the residual symmetries of the parent group.

In Section \ref{sec:BCDSeries} we extend this analysis to Slodowy slices of $BCD$ series algebras up to rank 4. We find a complete set of refined Hilbert series, by working with the Higgs branches of multi-flavoured alternating $O/USp$ quivers with appropriately balanced gauge nodes. As in the case of $BCD$ nilpotent orbits \cite{Hanany:2016gbz}, calculation of these Higgs branches requires taking ${\mathbb Z}_2$ averages over the $SO$ and $O^-$ forms of $O$ group characters. We also identify a limited set of Higgs branch constructions based on $D$ series Dynkin diagrams. We summarise the restricted set of Coulomb branch monopole constructions that are available for ${\cal S}_{{\cal N},\rho}$, which are based on alternating $\textit{SO}/USp$ linear quivers. We highlight apparent restrictions on $3d$ mirror symmetry between Higgs and Coulomb branches of $BCD$ quiver theories; these include the requirements that the nilpotent orbit $\cal O_{\rho}$ should be \textit{special}, and that the $O/USp$ quivers should not be ``bad" \cite{Gaiotto:2008ak} due to containing monopole operators with zero conformal dimension. We find matrix formulations for the Higgs branch generators of $BCD$ series Slodowy slices, and their relations, which explicate the mechanism of symmetry breaking and the residual symmetries of the parent group.

Taken together with other recent studies \cite{Hanany:2016gbz, Cabrera:2017ucb}, this analysis of Hilbert series is relevant for the understanding of $T_\sigma^\rho(G)$ theories of type $BCD$. It highlights the difference between orthogonal $O(n)$ and special orthogonal $SO(n)$ nodes and the surrounding problems associated with $3d$ mirror symmetry between orthosymplectic quivers.

The final Section summarises conclusions, discusses open questions and identifies areas for further work. Some notational conventions are detailed in Appendix \ref{apx:1}.

\section{Slodowy Slices}
\label{sec:Slodowy}

\subsection{Relationship to Nilpotent Orbits}
\label{sec:Slodowy1}

Each nilpotent orbit $\cal O_{\rho}$ of a Lie algebra $\mathfrak g$ is defined by the conjugacy class $\mathfrak g^X$  of nilpotent elements $X \in \mathfrak g$ under the group action \cite{Collingwood:1993fk}. Each nilpotent element $X$ forms part of a standard $\mathfrak{su(2)}$ triple $\{ X,Y,H \}$ and, following the Jacobson Morozov theorem, the conjugacy classes are in one to one correspondence with the equivalence classes of embeddings of $\mathfrak{su(2)}$ into $\mathfrak g$, described by some homomorphism $\rho$. The closure of each orbit $\cal {\bar O}_{\rho}$, can, as discussed in \cite{Hanany:2016gbz, Hanany:2017ooe}, be described as a moduli space, by a refined Hilibert series of representations of $G$, graded according to the degree of symmetrisation of the underlying nilpotent element.

The closures ${\cal {\bar O}}_{\rho}$ of the nilpotent orbits of $\mathfrak g$ form a poset, ordered according to their inclusion relations\footnote{See for example the Hasse diagrams in \cite{Kraft:1982fk}.}. The closure of the \textit{maximal} (also termed \textit{principal} or \textit{regular}) nilpotent orbit is called the nilpotent cone $\cal N$; it contains all the orbits $\cal {O}_{\rho}$ and has dimension $| {\cal N} |$ equal to that of the rootspace of $\mathfrak g$.  The poset of nilpotent orbits contains a number of canonical orbits. These include the trivial nilpotent orbit (described by the Hilbert series $1$ with dimension zero), a minimal (lowest dimensioned non-trivial) nilpotent orbit, a sub-regular orbit of dimension $\left| {\cal N} \right|-2$ and the maximal nilpotent orbit:
\begin{equation} 
\label{eq:SS_NO}
\{ 0\} = {{\cal O}_{trivial}} \subset {{\cal{\bar O}}_{minimal}} \ldots  \subset {{\cal{\bar O}}_{sub - regular}} \subset {{\cal{\bar O}}_{maximal}}=\cal N.
\end{equation}
All nilpotent orbits have an even (complex) dimension and are HyperK\"ahler cones.

The closure of the minimal nilpotent orbit of $\mathfrak g$ corresponds to the reduced single G-instanton moduli space \cite{Kronheimer:1990ay, Benvenuti:2010pq}. As discussed in \cite{Hanany:2016gbz}, the Hilbert series of the nilpotent cone has a simple expression in terms of the symmetrisations of the adjoint representation of $G$, modulo Casimir operators, or equivalently in terms of (modified) Hall Littlewood polynomials:

\begin{equation} 
\label{eq:SS1a}
\begin{aligned}
{g_{HS}^{\cal N}}& {{ = PE}}\left[ {\chi _{adjoint}^G{t^2} - \sum\limits_{i = 1}^r {t^{2{d_i}}} } \right],\\
{g_{HS}^{\cal N}}& = mHL_{singlet}^G \left[ {{t^2}} \right],
\end{aligned}
\end{equation}
where $t$ is a counting fugacity, $\chi _{adjoint}^G$ is the character of the adjoint representation and $\{d_1,\ldots,d_r\}$ are the degrees of the symmetric Casimirs of $G$, which is of rank $r$.

Slodowy slices are defined as \textit{slices} $\cal S_\rho\subseteq \mathfrak{g}$ that are \textit{transverse} in the sense of \cite{slodowy_1980} to the orbit $\mathcal O _\rho$. 
%
The varieties $\cal S_{\cal N,\rho}$ that concern the present study are slices inside the nilpotent cone $\mathcal N$. They can be constructed as:

\begin{equation} 
\label{eq:SS2}
{\cal S}_{{\cal N},\rho} \equiv \cal S_\rho \cap {{\cal N}}.
\end{equation}

Naturally, the slice $\cal S_{\cal N,\rho}$ transverse to the trivial nilpotent orbit is the entire nilpotent cone $\cal N$ and the slice $\cal S_{\cal N,\rho}$ transverse to the maximal nilpotent orbit is trivial. In between these limiting cases, however, the Slodowy slices do not match any nilpotent orbit. Consequently we have a complementary poset of Slodowy slices:

\begin{equation} 
\label{eq:SS3}
{{\cal N}} = {{{\cal S}}_{trivial}} > {{{\cal S}}_{minimal}} \ldots  > {{{\cal S}}_{sub - regular}} > {{{\cal S}}_{maximal}} = \{ 0\} .
\end{equation}
%

\subsection{Dimensions and Symmetry Groups}
\label{sec:SS_Dim}

The dimensions of a Slodowy slice ${{\cal S}_{{\cal N},\rho}}$ plus those of the nilpotent orbit $\cal {O}_{\rho}$ combine to the dimensions of the nilpotent cone $\cal {N}$:

\begin{equation} 
\label{eq:SS4}
\left| {{\cal S}_{{\cal N},\rho}} \right| + \left| {{{{\cal O}}_\rho }} \right| = \left| {{\cal N}} \right| = \left| {\mathfrak g} \right| - rank[ {\mathfrak g} ].
\end{equation}
The elements of the Slodowy slice ${{\cal S}_{{\cal N},\rho}}$ lie in a subalgebra $ \mathfrak f$, which is the centraliser of the nilpotent element $X \in \mathfrak g$, so that $\left[ {X,{ \mathfrak f}} \right] = 0$, and $ \mathfrak f$ is often termed the commutant of $ \mathfrak{su(2)}$ in $ \mathfrak g$. The structure of $ \mathfrak f$ and the dimensions of ${{\cal S}_{{\cal N},\rho}}$ and $\cal {O}_{\rho}$ can be determined by analysing the embedding of ${ \mathfrak {su(2)}} \rightarrow  \mathfrak g$.

Following \cite{Dynkin:1957um}, a homomorphism $\rho$ can be described by a root space map from $ \mathfrak g$ to $\mathfrak {su(2)}$, and this is conveniently encoded in a Characteristic set of Dynkin labels.\footnote{A Characteristic $G[\ldots]$ is distinct from highest weight Dynkin labels $[\ldots,\ldots]_G$.} The Characteristic $[q_1\ldots q_r]$ provides a map from the simple root fugacities $\{z_1,\ldots,z_r\}$ of $ \mathfrak g$ to the simple root fugacity $\{z\}$ of $\mathfrak {su(2)}$:

\begin{equation} 
\label{eq:SS5}
\rho \left[ {{q_1} \ldots {q_r}} \right]:\left\{ {{z_1}, \ldots ,{z_r}} \right\} \to \left\{ {{z^{\frac{{{q_1}}}{2}}}, \ldots ,{z^{\frac{{{q_r}}}{2}}}} \right\},
\end{equation}
where the labels $q_i \in \{0,1,2\}$. This induces corresponding weight space maps under which each representation of $G$ of dimension $N$ branches to representations $[n]$ of $SU(2)$ \textit{at some multiplicity} $m_n$. This branching is conveniently described using partition notation, $\left( |[N-1]|^{m_{N-1}}, \ldots ,|[ n ]|^{m_n}, \ldots ,1^{m_0} \right)$, which lists the dimensions of the $SU(2)$ irreps, using exponents to track multiplicities. These partitions are tabulated in \cite{Hanany:2016gbz} for the key irreps of Classical groups up to rank 5, identifying each nilpotent orbit by its Characteristic.

For example, the homomorphism $\rho$ with Characteristic $[202]$, which generates the 10 dimensional nilpotent orbit of of $A_3$, induces the following maps:
\begin{equation} 
\label{eq:SS6}
\begin{aligned}
\rho \left[ {202} \right]:& \left\{ {{z_1},{z_2},{z_3}} \right\} \to \left\{ {z,1,z} \right\}, & & \\
\rho \left[ {202} \right]:& \left[ {1,0,1} \right] \to \left[ 4 \right] + 3 \otimes \left[ 2 \right] + \left[ 0 \right]  & \iff & \chi _{adjoint}^{{A_3}} \to \left( {5,{3^3},1} \right),\\
\rho \left[ {202} \right]:& \left[ {1,0,0} \right] \to \left[ 2 \right] + \left[ 0 \right]   & \iff & \chi _{fundamental}^{{A_3}} \to \left( {3,1} \right).
\end{aligned}
\end{equation}
These $SU(2)$ partitions are invariant under the symmetry group $F \subseteq G$ of the Slodowy slice and hence the multiplicities encode representations of $F$.

Under the branching, the adjoint representation of $G$ decomposes to representations of the product group $SU(2) \otimes F$ with branching coefficients $a_{nm}$:

\begin{equation} 
\label{eq:SS7}
\begin{aligned}
\chi _{adjoint}^G & \to \bigoplus \limits_{[n][m]} {a_{nm}} \left(  {\chi _{[n]}^{SU(2)}\mathop \bigotimes \chi _{[m]}^F} \right).
\end{aligned}
\end{equation}
Other than for the trivial nilpotent orbit (in which the adjoint of $G$ branches to itself times an $SU(2)$ singlet), the adjoint of $SU(2)$ and the adjoint (if any) of $F$ each appear separately in the decomposition, so that $rank[G]\ge rank[F]\ge 0$. Along with the requirement that any multiplicities $m_n$ appearing in a partition must be dimensions of representations of $F$, this often makes it possible to determine the Lie algebra $\mathfrak f$ of the Slodowy slice directly from the partition data. In the example \ref{eq:SS6} the presence of a single $SU(2)$ singlet in the partition of the adjoint of $A_3$ entails that the symmetry group of the Slodowy slice to the $[202]$ orbit is simply $U(1)$.

The adjoint partition data also permits direct calculation of the complex dimensions of a Slodowy slice or nilpotent orbit, by summing multiplicities of SU(2) irreps or, equivalently, dimensions of $F$ irreps:
\begin{equation} 
\label{eq:SS8}
\begin{aligned}
\left| {{S_\rho }} \right| & = \sum\limits_{[n][m]} {{a_{nm}}\left| {\chi _{[m]}^F} \right|},\\
\left| {{{{\cal O}}_\rho }} \right| & = \left| G \right| - \left| {{\cal S}_{\rho}} \right|,\\
\left| {{\cal S}_{{\cal N},\rho}} \right| & = \left| {{S_\rho }} \right|{}- rank[G].
\end{aligned}
\end{equation}

\subsection{Hilbert Series}
\label{sec:SS_HS}

The branching of the adjoint representation of $G$ determines the generators of the Slodowy slice. If the decomposition \ref{eq:SS7} is known, the Hilbert series for the Slodowy slice can be derived from the HS of the nilpotent cone by substitution under a particular choice of grading. Consider the map $\tilde \rho$ of the adjoint that is obtained from \ref{eq:SS7} by the replacement of $SU(2)$ irreps by their highest weight fugacities $\chi _{[n]}^{SU(2)} \to {t^n}$, taking the particular choice of $t$ from \ref{eq:SS1a} as the counting variable:

\begin{equation} 
\label{eq:SS9}
\begin{aligned}
\tilde \rho :\chi _{adjoint}^G &\to \bigoplus \limits_{[n][m]} {a_{nm}}  { \chi _{[m]}^{{F}} {t^n}}.
\end{aligned}
\end{equation}
When the adjoint map \ref{eq:SS9} is applied to the generators of the nilpotent cone \ref{eq:SS1a}, the replacement of the $SU(2)$ representations $[n]$ by the counting fugacity $t^n$ entails that the resulting Hilbert series only transforms in the symmetry group of the Slodowy slice. Thus, $g_{HS}^{{\cal S}_{{\cal N},\rho}} =g_{HS}^{ {\left.{{\cal N}} \right|_{\tilde \rho }}}$,
%
or, written more explicitly:

\begin{equation} 
\label{eq:SS11}
\begin{aligned}
g_{HS}^{{\cal S}_{{\cal N},\rho}}(x,t) &=PE\left[ {{\left. {\chi _{adjoint}^G} \right|_{\tilde \rho}}{~} {t^2} - \sum\limits_{i = 1}^r {t^{2{d_i}}} } \right]\\
&= PE\left[ {\mathop  \bigoplus \limits_{[n][m]}^{} {a_{nm}}  \chi _{[m]}^F{t^{n + 2}} - \sum\limits_{i = 1}^r {{t^{2{d_i}}}} } \right].
\end{aligned}
\end{equation}
The expression \ref{eq:SS11} gives the \textit{refined} Hilbert series of the Slodowy slice in terms of its generators, which are representations of the Slodowy slice symmetry group $F$, at some counting degree in $t$, less its relations, which are set by the degrees of the Casimirs of $G$.\footnote{This construction for Slodowy slices is simpler, but equivalent to the Hall Littlewood method presented in \cite{Cremonesi:2014uva}.}

Importantly, an \textit{unrefined} Hilbert series, with representations of $F$ replaced by their dimensions, ${m_n} = \sum\limits_m {{a_{nm}}} |\chi _{[m]}^F|$, can be calculated directly from the adjoint partition under $\rho$, without knowledge of the precise details of the embedding \ref{eq:SS7}:

\begin{equation} 
\label{eq:SS12}
\begin{aligned}
g_{HS}^{{\cal S}_{{\cal N},\rho}} (1,t) & = PE\left[ {\sum\limits_n^{} {{m_n}} {t^{n+2}} - \sum\limits_{i = 1}^r {{t^{2{d_i}}}} } \right].
\end{aligned}
\end{equation}
Finally, the freely generated Hilbert Series for the canonical Slodowy slices ${\cal S}_{\rho}$ are related to those of their nilpotent intersections ${\cal S}_{{\cal N},\rho}$ simply by the exclusion of the Casimir relations:

\begin{equation} 
\label{eq:SS13}
\begin{aligned}
g_{HS}^{{S_\rho }}(x,t){} \equiv g_{HS}^{{S_{N,\rho }}}(x,t)~PE\left[ {\sum\limits_{i = 1}^r {{t^{2{d_i}}}} } \right]= PE\left[ {\mathop  \bigoplus \limits_{[n][m]}^{} {a_{nm}}  \chi _{[m]}^F{t^{n + 2}} } \right].
\end{aligned}
\end{equation}

In Sections \ref{sec:ASeries} and \ref{sec:BCDSeries} we set out the quiver constructions that provide a comprehensive method for identifying the decomposition \ref{eq:SS7} and for calculating the refined Hilbert series of the Slodowy slices ${\cal S}_{{\cal N},\rho}$. 

\subsection{Sub-Regular Singularities}
\label{sec:SS_Singular}

As shown in \cite{brieskorn1970singular,slodowy_1980}, the Slodowy slices of sub-regular orbits $ {{\cal S}_{{\cal N},subregular}} $ take the form of ADE type singularities, ${\mathbb C}^2 / \Gamma $, where $\Gamma$ is a finite group of type ADE. Under the nilpotent orbit grading by $t^2$ used herein, these take the forms in table \ref{table:SS1}. The intersection $ {{\cal S}_{{\cal N},subregular}} $ is an example of a transverse slice between adjacent nilpotent orbits; all such transverse slices of Classical algebras were classified by Kraft and Procesi in \cite{Kraft:1982fk}.
%
\begin{table}[htp]
\begin{tabular}{|c|c|c|c|}
\hline
$\text{Group}$&$ \text{Singularity}$&$ \text{Dimension} $&$ \text{Hilbert Series} $\\
\hline
$A_r$ & ${{{\hat A}_r} \equiv {{\mathbb C}^2}/{{\mathbb Z}_{r + 1}}}$&$ 2 $ & ${PE \left[ 2{t^{r + 1}} + {t^2} - {t^{2r + 2}} \right]} $\\
$B_r$ & ${{{\hat A}_{2r-1}} \equiv {{\mathbb C}^2}/{{\mathbb Z}_{2r }}}$&$ 2 $ & ${{PE \left[ 2{t^{2r}} + {t^2} - {t^{4r}} \right] }} $\\
$C_{r>1}$ & ${{{\hat D}_{r+1}} \equiv {{\mathbb C}^2}/{{ Dic}_{r - 1}}}$&$ 2 $ & ${PE \left[{{t^{2r - 2}} + {t^{2r}} + {t^4} - {t^{4r}}} \right]} $\\
$D_{r>2}$ & ${{{\hat D}_r} \equiv {{\mathbb C}^2}/{{ Dic}_{r-2}}}$&$ 2 $ & ${PE \left[ {{t^{2r - 4}} + {t^{2r - 2}} + {t^4} - {t^{4r - 4}}} \right]} $\\
\hline
\end{tabular}
\footnotesize \text{The dicyclic group of order $4k$ is denoted as $Dic_k$.}
\caption{Sub-regular Slodowy Slices of Classical Groups}
\label{table:SS1}
\end{table}

This known pattern of singularities amongst the Slodowy slices of subregular orbits, along with the known forms of trivial and maximal Slodowy slices and dimensions, provide consistency checks on the grading methods and constructions given herein.


\FloatBarrier

\section{$A$ Series Quiver Constructions}
\label{sec:ASeries}

\subsection{Quiver Types}
\label{subsec:AQuivers}

\begin{figure}[htbp]
\centering
\includegraphics[scale=0.5]{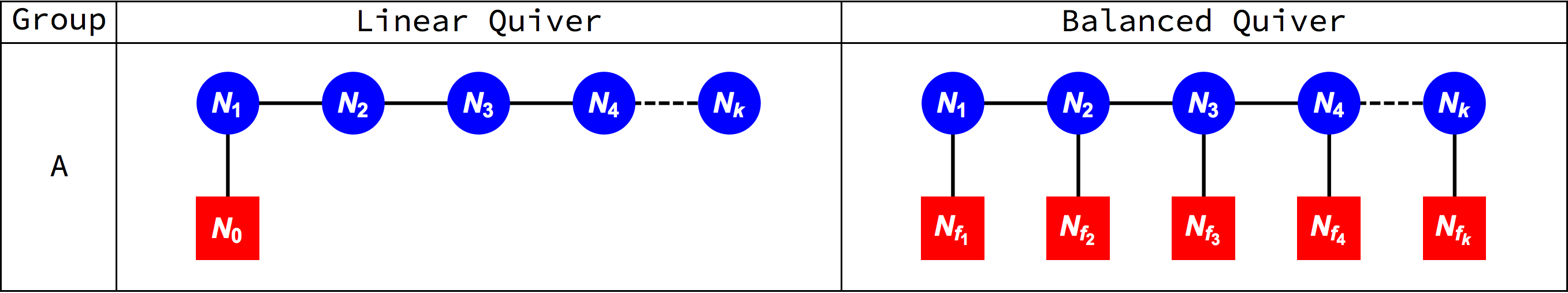}\\
\caption[$A$ Series Linear and Balanced Quiver Types.]{$A$ Series linear and balanced quiver types. In the canonical linear quiver, the unitary gauge nodes (blue) are in descending order with decrements in non-increasing order. In the balanced quiver, the unitary gauge nodes are all balanced by their attached gauge nodes (blue) and flavour nodes (red).}
\label{fig:A1}
\end{figure}

The constructions for the Slodowy slices of $A$ series nilpotent orbits draw upon the same two quiver types as the constructions for the closures of the nilpotent orbits. These are shown in figure \ref{fig:A1}:

\begin{enumerate}
\item Linear quivers based on partitions. These quivers ${\cal L}_A (\rho)$ consist of a $SU(N_0)$ flavour node connected to a linear chain of $U(N_{i})$ gauge nodes, where the decrements between nodes, $\rho_i = N_{i-1} - N_{i}$, constitute an ordered partition of $N_0$, $\rho \equiv \{\rho_1,\ldots,\rho_{{k}}\}$, where $\rho_i \ge \rho_{i+1}$ and $\sum \nolimits_{i = 1}^{k} {\rho _i} = N_0$.

\item Balanced quivers based on Dynkin diagrams. These quivers ${\cal B}_A ({\mathbf N_f})$ consist of a linear chain of $U(N_i)$ gauge nodes (in the form of an $A$ series Dynkin diagram), with each gauge node connected to a flavour node of rank $N_{f_i}$, where $N_{f_i} \ge 0$. The ranks of the gauge nodes are chosen such that each gauge node is balanced (as explained below), after taking account of any attached flavour nodes.

\end{enumerate}
On the Higgs branch, the flavour nodes of both types of quiver define an overall $S(\mathop \otimes \limits_i {U_{N_{f_i}}})$ global symmetry, while on the Coulomb branch, the global symmetry group follows from the Dynkin diagram formed by any balanced gauge nodes in the quiver.\footnote{The concept of balance was used in \cite{Gaiotto:2008sa}, in order to distinguish between (a) those Coulomb branch monopole operators that are ``good", with unit conformal dimension and act as root space operators, (b) those that are ``ugly" with half-integer conformal dimension and act as weight space operators, and (c) those that are ``bad" with zero or negative conformal dimension, which lead to divergences.}

The balance of a unitary gauge node is defined as the sum of the ranks of its adjacent gauge nodes, plus the number of attached flavours, less twice its rank:
\begin{equation} 
\label{eq:Aquivers1}
\begin{aligned}
\text{Balance(i)} \equiv {N_{f_i}} + \sum\limits_{j = i \pm 1} {N_j}  - 2{N_i} &.
\end{aligned}
\end{equation}
For the $A$ series balanced theories, the balance condition is ${\bf B} \equiv \{\text{Balance}(i)\}= \bf{0}$, and \ref{eq:Aquivers1} can be simplified as:
\begin{equation} 
\label{eq:Aquivers2}
\begin{aligned}
{\mathbf N_f} ={\mathbf A}  \cdot {\mathbf N},
\end{aligned}
\end{equation}
where the flavour and gauge nodes have been written as vectors ${\mathbf N_f} \equiv (N_{f_1},\ldots,N_{f_{k}})$ and ${\mathbf N} \equiv (N_{1},\ldots,N_{k})$, and $\mathbf A$ is the Cartan matrix of $A_{k}$.

$A$ series nilpotent orbits are in bijective correspondence with the partitions of $N$, and the linear quivers provide a complete set of Higgs branch constructions. The balanced quivers also provide a complete set of Coulomb branch constructions under the unitary monopole formula. Both types of quiver are thus in bijective correspondence with $A$ series orbits and can be related by $3d$ mirror symmetry \cite{Intriligator:1996ex}.

For Slodowy slices, the roles of these quiver types are reversed: the linear $A$ series quivers provide a complete set of Coulomb branch constructions, while the balanced $A$ series quivers provide a complete set of Higgs branch constructions.

When quivers of linear type are used to calculate Slodowy slices, via their Coulomb branches, the lack of balance of such quivers generally breaks the symmetry of $SU(N_0)$ to a subgroup, which becomes the isometry group of the Slodowy slice; this subgroup is in turn defined by the Dynkin diagram of the subset of gauge nodes in the linear quiver that remain balanced.

The identification of quivers for Slodowy slices follows directly from the partition data discussed in section \ref{sec:SS_Dim}. For the $A$ series, it is convenient to write the $SU(2)$ partition of the fundamental representation under $\rho$ as:

\begin{equation} 
\label{eq:Aquivers3}
\begin{aligned}
\rho {\left[ {1,0, \ldots } \right]_{A}} =\left( {{N^{{N_{{f_N}}}}}, \ldots ,{n^{{N_{{f_n}}}}}, \ldots ,{1^{{N_{{f_1}}}}}} \right), 
 \end{aligned}
\end{equation}
so that the multiplicities of partition elements, which may be zero, are mapped to the flavour vector ${\mathbf N_{f}}$. The \textit{linear} quiver ${\cal L}_A (\rho)$ can be extracted simply by writing $\rho [fund.]$ in long form. The ranks $\bf N$ of the gauge nodes of the \textit{balanced} quiver ${\cal B}_A ({\mathbf N_f}(\rho))$ can be found from $\mathbf N_f$ by inverting \ref{eq:Aquivers2}. Alternatively, the balanced quivers ${\cal B}_A ({\mathbf N_f}(\rho))$ can be obtained by applying $3d$ mirror symmetry transformations to the linear quivers ${\cal L}_A (\rho)$, and vice versa.

We can use the notation above to express the key relationships and dualities involving $A$ series quivers for the Slodowy slices of nilpotent orbits:

\begin{equation} 
\label{eq:Aquivers4}
\begin{aligned}
{{\cal S}}_{{{\cal N}},\rho }  = Higgs \left[ {{\cal B}_A( {{\mathbf N_f}( \rho)})} \right] &= Coulomb\left[ {{\cal L}_A( \rho)} \right],\\
{\bar {\cal O}}_\rho  = Higgs\left[ {{\cal L}_A(\rho ^T)} \right] &= Coulomb \left[ {{\cal B}_A( {{\mathbf N_f}( {{\rho ^T}} )} )} \right],\
 \end{aligned}
\end{equation}
or, taking the transpose of $\rho$:
\begin{equation} 
\label{eq:Aquivers4aa}
\begin{aligned}
{{\cal S}}_{{{\cal N}},{\rho^T} } = Higgs \left[ {{\cal B}_A( {{\mathbf N_f}( \rho ^T)})} \right] &= Coulomb\left[ {{\cal L}_A( \rho^T)} \right],\\
{\bar {\cal O}}_{\rho^T}= Higgs\left[ {{\cal L}_A(\rho) } \right] &= Coulomb \left[ {{\cal B}_A( {{\mathbf N_f}( {{\rho}} )} )} \right].\\
 \end{aligned}
\end{equation}
The quivers for $A$ series slices are related to the quivers for the underlying orbits simply by the transpose of the partition $\rho$, combined with exchange of Coulomb and Higgs branches. This transposition of partitions, which is an order reversing involution on the poset of $A$ series orbits, is known as the Lusztig-Spaltenstein map \cite{Baohua-Fu:2015nr}.

\begin{figure}[htbp]
\centering
\includegraphics[scale=0.5]{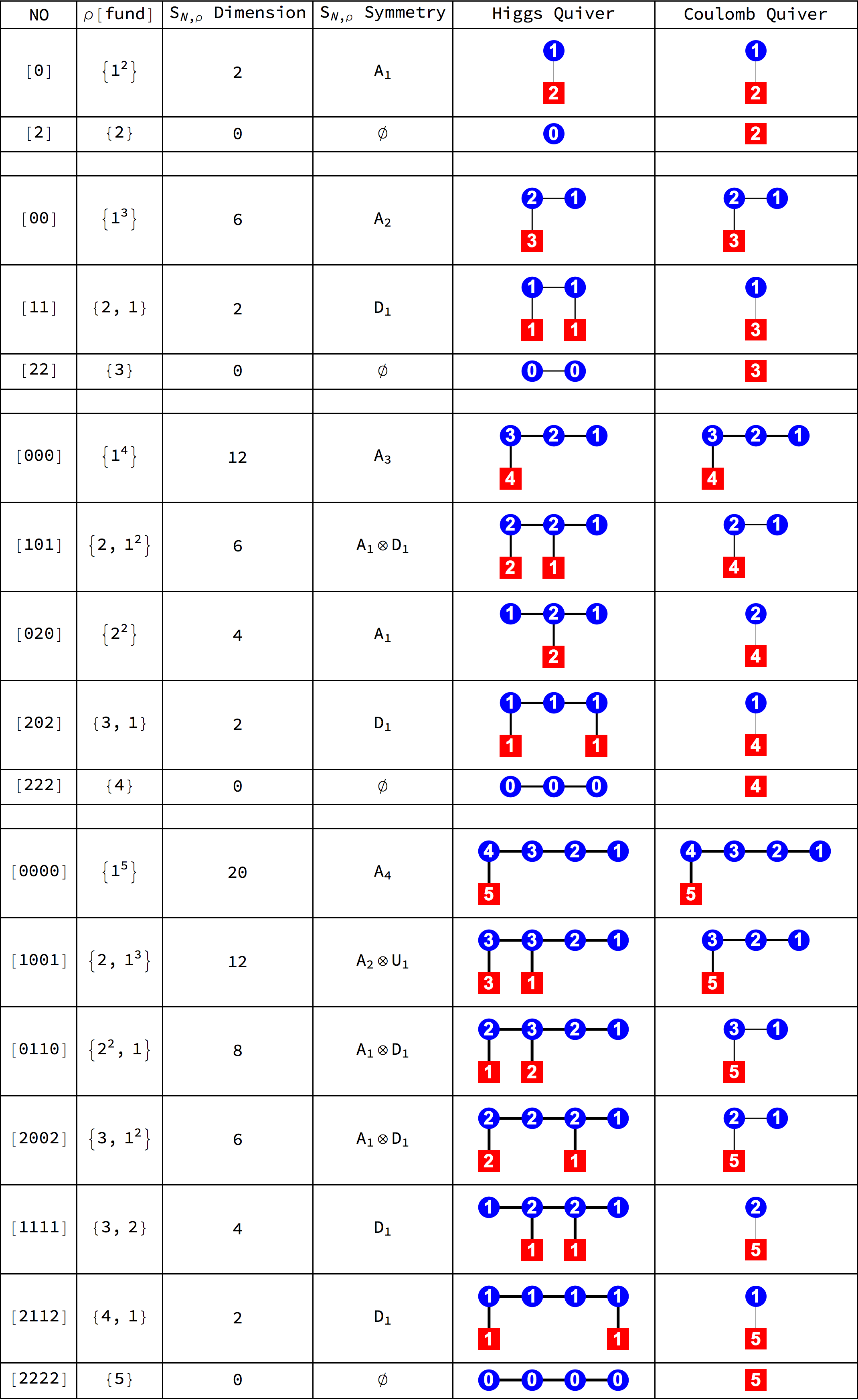}\\
\caption[Quivers for $A_1$ to $A_4$ Slodowy Slices.]{{Quivers for $A_1 to A_4$ Slodowy slices.} The Higgs quivers are of type ${{\cal B}_{A}} \left({\mathbf N_f}(\rho) \right)$ and the Coulomb quivers are of type ${{{\cal L}}_{A}}\left(\rho \right)$.}
\label{fig:A2}
\end{figure}

\begin{figure}[htbp]
\centering
\includegraphics[scale=0.5]{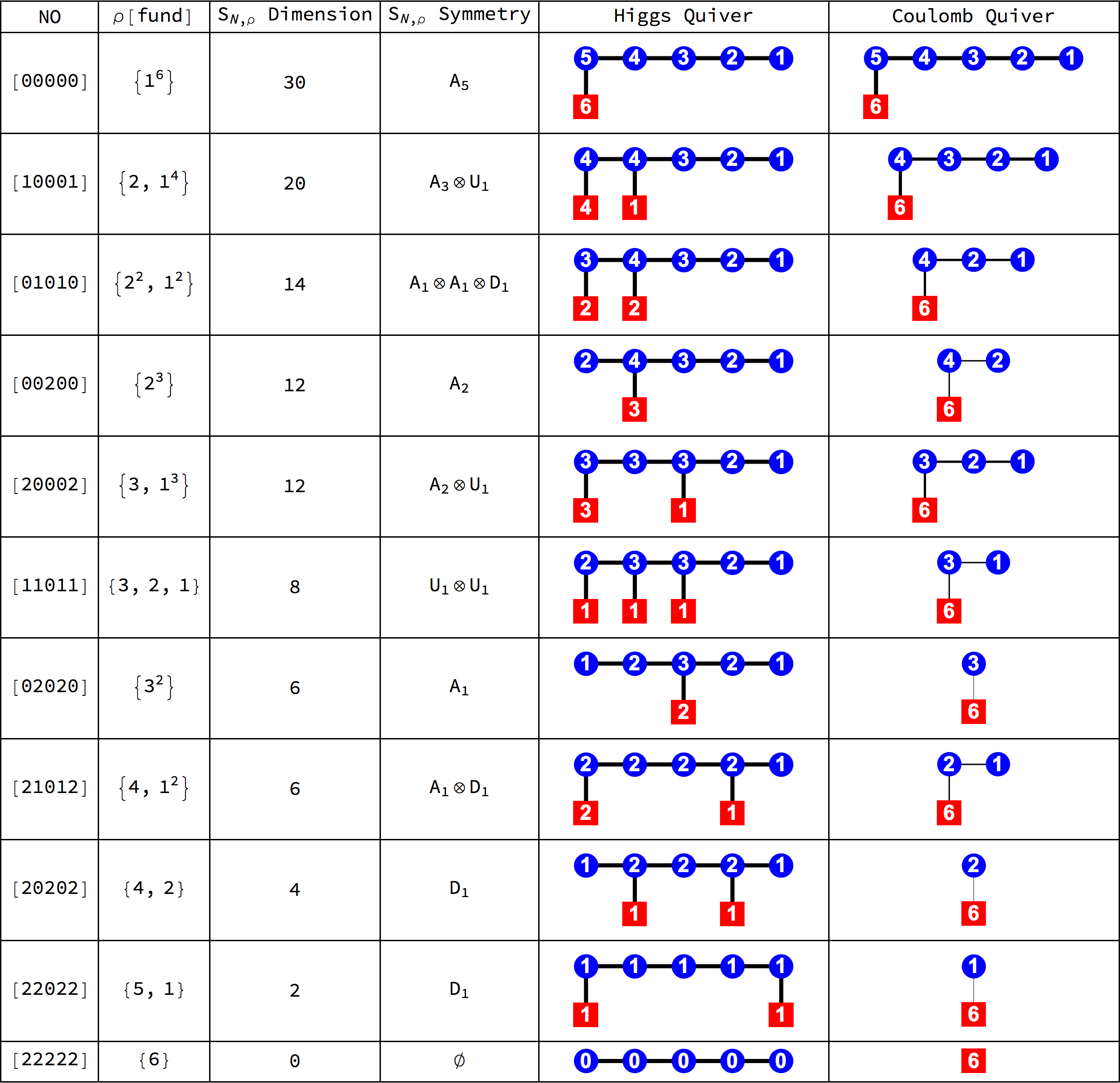}\\
\caption[Quivers for $A_5$ Slodowy Slices.]{{Quivers for $A_5$ Slodowy slices.} The Higgs quivers are of type ${{\cal B}_{A}} \left({\mathbf N_f}(\rho) \right)$ and the Coulomb quivers are of type ${{{\cal L}}_{A}}\left(\rho \right)$.}
\label{fig:A3}
\end{figure}

\FloatBarrier

These linear or balanced quiver types correspond to the limiting cases of $T_\sigma ^\rho \left( {SU\left( N \right)} \right)$ theories \cite{Gaiotto:2008ak, Cremonesi:2014uva}, where one of the partitions $\sigma$ or $\rho$ is taken as the trivial partition:

\begin{equation} 
\label{eq:Aquivers4a}
\begin{aligned}
{{\cal L}_A}(\rho )  \Leftrightarrow  T_\rho ^{(1,1, \ldots ,1)}\left( {SU\left( N \right)} \right),\\
{{\cal B}_A}({{\bf{N}}_f}(\rho )) \Leftrightarrow T_{(1,1, \ldots ,1)}^\rho \left( {SU\left( N \right)} \right).\\
 \end{aligned}
\end{equation}

Those quivers, whose Higgs or Coulomb branches yield Slodowy slices of $A$ series groups up to rank 5, are tabulated in figures \ref{fig:A2} and \ref{fig:A3}, labelled by the nilpotent orbit, giving the partition $\rho$ of the fundamental, the dimensions of the Slodowy slice, and the residual symmetry group\footnote{ We describe a $U(1)$ symmetry as $D_1$ if the characters $q^n$ of $U(1)$ irreps always appear paired with their conjugates in representations $(q^n+q^{-n})$.}. The balanced quivers used in the Higgs branch construction always have gauge nodes equal in number to the rank of $G= SU(N)$, while the linear quivers used in the Coulomb branch constructions always have a number of flavours equal to the fundamental dimension of $G = SU(N)$. The quivers ${\cal L}_A ((1^N))$ and ${\cal B}_A ({\mathbf N_f}(1^N))$ for the Higgs and Coulomb branch constructions of the Slodowy slice to the trivial nilpotent orbit are identical.

\subsection{Higgs Branch Constructions}
\label{subsec:AHiggs}

The calculation of Higgs branch Hilbert series from the balanced quivers draws on similar methods to those used in the calculation of the Higgs branches of the linear quivers for $A$ series nilpotent orbits, as elaborated in \cite{Hanany:2016gbz}. Pairs of bi-fundamental fields (and their complex conjugates) connect adjacent gauge nodes and, in addition, each non-trivial flavour node gives rise to a pair of bi-fundamental fields connected to its respective gauge node. The characters of all these fields are included in the PE symmetrisations. A HyperK\"ahler quotient is taken once for each gauge node, exactly as in the case of a linear quiver, and the Weyl integrations are then carried out over the gauge groups. The order of Weyl integrations can be chosen to facilitate computation.

The general Higgs branch formula for $A$ series Slodowy slices is:
\begin{equation} 
\small
\label{eq:Aquivers5}
\begin{aligned}
g_{HS}^{Higgs[{\cal B}_A({{\bf{N}}_{\bf{f}}}(\rho ))]} =&\\
 \oint\limits_{U\left( {{N_1}} \right) \otimes  \ldots U\left( {{N_{k}}} \right)}{d\mu}{~}& \prod\limits_{n = 1}^{k} {\frac{{PE\left[ {{{[ {fund.} ]}_{U( {{N_n}} )}} \otimes {{[ {anti.} ]}_{U( {{N_{{f_n}}}} )}} + {{[ {anti.} ]}_{U( {{N_n}} )}} \otimes {{[ {fund.} ]}_{U( {{N_{{f_n}}}} )}},t} \right]}}{{PE\left[ {{{[ {adjoint} ]}_{U( {{N_n}} )}},{t^2}} \right]}}} \\
 \times & \prod\limits_{n = 1}^{k - 1} {PE\left[ {{{[ {fund.} ]}_{U( {{N_n}} )}} \otimes {{[ {anti.} ]}_{U( {{N_{n + 1}}} )}} + {{[ {anti.} ]}_{U( {{N_n}} )}} \otimes {{[ {fund.} ]}_{U( {{N_{n + 1}}} )}},t} \right]},
 \end{aligned}
\end{equation}
where $ {d\mu}$ is the Haar measure for the ${U\left( {{N_1}} \right) \otimes  \ldots U\left( {{N_{k}}} \right)}$ product group. Note that the bifundamental fields are symmetrised with the fugacity $t$, while the HyperK\"ahler quotient (``HKQ") is symmetrised with $t^2$.

The Higgs branch formula can be simplified, by drawing on the dimensions of the bi-fundamentals and the gauge groups, to give a rule for the dimensions of an $A$ series Slodowy slice, and this can be simplified further by the balance condition \ref{eq:Aquivers2}:

\begin{equation} 
\label{eq:Aquivers6}
\begin{aligned}
\left| g_{HS}^{Higgs[{\cal B}_A ({{\bf{N}}_{\bf{f}}}(\rho ))]} \right| & = 2 {{\bf{N}}_f}\left( \rho  \right)\cdot {\bf{N}}\left( \rho  \right) - {\bf{N}}\left( \rho  \right) \cdot {\bf{A}} \cdot {\bf{N}}\left( \rho  \right)\\
& =  {{\bf{N}}_f}\left( \rho  \right)\cdot {\bf{N}}\left( \rho  \right).
 \end{aligned}
\end{equation}
For further details of the calculation methodology the reader is referred to the Plethystics Program Literature. The same Hilbert series can in principle also be obtained algebraically by working with matrix generators and relations, as in section \ref{subsec:Agenerators}.

\subsection{Coulomb Branch Constructions}
\label{subsec:ACoulomb}

The monopole formula, which was introduced in \cite{Cremonesi:2013lqa}, provides a systematic method for the construction of the Coulomb branches of particular SUSY quiver theories, being ${\cal N}=4$ superconformal gauge theories in $2+1$ dimensions. The Coulomb branches of these theories are HyperK\"ahler manifolds. The formula draws upon a lattice of monopole charges, defined by the linked system of gauge and flavour nodes in a quiver diagram.

Each gauge node carries adjoint valued fields from the SUSY vector multiplet and the links between nodes correspond to complex bi-fundamental scalars within SUSY hypermultiplets. The monopole formula generates the Coulomb branch of the quiver by projecting charge configurations from the monopole lattice into the root space lattice of $G$, according to the monopole flux over each gauge node, under a grading determined by the conformal dimension of each overall monopole flux $q$.

The \textit{conformal dimension} (equivalent to R-charge or the highest weight of the $SU(2)_R$ global symmetry) of a monopole flux is given by applying the following general schema \cite{Gaiotto:2008ak} to the quiver diagram:
\begin{equation}
\label{eq:mon0}
\Delta \left( q \right) = \underbrace {\frac{1}{2}\sum\limits_{i = 1}^r {\sum\limits_{{\rho _i} \in R_i}^{} {\left| {{\rho _i}(q)} \right|} } }_{\scriptstyle contribution~of~N = 4 \atop
\scriptstyle hyper~multiplets} - \underbrace {\sum\limits_{\alpha  \in \Phi_+ }^{}{\left| {\alpha (q)} \right|}}_{\scriptstyle contribution~of~N = 4\atop
\scriptstyle vector~multiplets}.
\end{equation}
The positive R-charge contribution in the first term comes from the bi-fundamental matter fields that link adjacent nodes in the quiver diagram. The second term captures a negative R-charge contribution from the vector multiplets, which arises due to symmetry breaking, whenever the monopole flux $q$ over a gauge node contains a number of different charges.

The calculation of Hilbert series for Coulomb branches of $A$ type quivers draws on the \textit{unitary} monopole formula, which follows from specialising \ref{eq:mon0} to unitary gauge groups. Each $U(N_i)$ gauge node carries a \textit{monopole flux} $q_i \equiv (q_{i,1}, \ldots ,q_{i,N_i})$ comprising one or more \textit{monopole charges} $q_{i,m}$. The fluxes are assigned the collective coordinate $q \equiv (q_1, \ldots, q_r)$. Each flavour node carries $N_{f_i}$ flavours of zero charge.\footnote{Flavour nodes may also carry non-zero charges, although these are not required by the Slodowy slice (or nilpotent orbit) constructions.}

With these specialisations, the conformal dimension $\Delta(q)$ associated with a flux $q$ yields the formula:
\begin{equation}
\label{eq:mon3}
\begin{aligned}
\Delta \left( {q } \right) = \frac{1}{2}\underbrace {\sum\limits_{j > i \ge 1}^r {\sum\limits_{m,n} {\left| {{q_{i,m}}{A_{ij}} - {q_{j,n}}{A_{ji}}} \right|} } }_{\text{gauge - gauge hypers}} 
& + \frac{1}{2} \underbrace {\sum\limits_{i} {\sum\limits_{m} {N_{f_i}  }{\left| {{q_{i,m}}} \right|} } }_{{\text{gauge   -   flavour hypers}}}\\
& - \underbrace {\sum\limits_{i = 1}^r {\sum\limits_{m > n}^{} {\left| {{q_{i,m}} - {q_{i,n}}} \right|} } }_{\text{gauge vplets}},
\end{aligned}
\end{equation}
where (i) the summations are taken over all the monopole charges within the flux $q$ and (ii) the linking pattern between nodes is defined by the $A_{ij}$ off-diagonal $A_r$ Cartan matrix terms, which are only non-zero for linked nodes.\footnote{For theories with simply laced quivers of $ADE$ type, $A_{ij} = 0$ or $-1$, for $i \neq j$.}

With conformal dimension defined as above, the \textit{unitary monopole formula} for a Coulomb branch HS is given by the schema \cite{Cremonesi:2013lqa}:

\begin{equation}
\label{eq:mon1}
g_{HS}^{{\rm{Coulomb}}}\left( {z,t^2} \right) \equiv \sum\limits_q {} {}P_q^{U\left( N \right)}\left( t^2 \right){z^q}{~}{t^{2 \Delta \left( q \right)}},
\end{equation}
where:
\begin{enumerate}

\item The limits of summation for the monopole charges are $\infty  \ge {q_{i,1}} \ge \ldots {q_{i,m}}  \ge \ldots {q_{i,{N_i}}} \ge  - \infty $ for $i=1,\ldots r$.

\item The monopole flux over the gauge nodes is counted by the fugacity $z \equiv (z_1, \ldots, z_r)$, where the $z_i$ are fugacities for the simple roots of $A_r$.

\item The monomial $z^q$ combines the monopole fluxes $q_i$ into total charges for each $z_i$ and is expanded as ${z^q} \equiv \prod\limits_{i = 1}^r {z_i^{\sum\limits_{m = 1}^{{N_i}} {{q_{i,m}}} }}$.

\item The term $P_{q}^{U\left( N \right)}$ encodes the degrees $d_{i,j}$ of the Casimirs of the residual $U(N)$ symmetries that remain at the gauge nodes under a monopole flux $q$:

\begin{equation}
\label{eq:mon2}
P_{q}^{U\left( N \right)}(t^2) \equiv \prod\limits_{i,j} {\frac{1}{{\left( {1 - {t^{{2 d_{i,j}(q)}}}} \right)}}}  = \prod\limits_{i = 1}^r {\prod\limits_{j = 1}^{{N_i}} {\prod\limits_{k = 1}^{{\lambda _{ij}}\left( {{q_i}} \right)} {\frac{1}{{1 - {t^ {2k}}}}} } }.
\end{equation}

Recalling that a $U(N)$ group has Casimirs of degrees 1 through $N$, the residual symmetries can be determined as in \cite{Cremonesi:2013lqa}.\footnote{We construct a partition of $N_i$ for each node, which counts how many of the charges $q_{i,m}$ are equal, such that $\lambda(q_i)=(\lambda_{i,1},\ldots,\lambda_{i,N_i})$, where $\sum\limits_{m = 1}^{{N_i}} {{\lambda _{i,m}}}  = {N_i}$ and ${\lambda _{i,m}} \ge {\lambda _{i,m+ 1}} \ge 0 $. The non-zero terms $\lambda_{i,j}$ in the partition give the ranks of the residual $U(N)$ symmetries associated with each node, so that it is a straightforward matter to compound the terms in the degrees of Casimirs. For example, if $q_{i,m}= q_{i,n}$ for all $m ,n$, then $\{d_{i,1},\ldots d_{i,N_i}\}=\{1,\ldots, N_i$\} and if $q_{i,m}\neq q_{i,n}$ for all $m, n$, then $\{d_{i,1},\ldots d_{i,N_i}\}=\{1,\ldots, 1\}$.} Alternatively, the residual symmetries for a flux $q_i$ can be fixed from the sub-group of $U(N_i)$ identified by the Dynkin diagram formed by those monopole charges that equal their successors $\{q_{i,m}: q_{i,m}=q_{i,m+1}\}$, (or equivalently, correspond to zero Dynkin labels).

\end{enumerate}
The exact calculation of a Coulomb branch HS can be carried out by evaluating \ref{eq:mon1} as a geometric series over each sub-lattice of monopole charges $q$, for which both conformal dimension $\Delta(q)$ and the symmetry factors $P_{q}^{U\left( N \right)}$ are linear (rather than piecewise or step) functions, and then summing the many resulting polynomial quotients. These sub-lattices of monopole charges form a hypersurface and care needs to be taken to avoid duplications at edges and intersections.

\subsection{Hilbert Series}
\label{subsec:AHilbert}

The Hilbert series of the Slodowy slices of algebras $A_1$ to $A_4$, calculated as above, are summarised in table \ref{tab:A1}. Both the Higgs and Coulomb branch calculations lead to identical refined Hilbert series, up to choice of CSA coordinates or fugacities. 

The Hilbert series are presented in terms of their generators, or $PL[HS]$, using character notation $[n_1,\ldots, n_r]$ to label $A_r$ irreps. Symmetrisation of these generators using the $PE$ recovers the refined Hilbert series. The underlying adjoint maps  \ref{eq:SS9} can readily be recovered from the generators by inverting \ref{eq:SS11}. The HS can be unrefined by replacing representations of the global symmetry groups by their dimensions.

\begin{sidewaystable}[htp]
\centering
\begin{tabular}{|c|c|c|c|c|}
\hline
${\begin{array}{c} \text{Nilpotent}\\\text{Orbit} \end{array}}$
&${\begin{array}{c} \text{Dimension}\\ {| {\cal S}_{\cal N,\rho}|} \end{array}}$
&${\begin{array}{c}  \text{Symmetry} \\ F\end{array}}$
&$ \text{Generators of HS} \equiv \text{PL[HS]}$
&$ \text{Unrefined HS} $\\
\hline
$[0]$&$ 2  $&
$A_1$& 
$  [2]t^2-t^4   $&$\frac {(1 - t^4)}{(1 - t^2)^3}    $\\
$[2]$&$ 0  $& $\emptyset$ & 
$  0   $&$  1    $\\
\hline

$[00]$&$ 6  $& $A_2$& 
$  [1,1]t^2-t^4-t^6   $&$ \frac{ (1 - t^4)}{(1 - t^2)^3 }   $\\
$[11]$&$ 2  $& $D_1$& 
$ t^2+(1)_{q_1/q_2} t^3-t^6   $&$ \frac{ (1 - t^6)}{(1 - t^2) (1 - t^3)^2}   $\\
$[22]$&$ 0  $&  $\emptyset$& 
$ 0   $&$ 1    $\\
\hline

$[000]$&$ 12  $& $A_3$& 
$  [1,0,1]t^2-t^4-t^6-t^8   $&$ \frac{(1 - t^4) (1 - t^6) (1 - t^8)}{(1 - t^2)^{15}}   $\\
$[101]$&$ 6  $& $A_1 \otimes D_1$& 
$t^2+[2]t^2+[1](1)_{q_1/q_2} t^3-t^6-t^8   $&$ \frac {(1 - t^6) (1 - t^8)}{(1 - t^2)^4 (1 - t^3)^4}   $\\
$[020]$&$ 4  $& $A_1$&
$ [2]t^2+[2]t^4-t^6-t^8 $&$\frac {(1 - t^6) (1 - t^8)}{(1 - t^2)^3 (1 - t^4)^3}  $\\
$[202]$&$ 2  $& $D_1$&
$t^2+(1)_{q_1/q_3}  t^4-t^8  $&$ \frac {(1 - t^8)}{(1 - t^2) (1 - t^4)^2} $\\
$[222]$&$ 0  $&  $\emptyset$&
$ 0   $&$ 1    $\\
\hline

$[0000]$&$ 20  $& $A_4$&
$  [1,0,0,1]t^2-t^4-t^6-t^8-t^{10}   $&$\frac {(1 - t^4) (1 - t^6) (1 - t^8) (1 - t^{10})}{(1 - t^2)^{24}} $\\
$[1001]$&$ 12  $& $A_2\otimes U(1)$& 
$t^2+[1,1]t^2+[1,0]{q_1/q_2} t^3+[0,1]{q_2/q_1} t^3-t^6-t^8-t^{10}   $&$\frac{(1 - t^6) (1 - t^8) (1 - t^{10})}{(1 - t^2)^9 (1 - t^3)^6} $\\
$[0110]$&$ 8  $& $A_1 \otimes D_1$& 
$t^2+ [2]t^2+[2]t^4+[1](1)_{q_1/q_2} t^3-t^6-t^8-t^{10} $&$\frac{(1 - t^6) (1 - t^8) (1 - t^{10})}{(1 - t^2)^4 (1 - t^3)^4 (1 - t^4)^3}  $\\
$[2002]$&$ 6  $& $A_1 \otimes D_1$& 
$t^2+[2]t^2+[1](1)_{q_1/q_3} t^4-t^8-t^{10}   $&$\frac {(1 - t^8) (1 - t^{10})}{(1 - t^2)^4 (1 - t^4)^4} $\\
$[1111]$&$ 4  $& $D_1$& 
$t^2+(1) t^3+t^4+(1)_{q_2/q_3} t^5-t^8-t^{10} $&$\frac {(1 - t^8) (1 - t^{10})}{(1 - t^2) (1 - t^3)^2 (1 - t^4) (1 - t^5)^2} $\\
$[2112]$&$ 2  $& $D_1$& 
$t^2+(1)_{q_1/q_4} t^5-t^{10}   $&$ \frac{(1 - t^{10})}{(1 - t^2) (1 - t^5)^2}  $\\
$[2222]$&$ 0  $&  $\emptyset$& 
$ 0   $&$ 1    $\\
\hline
\end{tabular}
\text{N.B. $(n)_q$ denotes the character of the $D_1 \equiv SO(2)$ reducible representation $q^n+q^{-n}$ of $U(1)$.}
\caption{Hilbert Series for Slodowy Slices of $A_1$, $A_2$, $A_3$ and $A_4$.}
\label{tab:A1}
\end{sidewaystable}

\begin{sidewaystable}[htp]
\centering
\begin{tabular}{|c|c|c|c|c|}
\hline
${\begin{array}{c} \text{Nilpotent}\\\text{Orbit} \end{array}}$
&${\begin{array}{c} \text{Dimension}\\ {| {\cal S}_{\cal N,\rho}|} \end{array}}$
&${\begin{array}{c}  \text{Symmetry} \\ F\end{array}}$
&$ \text{Generators of HS} \equiv \text{PL[HS]}$
&$ \text{Unrefined HS} $\\
\hline
$[00000]$&$ 30  $& $A_5$&
$  [1,0,0,0,1]t^2 - t^4-t^6  -t^8-t^{10} - t^{12}$
&$\frac{(1-t^4) (1-t^6) (1-t^8) (1-t^{10}) (1-t^{12})}{(1-t^2)^{35}} $\\
\hline
$[10001]$&$ 20  $& $A_3\otimes U(1)$& 
$\begin{array}{c}t^2+[1,0,1]t^2+([1,0,0]q_1/q_2+[0,0,1]q_2/q_1)t^3\\
-t^6-t^8-t^{10}-t^{12} \end{array}$
&$\frac{(1-t^6) (1-t^8) (1-t^{10}) (1-t^{12})}{(1-t^2)^{16} (1-t^3)^8} $\\
\hline
$[01010]$&$ 14 $& $A_1 \otimes A_1 \otimes D_1$& 
$\begin{array}{c}t^2+[2][0]t^2+[0][2]t^2+[1][1](1)_{q_1/q_2}t^3\\+[2][0]t^4-t^6-t^8-t^{10}-t^{12}\end{array}$
&$\frac{(1-t^6) (1-t^8) (1-t^{10}) (1-t^{12})}{(1-t^2)^7 (1-t^3)^8 (1-t^4)^3} $\\
\hline
$[00200]$&$ 12 $& $A_2$& 
$[1,1]t^2+[1,1]t^4-t^6-t^8-t^{10}-t^{12} $
&$\frac{ (1-t^6 )  (1-t^8 )  (1-t^{10} )  (1-t^{12} )}{ (1-t^2 )^8  (1-t^4 )^8} $\\
\hline
$[20002]$&$ 12$& $A_2 \otimes U(1)$& 
$\begin{array}{c}t^2+[1,1]t^2+[1,0] q_1/q_3 t^4+[0,1]q_3/q_1 t^4\\-t^8-t^{10}-t^{12}  \end{array} $
&$\frac{ (1-t^8 )  (1-t^{10} )  (1-t^{12} )}{ (1-t^2 )^9  (1-t^4 )^6} $\\
\hline
$[11011]$&$ 8$& $U(1)  \otimes U(1) $& 
$\begin{array}{c} 2{t^2} + ({(1)_{q1/q2}} + {(1)_{q2/q3}})){t^3} + {t^4} + {(1)_{{q_1}/{q_3}}}{t^4} \\+ (1)_{q_2/q_3}{t^5} - {t^8} - t^{10} - t^{12} \end{array} $
&$\frac{ (1-t^8 )  (1-t^{10} )  (1-t^{12} )}{ (1-t^2 )^2  (1-t^3 )^4  (1-t^4 )^3  (1-t^5 )^2} $\\
\hline
$[02020]$&$ 6  $& $A_1$& 
$[2]t^2+[2]t^4+[2]t^6-t^8-t^{10}-t^{12} $
&$\frac{ (1-t^8 )  (1-t^{10} )  (1-t^{12} )}{ (1-t^2 )^3  (1-t^4 )^3  (1-t^6 )^3}  $\\
\hline
$[21012]$&$ 6  $& $A_1 \otimes D_1  $& 
$t^2+[2]t^2+[1](1)_{q_1/q_4} t^5-t^{10}-t^{12}  $
&$\frac{ (1-t^{10} )  (1-t^{12} )}{ (1-t^2 )^4  (1-t^5 )^4} $\\
\hline
$[20202]$&$ 4  $& $D_1$& 
$t^2+t^4+(1)_{q_2/q_4}t^4+(1)_{q_2/q_4} t^6-t^{10}-t^{12}  $
&$\frac{ (1-t^{10} )  (1-t^{12} )}{ (1-t^2 )  (1-t^4 )^3  (1-t^6 )^2}  $\\
\hline
$[22022]$&$ 2  $& $D_1$& 
$t^2+(1)_{q_1/q_5} t^6-t^{12}$
&$\frac{1-t^{12}}{ (1-t^2 )  (1-t^6 )^2}  $\\
\hline
$[22222]$&$ 0  $&  $\emptyset$& $ 0   $&$ 1    $\\
\hline
\end{tabular}
\text{N.B. $(n)_q$ denotes the character of the $D_1 \equiv SO(2)$ reducible representation $q^n+q^{-n}$ of $U(1)$.}
\caption{Hilbert Series for Slodowy Slices of $A_5$.}
\label{tab:A5}
\end{sidewaystable}
Several observations can be made about the Hilbert series. 

\begin{enumerate}

\item As expected, (i) the Slodowy slice to the trivial nilpotent orbit $\mathcal{S}_{\mathcal{N},(1^N)}$ has the same Hilbert series as the nilpotent cone, (ii) the slice to the sub-regular orbit $\mathcal {S}_{\mathcal {N},(N-1,1)}$ has the Hilbert series of a Kleinian singularity of type $\hat A_{N-1}$, and (iii) the slice to the maximal nilpotent orbit $\mathcal {S}_{\mathcal {N},(N)}$ is trivial.

\item As expected, the Slodowy slices $\mathcal S_{\mathcal N,\rho}$ are all complete intersections.

\item The global symmetry groups of the Slodowy slice generators include mixed $SU$ and unitary groups, and descend in rank as the dimension of the Slodowy slice reduces. Sometimes different Slodowy slices share the same symmetry group, with inequivalent embeddings of $F$ into $G$.

\item Complex representations always appear combined with their conjugates to give real representations.

\item The adjoint maps often contain singlet generators at even powers of $t$ up to the (twice the) degree of the highest Casimir of $\mathfrak g$; these generators may be cancelled by one or more Casimir relations.

\end{enumerate}
Many of these observations have counterparts amongst the Slodowy slices of $BCD$ series, although these also raise several new intricacies, as will be seen in section \ref{sec:BCDSeries}.



\subsection{Matrix Generators for Unitary Quivers}
\label{subsec:Agenerators}

A Hilbert series over the class functions of a Classical group can be viewed in terms of matrix generators (or operators), and this perspective makes it possible to identify the generators of a Slodowy slice directly from the partition data or its Higgs branch quiver.

\subsubsection{Fundamental Decomposition}
\label{sec:AFD}

From \ref{eq:Aquivers3}, it follows that the character of the fundamental representation of $G$ decomposes into fundamental representations of a unitary product group:

\begin{equation} 
\label{eq:Agens1}
\begin{aligned}
\rho :\chi _{fund.}^G \to {  \bigoplus \limits_{[n]}}{[n]_{\rho}}{~}\chi _{[ {fund.} ]}^{{SU_{{N_{{f_{n+1}}}} }}}{q_{n+1} },
\end{aligned}
\end{equation}
where ${[n]_{\rho}}$ are irreps of the $SU(2)$ associated with the nilpotent orbit embedding $\rho$, and the $U(1)$ charges $q_i$ on the flavour nodes satisfy the overall gauge condition $\prod\limits_{i=1}^k {i} {N_{f_i}}{q_i} = 1$.\footnote{This corresponds to viewing the fields in a centre of mass frame.} Once this decomposition has been identified, the mapping of the adjoint of $G$ into matrix generators follows, by taking the product of the fundamental and antifundamental characters, and eliminating a singlet. This can be checked against the adjoint partition $\rho :\chi _{adjoint}^G$.

\FloatBarrier

\subsubsection{Generators from Quiver Paths}
\label{sec:OfQA}

Alternatively the operators can be read from a quiver of type ${B_A}({{\bf{N}}_{\bf{f}}}(\rho ))$, following the prescription:

\begin{enumerate}

\item Draw the chiral multiplets explicitly as arrows in the quiver:
\be
	\node{\fdu{}{N_{f_1}}}{\,\,N_1\,\,}\lrarr\node{\fdu{}{N_{f_2}}}{\,\,N_2\,\,}\lrarr\node{\fdu{}{N_{f_3}}}{\,\,N_3\,\,}\lrarr\cdots\lrarr\node{\fdu{}{N_{f_k}}}{\,\,N_k\,\,}
\ee

\item Every path in the quiver that starts and ends on a flavor node corresponds to an operator in the chiral ring of the Higgs branch.

\item There is a one to one correspondence between paths that appear as generators in the PL[HS] of the Higgs branch and the paths of the type $\mathcal{P}_{ij}(a)$, defined as below.

\item The operator $\mathcal P_{ij}(a)$ transforms under the fundamental representation of $U(N_{f_i})$ and the antifundamental representation of $U(N_{f_j})$ and sits on an irrep of $SU(2)_R$ with spin $s=A/2$, where $A$ is the number of arrows in the path that defines $\mathcal P_{ij}(a)$. This means that it appears in the plethystic logarithm of the refined Hilbert series as the character of the corresponding representation multiplied by the fugacity $t^{A}$.

\item Therefore, there is a one to one correspondence between operators $\mathcal P_{ij}(a)$ and irreducible representations in the decomposition of the adjoint representation of $A_k$ in  \ref{eq:SS9}.

\end{enumerate}

\begin{definition} {\it $\mathbf{\mathcal P_{ij}(a)}$: Let $\mathcal P_{ij}(a)$ be an operator $\mathcal P_{ij}(a)$ with $i,j \in \{1,2,\dots , k \}$ and $a \in \{1,2,\dots, min(i,j)\}$. $\mathcal{P}_{ij}(1)$ is defined as the operator formed by products of operators represented by arrows in the shortest possible path that starts at node $N_{f_i}$ and ends at node $N_{f_j}$ (note that $i$ and $j$ could be equal). $\mathcal{P}_{ij}(2)$ represents a path that differs from $\mathcal{P}_{ij}(1)$ only in that it has been extended to incorporate the arrows between the gauge nodes $N_{min(i,j)}$ and $N_{min(i,j)-1}$. $\mathcal{P}_{ij}(3)$ differs from $\mathcal{P}_{ij}(2)$ in that it also includes arrows between the gauge nodes $N_{min(i,j)-1}$ and $N_{min(i,j)-2}$. In this way $\mathcal P_{ij}(a)$ is defined recursively as an extension of $\mathcal P_{ij}(a-1)$.}
\end{definition}

\paragraph {Example 1.} Let us start with the balanced $A_3$ quiver based on the fundamental partition $\rho=(2,1^2)$, whose Higgs branch is the the Slodowy slice $\CS_{{\cal N},(2,1^2)}$ to the nilpotent orbit $A[101]$. The quiver is:

\be \label{eq:Agens2}
{B_A}({{\bf{N}}_{\bf{f}}}(2,1^2 )) ={~}  \node{\wver{}{\,\,2}}{2}- \node{\wver{}{\,\,1}}{2}- \node{}{1}.
\ee
%
%
From table \ref{tab:A1}, the Hilbert series  is:
\be\label{eq:Agens3}
	g_{HS}^{Higgs[{B_A}({{\bf{N}}_{\bf{f}}}(2,{1^2}))    ] } = PE[t^2+[2]t^2+[1](q+1/q)t^3-t^6-t^8].
\ee
To obtain this using the prescription in section \ref{sec:AFD}, we first identify the fugacity map for the group decomposition using \ref{eq:Agens1}:

\ba \label{eq:Agens4}
		SU(4) & \rightarrow SU(2)_{\rho} \otimes SU(2)\otimes U(1),\\
		[1,0,0] & \rightarrow [1]_{\rho} q^{1/2} + [1] q^{-1/2},\\
		[0,0,1] & \rightarrow [1]_{\rho} q^{-1/2} +[1] q^{1/2},\\
		[1,0,1] & \rightarrow ([2]+1)[0]_{\rho} + [1](q+1/q) [1]_{\rho}+[2]_{\rho}.\\
\ea		
%
Next the irreps $[n]_{\rho}$ of $SU(2)_{\rho}$ are mapped to the fugacity $t^{n+2}$, giving the generators:

\ba \label{eq:Agens5}
[1,0,1] & \rightarrow [2] t^2+ t^2+ [1](q+1/q) t^3 +t^4.\\
\ea
Subtracting the relations $-t^4-t^6-t^8$, corresponding to Casimirs of $A_3$, we obtain:

\be \label{eq:Agens6}
	PL[g_{HS}^{Higgs[{{\cal B}_{A} ({ \bf N_f}   (2,1^2)  )      }]}]=[2]t^2+t^2+[1](q+1/q)t^3-t^6-t^8.
\ee
The generators in \ref{eq:Agens6} can be understood as operators from paths in the quiver \ref{eq:Agens2}:

\begin{table}[htp]
\centering
\begin{tabular}{|c|c|c|}
\hline
$   {\cal P}_{ij}(a)         $&     Quiver Path        &       Generator       \\
\hline
$       {\cal P}_{1,1}(1)             $&$      \ \ \ \node {\fdu {}{\,\,2}}{2}\ \node{}{2} \ \  \node{}{1}         $&$        [2] t^2       $\\
\hline
$      {\cal P}_{2,2}(1)              $&$           \ \ \ \node{}{2}\ \node {\fdu {}{\,\,1}}{2}\ \ \node{}{1}    $&$          t^2     $\\
\hline
$      {\cal P}_{1,2}(1)              $&$           \ \ \ \node{\fd{}{\,\,2}}{2} \rarr \node{\fu {}{\,\,1}}{2}\ \node{}{1}   $&$     [1] q t^3         $\\
\hline
$        {\cal P}_{2,1}(1)            $&$        \ \ \ \node{\fu{}{\,\,2}}{2} \larr \node{\fd {}{\,\,1}}{2}\ \node{}{1}       $&$        [1] {q^{-1}} t^3       $\\
\hline
$       {\cal P}_{2,2}(2)             $&$      \ \ \ \node{}{2} \lrarr \node{\fdu {}{\,\,1}}{2}\ \node{}{1}      $&$      t^4         $\\
\hline
\end{tabular}
\caption{Generators for Slodowy Slice to $A[101]$.}
\label{tab:A2}
\end{table}
 %
%
The irrep of each generator corresponds with the flavor nodes where the path starts and ends. The $U(1)$ fugacity $q \equiv q_1/q_2$. The exponent of the fugacity $t$ corresponds to the length of the path. 

\paragraph {Example 2.} Now consider the balanced quiver based on the $A_4$ partition $(3,2)$, whose Higgs branch is the the Slodowy slice $\CS_{{\cal N},(3,2)}$ to the nilpotent orbit $A[1111]$:

\be
{{\cal B}_{A}{({\bf N_f}(3,2)})} ={~} \node{}{1}- \node{\wver{}{\,\,1}}{2}- \node{\wver{}{\,\,1}}{2}- \node{}{1}.
\ee
The group decomposition is:

\be \label{eq:ex2}
	SU(5) \rightarrow SU(2)_{\rho} \otimes S(U(1)\otimes U(1)).
\ee
The paths in the quiver can be used to predict the generators in table \ref{tab:A3}.
\begin{table}[htp]
\centering
\begin{tabular}{|c|c|c|}
\hline
$   {\cal P}_{ij}(a)         $&     Quiver Path        &       Generator       \\
\hline
$      {\cal P}_{2,2}(1)              $&$      \node{}{1}\ \ \node{\fdu{}{\,\,1}}{2}\ \ \node{}{2}\ \ \node{}{1}  $&$          t^2     $\\
\hline
$      {\cal P}_{3,3}(1)              $&$        \node{}{1}\ \ \node{}{2}\ \ \node{\fdu{}{\,\,1}}{2}\ \ \node{}{1}   $&$    t^2       $\\
\hline
$        {\cal P}_{2,3}(1)            $&$     \node{}{1}\ \ \node{\fd{}{\,\,1}}{2}\rarr \node{\fu{}{\,\,1}}{2}\ \ \node{}{1}   $&$ {q_2/q_3} t^3  $\\
\hline
$        {\cal P}_{3,2}(1)            $&$   \node{}{1}\ \ \node{\fu{}{\,\,1}}{2}\larr \node{\fd{}{\,\,1}}{2}\ \ \node{}{1}  $&$ {q_3/q_2} t^3  $\\
\hline
$        {\cal P}_{2,2}(2)            $&$  \node{}{1}\lrarr \node{\fdu{}{\,\,1}}{2}\ \ \node{}{2}\ \ \node{}{1}  $&$ t^4 $\\
\hline
$        {\cal P}_{3,3}(2)            $&$ \node{}{1}\ \ \node{}{2}\lrarr \node{\fdu{}{\,\,1}}{2}\ \ \node{}{1}  $&$ t^4 $\\
\hline
$        {\cal P}_{2,3}(2)            $&$  \node{}{1}\lrarr \node{\fd{}{\,\,1}}{2}\rarr \node{\fu{}{\,\,1}}{2}\ \ \node{}{1}   $&$ {q_2/q_3} t^5  $\\
\hline
$        {\cal P}_{3,2}(2)            $&$\node{}{1}\rlarr \node{\fu{}{\,\,1}}{2}\larr \node{\fd{}{\,\,1}}{2}\ \ \node{}{1} $&$ {q_3/q_2} t^5  $\\
\hline
$        {\cal P}_{3,3}(3)            $&$\node{}{1}\lrarr \node{}{2}\lrarr \node{\fdu{}{\,\,1}}{2}\ \ \node{}{1}  $&$ t^6 $\\
\hline
\end{tabular}
\caption{Generators for Slodowy Slice to $A[1111]$.}
\label{tab:A3}
\end{table}
Subtracting relations $-\sum_{i=1}^{5}t^{2i}$, corresponding to the special condition in \ref{eq:ex2}, which eliminates one of the $U(1)$ symmetries, and the Casimirs of $A_4$, and substituting $q$ for ${q_2/q_3}$ gives the expected $PL[HS]$:
\be
	PL[g_{HS}^{Higgs[{{\cal B}_{A} ({ \bf N_f} (3,2))}]}]=t^2+\lb q+\frac{1}{q}\rb t^3+t^4+\lb q+\frac{1}{q}\rb t^5-t^8-t^{10},
\ee
in accordance with table \ref{tab:A1}.

\subsubsection{Matrices and Relations}
\label{Arelations}

In this section we offer a reinterpretation of the previous results for Slodowy slices $\mathcal S_{\mathcal N,\rho}$ as sets of matrices that satisfy specific relations. The aim of this analysis is to build a bridge between the algebraic definition of the nilpotent cone $\mathcal S_{\mathcal N,(1^N)} = \mathcal N$ and that of the Kleinian singularity $\mathcal S_{\mathcal N,(N-1,1)} = \mathbb{C}^2/\mathbb{Z}_N$. 

First, let us remember that the Kleinian singularity $\mathcal S_{\mathcal N,(N-1,1)} = \mathbb{C}^2/\mathbb{Z}_N$ can be defined as the set of points parametrized by three complex variables $x,y,z\in \mathbb{C}$, subject to one relation:
\begin{equation}
	x^N=yz.
\end{equation}

Secondly, the nilpotent cone $\mathcal S_{\mathcal N,(1^N)} = \mathcal N$ can be defined as a set of complex variables arranged in a $N\times N$ matrix $M\in \mathbb{C}^{N\times N}$, subject to the following relations:
\begin{equation}\label{eq:nilpCone}
		\tr(M^p)=0\ \ \ \forall p =1,2,\dots, N.
\end{equation}

We want to show that a Slodowy slice $\mathcal S_{\mathcal N,\rho}$ can be viewed as an intermediate case between these two descriptions. In order to do this we build examples of varieties described by sets of complex matrices, choose relations among them and compute the (unrefined) Hilbert series of their coordinate rings, utilizing the algebraic software {\it Macaulay2} \cite{M2}. These Hilbert series can be checked to be the same as those in table \ref{tab:A1}.

The specific matrices that generate the coordinate rings are chosen according to the operators $\mathcal P_{ij}(a)$ found in the balanced quivers $\mathcal B_A(\mathbf N_f(\rho))$. For example, let us study the balanced quiver whose Higgs branch is the Slodowy slice $\mathcal S_{\mathcal N,(2,1^3)}$:
\begin{equation}
	\mathcal B_A(\mathbf N_f(2,1^3)) = \ \node{\wver{}{\,\, 3}}{3} - \node{\wver{}{\,\, 1}}{3} - \node{}{2} - \node{}{1}.
\end{equation}
One can assemble the generators $\Pija$ into three different complex matrices $M$, $A$ and $B$ of dimensions $3\times 3$, $3\times 1$ and $1\times 3$ respectively. Let us show how this can be done explicitly. There are five paths of the form $\Pija$: $\mathcal P_{11}(1)$, $\mathcal P_{22}(1)$, $\mathcal P_{22}(2)$, $\mathcal P_{12}(1)$, $\mathcal P_{21}(1)$. Out of these five sets of operators $P_{22}(1)$ can be removed by the relation $-t^2$ that removes the center of mass and $P_{22}(2)$ by the first Casimir invariant relation $-t^4$. This means that there is a remaining set of generators transforming in the following irreps:
\begin{equation}
	\begin{aligned}
		\mathcal P_{11}(1) &\rightarrow \ ([1,1] + [0,0])t^2,\\
		\mathcal P_{12}(1) &\rightarrow \ ([1,0]q)t^3,\\
		\mathcal P_{21}(1) &\rightarrow \ ([0,1]\frac{1}{q})t^3.\\
	\end{aligned}
\end{equation}

One can now assemble these generators in three complex matrices that transform in the usual way under the global symmetry $U(3)$:
\begin{equation}
	\begin{aligned}
		([1,1] + [0,0])t^2 &\rightarrow \ M_{3\times 3},\\
		 ([1,0]q)t^3 &\rightarrow \ A_{1\times 3},\\
		 ([0,1]\frac{1}{q})t^3 &\rightarrow \ B_{3\times 1}.\\
	\end{aligned}
\end{equation}

The chiral ring is then parametrized by the set of all matrices $\{M,A,B\}$, subject to one relation at order $t^6$, another relation at order $t^8$ and a final relation at order $t^{10}$. These relations are invariant under the global $U(3)$ symmetry. One can choose the following set of relations:
\begin{align}
	\tr(M^3)&=AB,\\
	\tr(M^4)&=AMB,\\
	\tr(M^5)&= AM^2 B.
\end{align}
Note that these look like corrections to the equations of the nilpotent cone \ref{eq:nilpCone}. The Hilbert series of the coordinate ring is then computed using {\it Macaulay2} to be:
\begin{equation}
	HS=\frac{(1-t^6)(1-t^8)(1-t^{10})}{(1-t^2)^9(1-t^3)^6}.
\end{equation}
This is the same Hilbert series as that of the variety $\slod{2,1^3}$ computed in table \ref{tab:A1}.

In tables \ref{tab:A456} and \ref{tab:A7} we provide a set of algebraic varieties described by matrices such that their HS have been computed to be identical to those of the corresponding Slodowy slices $\CS_{\mathcal N,\rho}$. Note that we rewrite the Kleinian singularity in terms of $1\times 1$ matrices, to clarify the connection with the algebraic description of the other Slodowy slices.

\begin{table}[h]
	\centering
	\begin{tabular}{|c|c|c|c|c|}
	\hline
	Orbit & Partition & Dimension & Generators; Degree & Relations \\
	 \hline
	\multirow{4}{*}{[0]}&\multirow{4}{*}{$(1^2)$}&\multirow{4}{*}{2}& $\begin{array}{rc} M_{2\times 2}; & 2 \\  \end{array}$&$\begin{array}{rl}tr(M) &= 0 \\ tr(M^2) &=0  \end{array}$\\ \cline{4-5}
    	&&&$\begin{array}{rc} M_{1\times 1}; & 2 \\  {A}_{1\times 1} ; & 2 \\ {B}_{1\times 1} ; & 2\end{array}$&$\begin{array}{rl}tr(M^2) &= {AB}  \end{array}$\\ \hline
	[2] & $(2)$ & 0 & - & - \\ 
	\hline
	\hline
	[00] & $(1^3)$ & 6 & $\begin{array}{rc} M_{3\times 3}; & 2\end{array}$ & $\begin{array}{rl}tr(M) &=0 \\ tr(M^2) &=0 \\ tr(M^3) &=0   \end{array}$  \\ \hline
	[11] & $(2,1)$ & 2 & $\begin{array}{rc}M_{1\times 1} ; & 2 \\ {A}_{1\times 1} ; & 3 \\ {B}_{1\times 1} ; & 3 \\\end{array}$ & $\begin{array}{rl} tr(M^3) & ={AB}  \\  \end{array}$  \\ \hline
	[22] & $(3)$ & 0 & - & - \\ 
	\hline
	\hline
	[000] & $(1^4)$ & 12 & $\begin{array}{rc} M_{4\times 4}; & 2\end{array}$ & $\begin{array}{rl}tr(M) &=0 \\ tr(M^2) &=0 \\ tr(M^3) &=0 \\ tr(M^4) &=0   \end{array}$  \\ \hline
	[101] & $(2,1^2)$ & 6 & $\begin{array}{rc} M_{2\times 2}; & 2  \\ {A}_{1\times 2} ; & 3 \\  B _{2\times 1} ; & 3\\ \end{array}$ & $\begin{array}{rl} tr(M^3) &= AB \\ tr(M^4) &=AMB   \end{array} $ \\ \hline
	[020] &  $(2^2)$ & 4 & $\begin{array}{rc}M_{2\times 2} ; & 2 \\ N_{2\times 2} ; & 4 \\ \end{array}$ & $\begin{array}{rl} tr(M) &= 0 \\tr(N) &= 0 \\tr(M^3) &= tr(MN) \\ tr(M^4)&=tr(N^2) \end{array}$  \\ \hline
	[202] & $(3,1)$ & 2 & $\begin{array}{rc}M_{1\times 1} ; & 2 \\ {A}_{1\times 1} ; & 4 \\ {B}_{1\times 1} ; & 4 \\\end{array}$ & $\begin{array}{rl} tr(M^4) & ={AB}  \\  \end{array}$  \\ \hline
	[222] & $(4)$ & 0 & - & - \\ \hline
	\end{tabular}
	\caption[$A_1$, $A_2$ and $A_3$ Slodowy Slice Varieties]{$A_1$, $A_2$ and $A_3$ varieties, generated by complex matrices $M$, $A$ and $B$ and their relations, with Hilbert series calcuated by {\it Macaulay2} to match Slodowy slices $\CS_{\mathcal N,\rho}$. Note that $\CS_{\mathcal N,(1^2)}$ has two alternative descriptions, one as the nilpotent cone and one as the subregular Kleinian singularity.}
	\label{tab:A456}
\end{table}

\begin{table}[h]
	\centering
	\begin{tabular}{|c|c|c|c|c|}
	\hline
	Orbit & Partition & Dimension & Generators; Degree & Relations \\ \hline
	[0000] & $(1^5)$ & 20 & $\begin{array}{rc} M_{5\times 5}; & 2\end{array}$ & $\begin{array}{rl}tr(M) &=0 \\ tr(M^2) &=0 \\ tr(M^3) &=0 \\ tr(M^4) &=0 \\ tr(M^5) &=0 \\  \end{array}$  \\ \hline
	[1001] & $(2,1^3)$ & 12 & $\begin{array}{rc} M_{3\times 3}; & 2  \\ {A}_{1\times 3} ; & 3 \\  B _{3\times 1} ; & 3\\ \end{array}$ & $\begin{array}{rl} tr(M^3) &= AB \\ tr(M^4) &=AMB \\ tr(M^5) &={AM^2B}  \\  \end{array} $ \\ \hline
	[0110] & $(2^2,1)$ & 8 & $\begin{array}{rc}M_{2\times 2} ; & 2 \\ {A}_{1\times 2} ; & 3 \\ {B}_{2\times 1} ; & 3 \\ N_{2\times 2} ; & 4 \\ \end{array}$ & $\begin{array}{rl}  tr(M^3) &= {AB} \\ tr(M^4) + tr(N^2) &={AMB} \\ tr(M^5) &={A(M^2+N)B}  \\  tr(N)&=0 \\  \end{array}$  \\ \hline
	[2002] & $(3,1^2)$ & 6 & $\begin{array}{rc}M_{2\times 2} ; & 2 \\ {A}_{1\times 2} ; & 4 \\ {B}_{2\times 1} ; & 4  \\ \end{array}$ & $\begin{array}{rl}tr(M^4) &= {AB} \\ tr(M^5) &= {AMB} \\  \end{array}$  \\ \hline
	[1111] & $(3,2)$ & 4 & $\begin{array}{rc}M_{1\times 1} ; & 2 \\ {A}_{1\times 1} ; & 3 \\ {B}_{1\times 1} ; & 3 \\ N_{1\times 1} ; & 4 \\ {C}_{1\times 1} ; & 5 \\ {D}_{1\times 1} ; & 5 \\\end{array}$ & $\begin{array}{rl} tr(M^4) + tr(N^2) &={A MB} + {AD} \\
	& +{BC} \\ tr(M^5) &=  {CD}\end{array}$  \\ \hline
	[2112] & $(4,1)$ & 2 & $\begin{array}{rc}M_{1\times 1} ; & 2 \\ {A}_{1\times 1} ; & 5 \\ {B}_{1\times 1} ; & 5 \\\end{array}$ & $\begin{array}{rl} tr(M^5) & ={AB}  \\  \end{array}$  \\ \hline
	[2222] & $(5)$ & 0 & - & - \\ \hline
	\end{tabular}
	\caption[$A_4$ Slodowy Slice Varieties]{$A_4$ varieties, generated by complex matrices $M$, $A$ and $B$ and their relations, with Hilbert series  calcuated by {\it Macaulay2} to match Slodowy slices $\CS_{\mathcal N,\rho}$.}
	\label{tab:A7}
	\end{table}

\FloatBarrier


\section{$BCD$ Series Quiver Constructions}
\label{sec:BCDSeries}

\subsection{Quiver Types}
\label{subsec:BCDQuivers}

The constructions for the Slodowy slices of $BCD$ algebras draw upon a different set of quiver types to the A series.

\begin{enumerate}
\item Linear orthosymplectic quivers. These quivers ${{\cal L}_{B/C/D}} (\sigma)$ consist of a $B$, $C$ or $D$ series flavour node of vector irrep dimension $N_0$  connected to an alternating linear chain of $(S)O/USp({N_i})$ gauge nodes of non-increasing vector dimension. For a subset of these linear quivers, the decrements, $\sigma_i = N_{i-1} - N_{i}$, between nodes constitute an ordered partition of $N_0$, $\sigma \equiv \{\sigma_1,\ldots,\sigma_{{k}}\}$, where $\sigma_i \ge \sigma_{i+1}$ and $\sum\nolimits_{i =1}^{k} {{\sigma _i}}  = N_0$. More generally, however, the $\sigma_i$ form a sequence of non-negative integers, subject to $\sum\nolimits_{i=1}^{k} {\sigma _i}  = N_0$, and to selection rules, such that $USp$ nodes of odd dimension do not arise.

\item Balanced orthosymplectic quivers. These quivers ${{\cal B}_{B/C/D}} ({\mathbf N_f})$ consist of an alternating linear chain of $O/USp(N_{i})$ nodes, with each gauge node connected to a flavour node $O/USp(N_{f_i})$, where $N_{f_i} \ge 0$. The ranks of the gauge nodes are chosen such that, taking account of any attached flavour nodes, each gauge node inherits its balance ${\bf B}$ (via \ref{eq:BCDquivers1}) from that of a canonical quiver (as defined below).

\item Dynkin diagram quivers. These quivers ${\cal D}_G ({\mathbf N_f})$ consist of a chain of $U(N_i)$ gauge nodes in the form of a simply laced Dynkin diagram, with each gauge node connected to $N_{f_i}$ flavours, where $N_{f_i} \ge 0$. ${\mathbf N_f}$ matches the Characteristic $G[\ldots]$ of a nilpotent orbit, and the ranks of the gauge nodes are chosen such that each is balanced (similarly to the A series quivers in section \ref{subsec:AQuivers}). These constructions are limited to certain Slodowy slices of $ADE$ algebras, as the Higgs branch construction is not available on non-simply laced Dynkin diagrams.
\end{enumerate}
Recall, the  \textit{nilpotent orbits} of a $BCD$ algebra correspond to a subset of the partitions $\rho$ of $N$, once these have been subjected to selection rules,\footnote{In a valid $B$ or $D$ partition $\rho$ each even integer appears at an even multiplicity; in a valid $C$ partition each odd integer appears at an even multiplicity \cite{Collingwood:1993fk}.} and linear quivers ${{\cal L}_{B/C/D}} (\rho^T)$ provide a complete set of Higgs branch constructions.
Also, balanced quivers ${{\cal B}_{B/C/D}} ({\mathbf N_f})$ provide Coulomb branch constructions, using the $O/USp$ monopole formula, for the unrefined Hilbert series of certain nilpotent orbits of orthogonal groups, as discussed in \cite{Cabrera:2017ucb}. The linear and balanced quivers can partially be related by $3d$ mirror symmetry, as discussed further in section \ref{sec:Conclusions}.
Many of these linear quivers have ``Higgs equivalent" quivers, ${{\cal L}_{B/C/D}} (\sigma)$, with a different choice of orthogonal gauge node dimensions, but the same Higgs branches; these are generally described by sequences $\sigma$ rather than partitions $\rho^T$: a $USp-O-USp$ subchain with the sub-partition $(\ldots, n,n,\ldots )$ has the Higgs equivalent sequence $(\ldots, \sigma_i,\sigma_{i+1},\ldots ) = (\ldots,n-1, n+1,\ldots )$, in which the vector dimension of the central $O$ node is increased by 1 \cite{Hanany:2016gbz}.

\begin{figure}[htbp]
\includegraphics[scale=0.42]{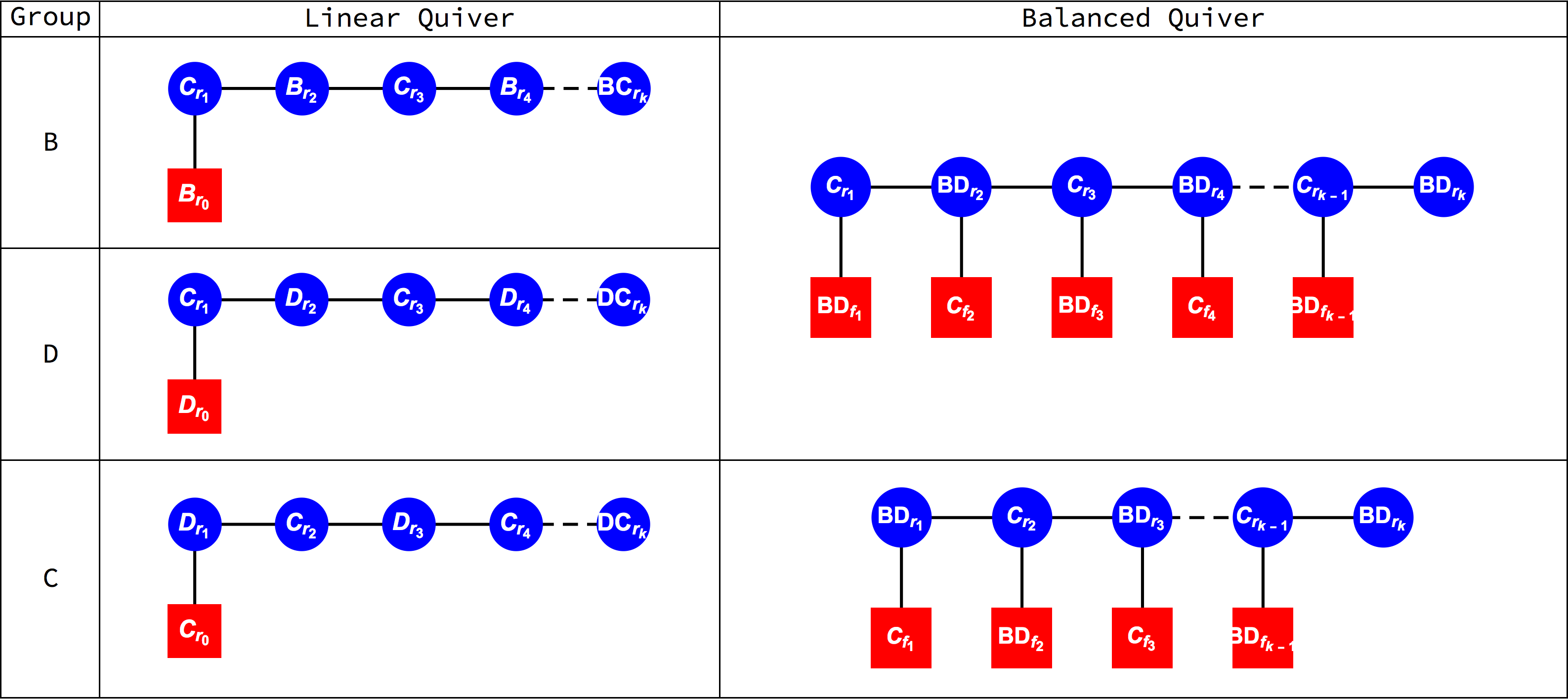}\\
\caption[$BCD$ Series Linear and Balanced Quiver Types]{$BCD$ linear and balanced quiver types. In the linear quivers ${\cal L}_{BC}$, ${\cal L}_{CD}$ and ${\cal L}_{DC}$, the ranks and fundmental dimensions of the gauge nodes (blue) are in non-increasing order L to R and the quivers are in the form of alternating $B-C$ or $D-C$ chains. In the balanced quivers, ${\cal B}_{B/C/D}$, the gauge nodes (blue) inherit their balance, taking account of attached gauge and flavour nodes (red), from a quiver for the nilpotent cone. Nodes labelled $C_r$ represent the group $USp(2r)$. Nodes labelled $B_r$ and $D_r$ represent $SO/O(2r+1)$ and $SO/O(2r)$ respectively.  Nodes labelled $BC$, $BD$ or $DC$ indicate a group of one of the two types, subject to the alternation rule and to balance.}
\label{fig:BCD1}
\end{figure}

Returning to \textit{Slodowy slices}, the roles of these quiver types are essentially reversed: balanced quivers  ${{\cal B}_{B/C/D}}$ provide a \textit{complete} set of Higgs branch \textit{refined} Hilbert series constructions, while linear quivers ${{\cal L}_{B/C/D}}$ provide Coulomb branch constructions for the \textit{unrefined} HS of \textit{certain} Slodowy slices. Within the general classes of linear and balanced quiver types, those that are most relevant to the construction of Slodowy slices are shown in figure \ref{fig:BCD1}.

We refer to the quivers of type ${\cal L}_{BC}$, ${\cal L}_{CD}$ or $ {\cal L}_{DC}$, which contain pure $B-C$, $C-D$ or $D-C$ chains, as \textit{canonical} linear quivers. On the Higgs branch, the flavour nodes (of either type of quiver) identify the overall global symmetry, although it is necessary to distinguish within the $B$ and $D$ series between $O$ and $SO$ groups. However, it is not easy to identify the global symmetry of the Coulomb branch of a $O/USp$ quiver.

It is important to explain how the specific quivers used in the construction of the Hilbert series for $BCD$ Slodowy slices arise from the partition of the vector representation of $G$ under the homomorphism $\rho$. 

The balanced quivers ${{\cal B}_{B/C/D}} ({\mathbf N_f(\rho)})$ are found via a modification of the $A$ series method explained in section \ref{subsec:AQuivers}. Firstly, the $SU(2)$ partition of a $BCD$ series vector representation under $\rho$ can be used to define a vector ${\mathbf N_f}(\rho)$ of alternating $O/USp$ flavour nodes, similarly to \ref{eq:Aquivers3}:

\begin{equation} 
\label{eq:BCDquivers3}
\begin{aligned}
\rho {\left[ {1,0, \ldots } \right]_{{B/C/D}}} =\left( {{N^{{N_{{f_N}}}}}, \ldots ,{n^{{N_{{f_n}}}}}, \ldots ,{1^{{N_{{f_1}}}}}} \right).
 \end{aligned}
\end{equation}
Next, consider linear quivers, whose Higgs branches match the nilpotent cone $\cal N$. In the case of $BCD$ groups, these quivers can be chosen, using Higgs equivalences, to be of canonical type. The balances ${\mathbf B}$ of their gauge nodes can be calculated by applying \ref{eq:Aquivers1} to vectors ${\mathbf N_f}$ and ${\mathbf N}$  defined from the vector/fundamental dimensions of the fields, as shown in table \ref{tab:BCD1}.

\begin{table}[htp]
\begin{center}
\begin{tabular}{|c|c|c|c|}
\hline
Group & Canonical Linear Quiver for $\cal N$ & $ \begin{array}{c} \text{Gauge Node}\\ \text{Balance}  \end{array} $\\
\hline
$ A $ & $\ \node{\wver{}{\,\, SU(N)}}{U(N-1)}-\node{}{U(N-2)}- \ldots - \node{}{U(2)}-\node{}{U(1)}$ &$0$ for all\\
\hline
$ B $ & $\ \node{\wver{}{\,\, SO(2n+1)}}{USp(2n)}-\node{}{O(2n-1)}- \ldots - \node{}{USp(2)}-\node{}{O(1)}$ &$0$ for all\\
\hline
$ C $&$\ \node{\wver{}{\,\, USp(2n)}}{O(2n)}-\node{}{USp(2n-2)}- \ldots - \node{}{USp(2)}-\node{}{O(2)}$ &$\left\{ \begin{array}{c} USp: + 2\\ O({even}): - 2 \end{array} \right. $ \\
\hline
$ D $&$\ \node{\wver{}{\,\, SO(2n)}}{USp(2n-2)}-\node{}{O(2n-2)}- \ldots - \node{}{USp(2)}-\node{}{O(2)}$&$\left\{ \begin{array}{c} USp: + 2\\ O({even}): - 2 \end{array} \right. $ \\
\hline
\end{tabular}
\end{center}
\caption{Higgs Branch Quivers for Nilpotent Cones.} 
\label{tab:BCD1}
\end{table}
%
These canonical quivers obey the generalisation of \ref{eq:Aquivers2}:
\begin{equation} 
\label{eq:BCDquivers1}
\begin{aligned}
{\mathbf N_f} ={\mathbf A}  \cdot {\mathbf N} + {\mathbf B},
\end{aligned}
\end{equation}
Whereas ${\mathbf B}=(0,\ldots,0)$ for the $A$ and $B$ series canonical quivers, ${\mathbf B}=(-2,2,\ldots,-2)$ for the $C$ series and ${\mathbf B}=(2,-2,\ldots,-2)$ for the $D$ series canonical quivers.

\begin{figure}[htbp]
\begin{center}
\includegraphics[scale=0.45]{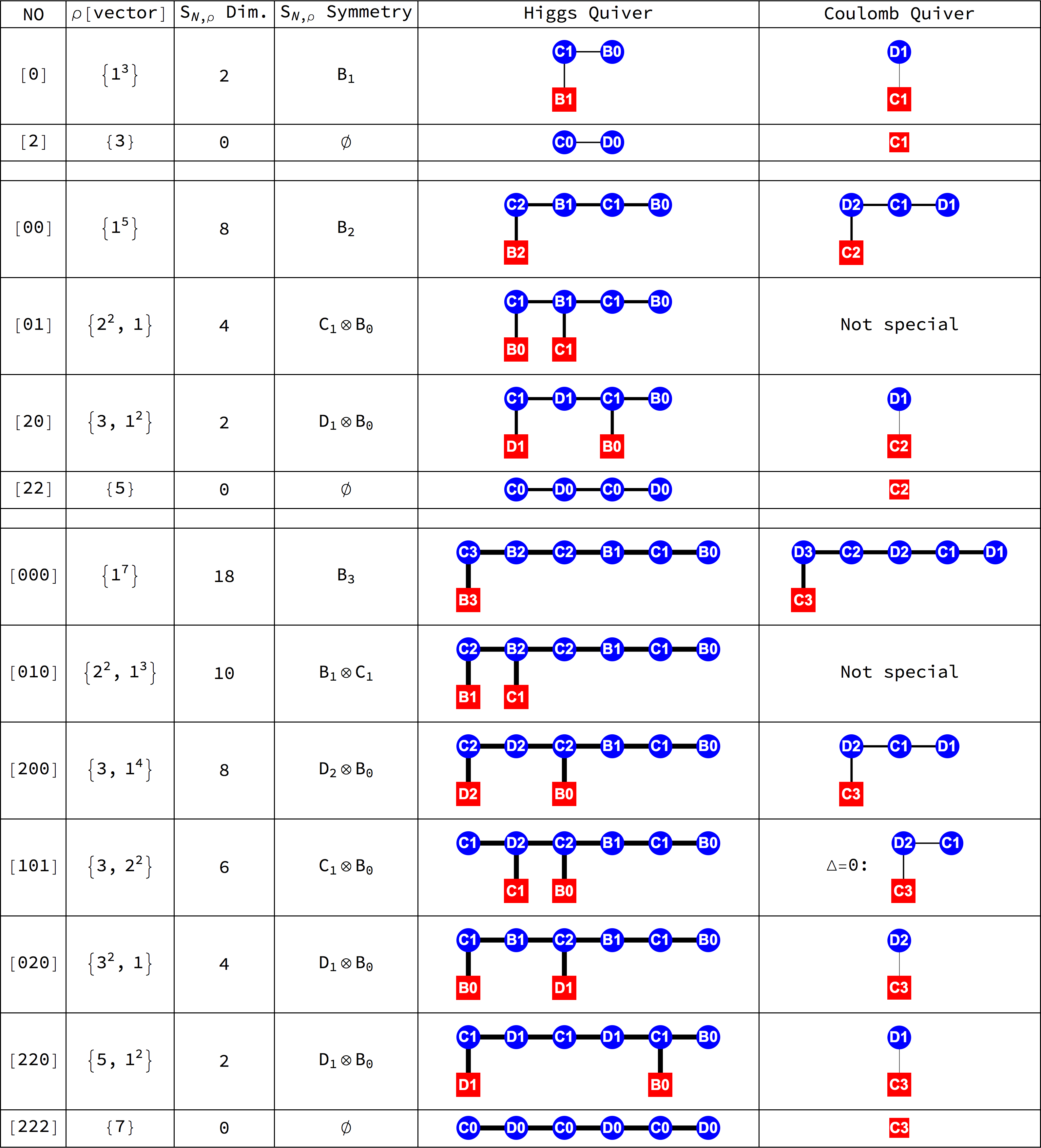}\\
\caption[Quivers for $B_1$ to $B_3$ Slodowy Slices.]{{Quivers for ${ B_1}$ to ${ B_3}$ Slodowy Slices}. The Higgs quivers are of type ${{\cal B}_{B/C/D}} \left({\mathbf N_f}(\rho) \right)$ and the Coulomb quivers are of type ${{\cal L}_{CD}}\left( {{{\left. {d_{BV}{\left( \rho  \right)}^T} \right|}_{CD}}} \right)$. Gauge nodes of $B$ or $D$ type are evaluated as $O$ nodes on the Higgs branch and $SO$ nodes on the Coulomb branch. $\Delta=0$ indicates a diagram for which the monopole formula contains zero conformal dimension.}
\label{fig:BCD3a}
\end{center}
\end{figure}

\begin{figure}[htbp]
\begin{center}
\includegraphics[scale=0.35]{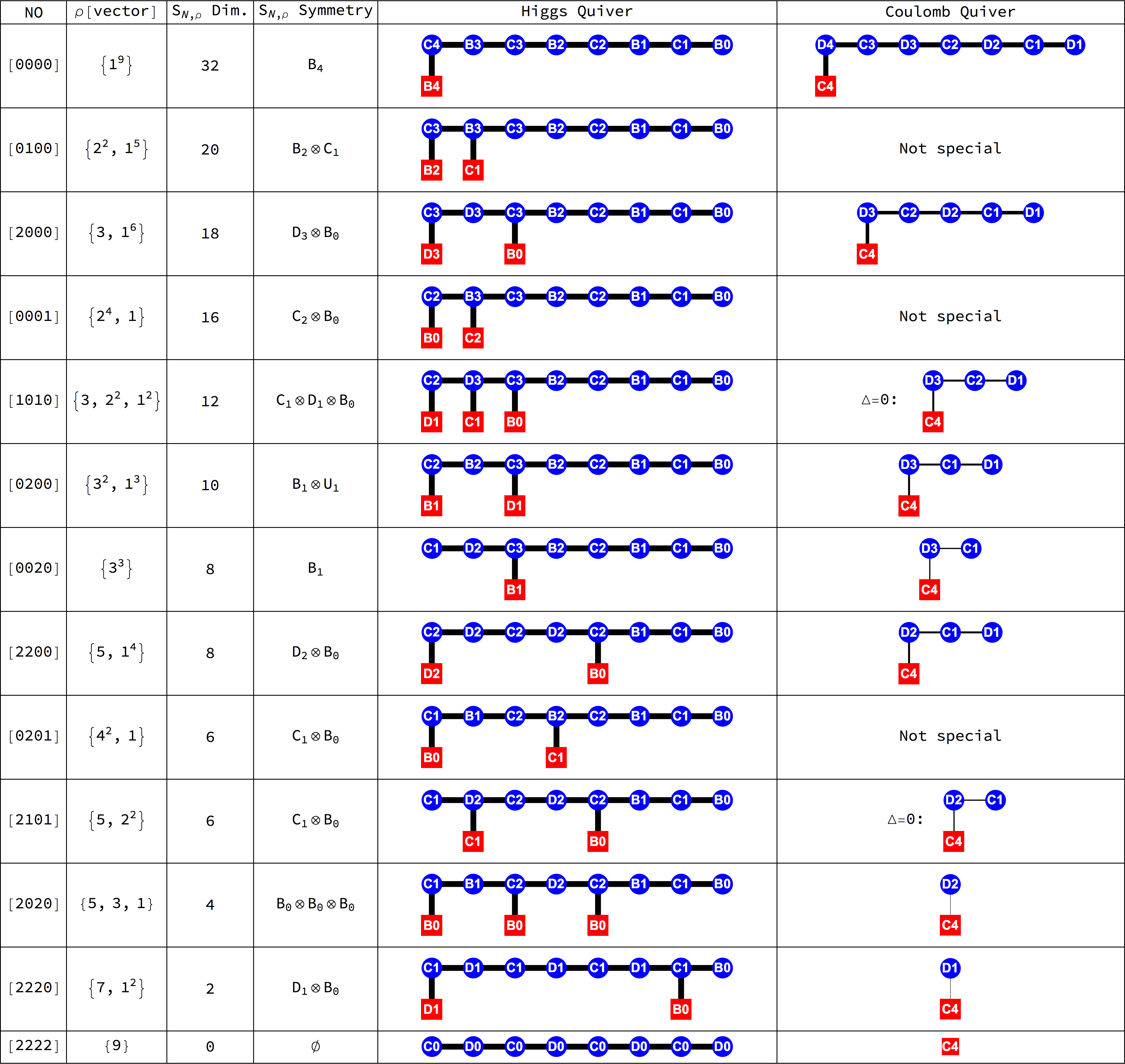}\\
\caption[Quivers for $B_4$ Slodowy Slices.]{{Quivers for $B_4$ Slodowy slices.} The Higgs quivers are of type ${{\cal B}_{B/C/D}} \left({\mathbf N_f}(\rho) \right)$ and the Coulomb quivers are of type ${{{\cal L}}_{CD}}\left( {{{\left. {d_{BV}^{}{{\left( \rho  \right)}^T}} \right|}_{CD}}} \right)$. Gauge nodes of $B$ or $D$ type are evaluated as $O$ nodes on the Higgs branch and $SO$ nodes on the Coulomb branch. $\Delta=0$ indicates a diagram for which the monopole formula contains zero conformal dimension.}
\label{fig:BCD3b}
\end{center}
\end{figure}

\begin{figure}[htbp]
\begin{center}
\includegraphics[scale=0.40]{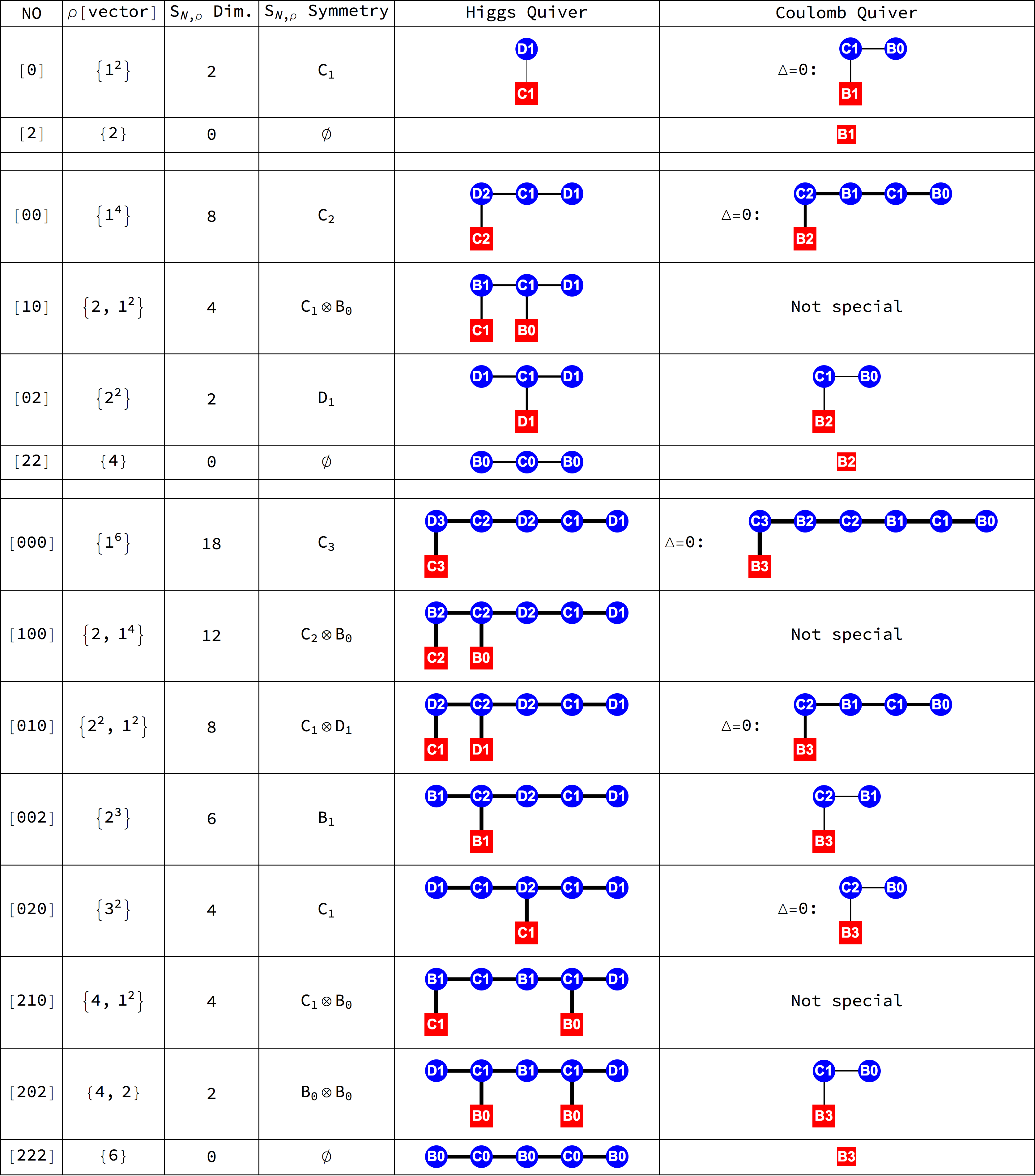}\\
\caption[Quivers for $C_1$ to $C_3$ Slodowy Slices.]{{Quivers for $C_1$ to $C_3$ Slodowy slices.} The Higgs quivers are of type ${{\cal B}_{B/C/D}} \left({\mathbf N_f}(\rho) \right)$ and the Coulomb quivers are of type ${{{\cal L}}_{BC}}\left( {{{\left. {d_{BV}^{}{{\left( \rho  \right)}^T}} \right|}_{BC}}} \right)$. Gauge nodes of $B$ or $D$ type are evaluated as $O$ nodes on the Higgs branch and $SO$ nodes on the Coulomb branch. $\Delta=0$ indicates a diagram for which the monopole formula contains zero conformal dimension.}
\label{fig:BCD4a}
\end{center}
\end{figure}

\begin{figure}[htbp]
\begin{center}
\includegraphics[scale=0.35]{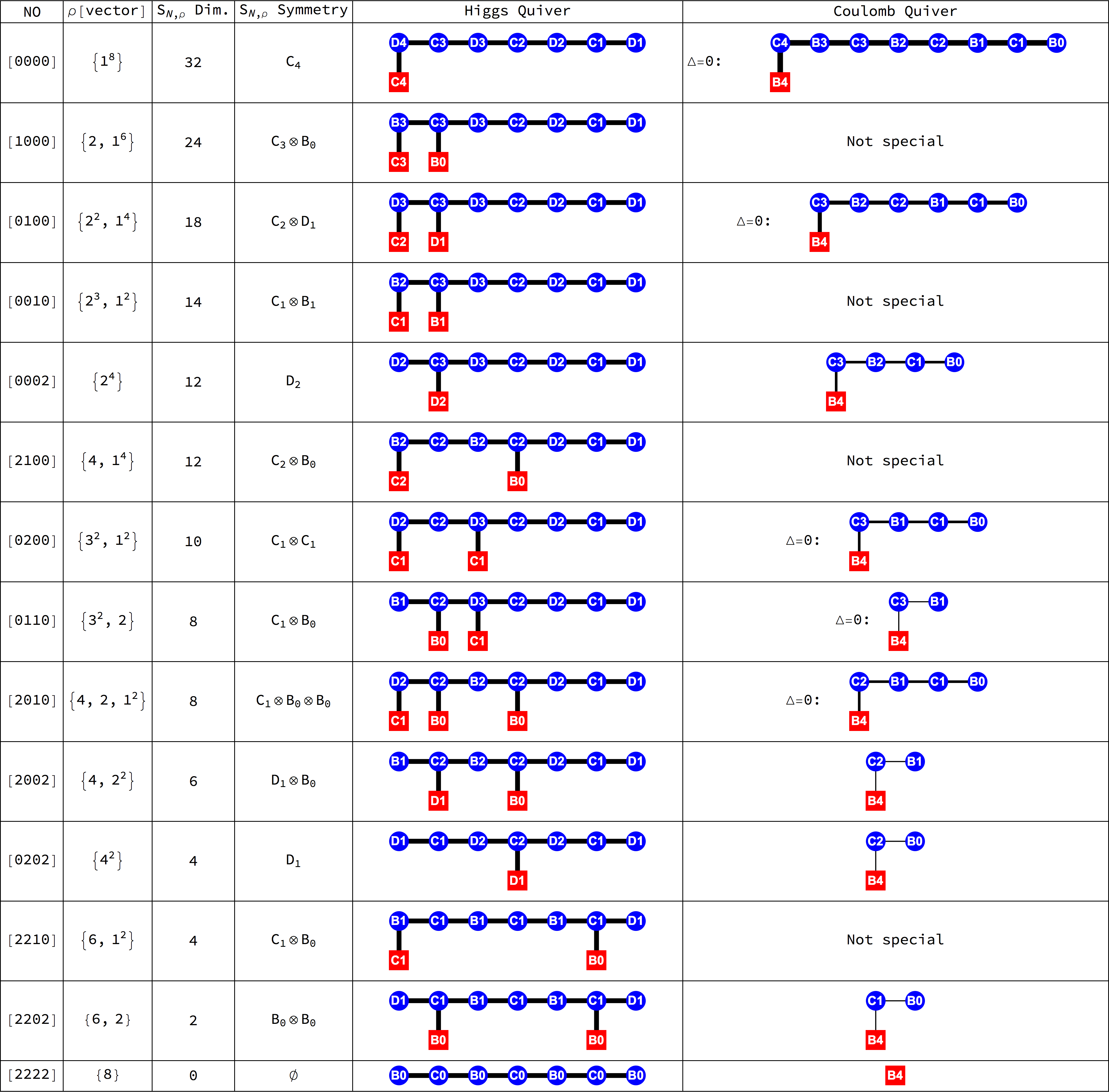}\\
\caption[Quivers for $C_4$ Slodowy Slices.]{{Quivers for $C_4$ Slodowy slices.} The Higgs quivers are of type ${{\cal B}_{B/C/D}} \left({\mathbf N_f}(\rho) \right)$ and the Coulomb quivers are of type ${{{\cal L}}_{BC}}\left( {{{\left. {d_{BV}^{}{{\left( \rho  \right)}^T}} \right|}_{BC}}} \right)$. Gauge nodes of $B$ or $D$ type are evaluated as $O$ nodes on the Higgs branch and $SO$ nodes on the Coulomb branch. $\Delta=0$ indicates a diagram for which the monopole formula contains zero conformal dimension.}
\label{fig:BCD4b}
\end{center}
\end{figure}

\begin{figure}[htbp]
\begin{center}
\includegraphics[scale=0.4]{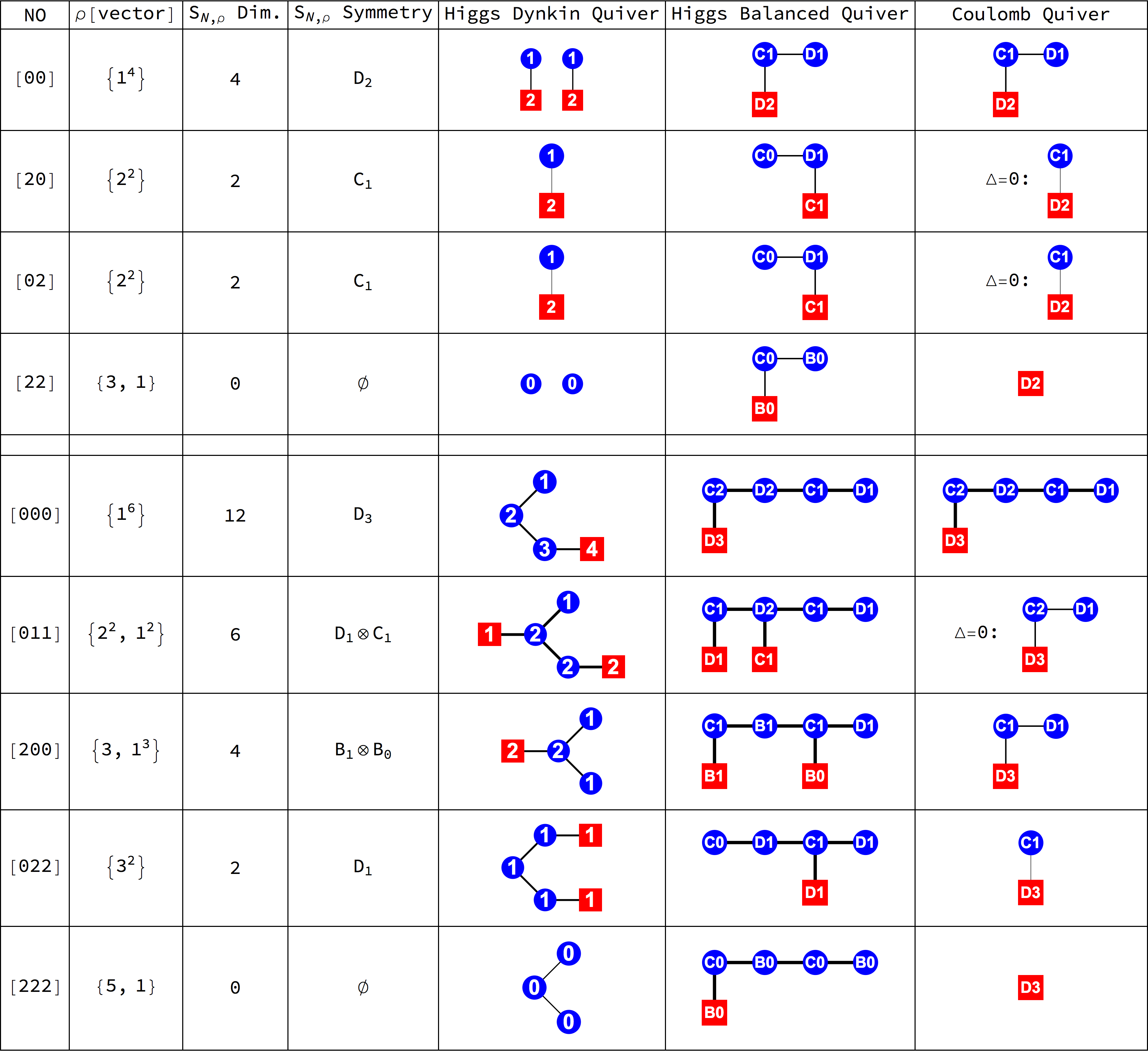}\\
\caption[Quivers for $D_2$ to $D_3$ Slodowy Slices.]{{Quivers for $D_2$ to $D_3$ Slodowy slices.} The Higgs balanced quivers are of type ${{\cal B}_{B/C/D}} \left({\mathbf N_f}(\rho) \right)$ and the Coulomb quivers are of type ${{{\cal L}}_{DC}}\left( {{{\left. {d_{BV}^{}{{\left( \rho  \right)}^T}} \right|}_{DC}}} \right)$. The Dynkin type quivers ${\cal D}_D ({\mathbf N_f}')$ are identified by $A$ series isomorphisms. Gauge nodes of $B$ or $D$ type are evaluated as $O$ nodes on the Higgs branch and $SO$ nodes on the Coulomb branch. $\Delta=0$ indicates a diagram for which the monopole formula contains zero conformal dimension.}
\label{fig:BCD5a}
\end{center}
\end{figure}

\begin{figure}[htbp]
\begin{center}
\includegraphics[scale=0.35]{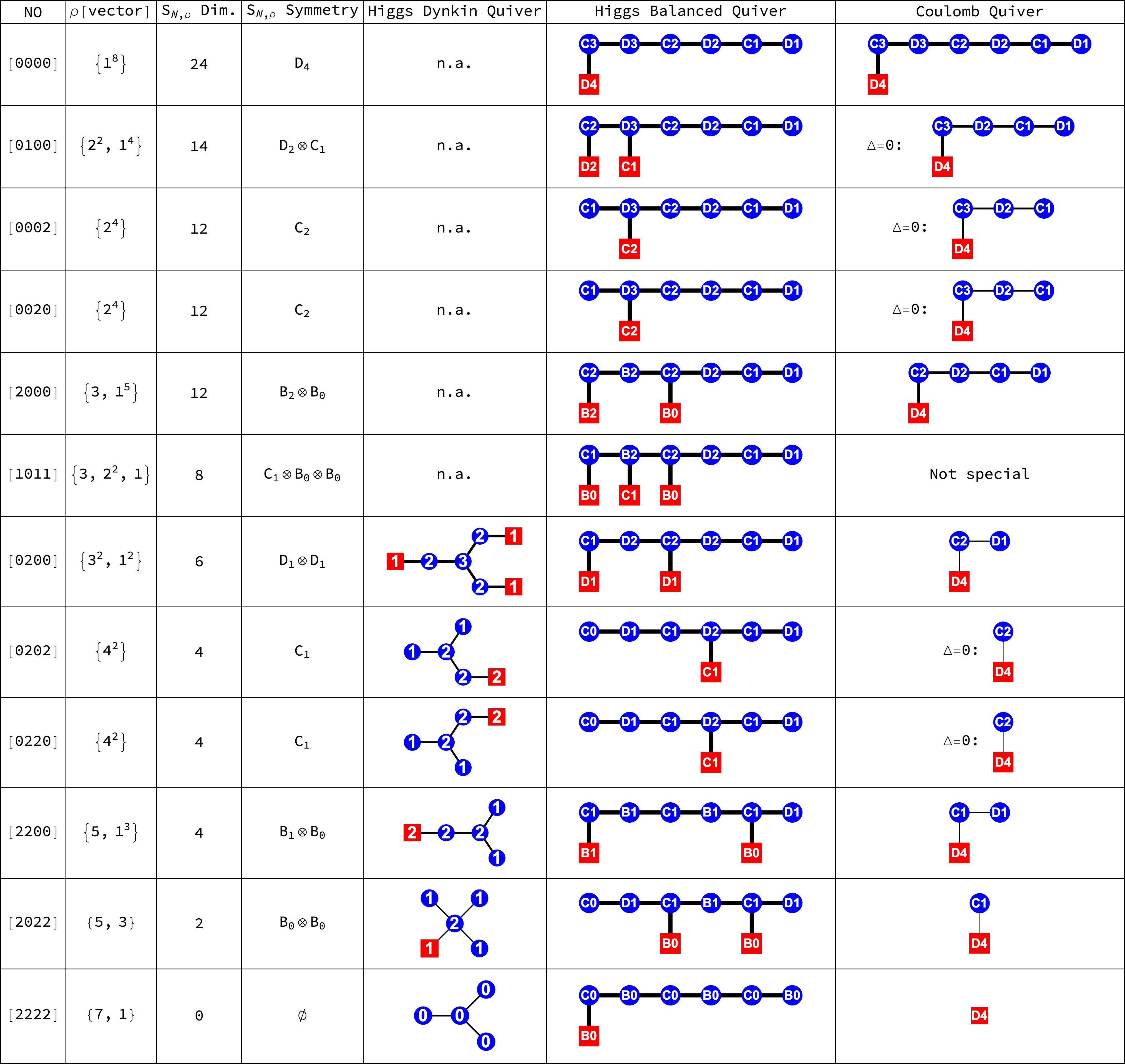}\\
\caption[Quivers for $D_4$ Slodowy Slices.]{{Quivers for $D_4$ Slodowy slices.} The Higgs balanced quivers are of type ${{\cal B}_{B/C/D}} \left({\mathbf N_f}(\rho) \right)$ and the Coulomb quivers are of type ${{{\cal L}}_{DC}}\left( {{{\left. {d_{BV}^{}{{\left( \rho  \right)}^T}} \right|}_{DC}}} \right)$. The Dynkin quivers of type ${\cal D}_D ({\mathbf N_f}([{d_{BV}}( \rho ) ] ))$, are those that have Higgs branches matching the balanced quivers. Gauge nodes of $B$ or $D$ type are evaluated as $O$ nodes on the Higgs branch and $SO$ nodes on the Coulomb branch. $\Delta=0$ indicates a diagram for which the monopole formula contains zero conformal dimension.}

\label{fig:BCD5b}
\end{center}
\end{figure}

By fixing ${\mathbf B}$, the gauge node balance condition \ref{eq:BCDquivers1} can be extended from $\cal N$ to general Slodowy slices $\mathcal{S}_{\mathcal{N},\rho}$, permitting the calculation of each gauge node vector ${\mathbf N}$ from its flavour node vector ${\mathbf N_f}$. In effect, the quivers ${{\cal B}_{B/C/D}} ({\mathbf N_f(\rho)})$ descend from the canonical linear quivers for $\cal N$, through a series of transitions that leave the balance vector ${\mathbf B}$ invariant. These canonically balanced quivers provide Higgs branch constructions for $BCD$ Slodowy slices. They are tabulated in figures \ref{fig:BCD3a} to \ref{fig:BCD5b}, along with the partitions of the fundamental, the dimensions of the Slodowy slices, and their residual symmetry groups.\footnote{Note that other quivers whose Higgs branches match $\cal N$ could be taken to define ${\mathbf B}$; each leads to a different family of quivers, whose Higgs branches match the Slodowy slice Hilbert series. The canonical choice, however, best illustrates the Higgs-Coulomb quiver dualities.}

On the other hand, the identification of linear quivers ${{\cal L}_{B/C/D}} (\sigma)$ for Coulomb branch constructions of $BCD$ series Slodowy slices poses a number of complications.

\begin{enumerate}

\item There is no bijection between partitions of $N$ and nilpotent orbits of $O(N)$ or $USp(N)$. So the quiver ${{\cal L}_{B/C/D}(\rho)}$ is valid only for partitions $\rho$ of \textit{special} nilpotent orbits; in the other cases ${{\cal L}_{B/C/D}(\rho)}$ (unlike ${{\cal L}_{B/C/D}(\rho^T)}$) would contain $USp(N)$ vectors of odd dimension $N$.


\item In the case of Coulomb branch constructions, GNO duality \cite{Goddard:1976qe} is relevant. This indicates that, since the non-simply laced $B$ and $C$ groups are GNO dual to each other, partitions of $B$ type will be necessary to produce quivers whose Coulomb branches generate Slodowy slices of $C$ algebras, and vice versa.

\item A quiver ${{\cal L}_{B/C/D}( {\rho ^T})}$ may have several Higgs equivalent quivers ${{\cal L}_{B/C/D}( {\sigma})}$, in which $\sigma$ is a sequence of non-negative integers, rather than an ordered partition. Such quivers have the same Higgs branch refined HS, but generally have different ranks of gauge groups, and therefore different Coulomb branch dimensions.

\item Any candidate quiver for a Slodowy slice must have the correct Hilbert series dimension. Since the Coulomb branch  monopole construction leads to a HS with complex dimension equal to twice the sum of the gauge group ranks in the quiver, this limits the candidates amongst Higgs equivalent quivers.

\item The Coulomb branches of quivers with $O$ gauge groups differ from those with $SO$ gauge groups; a correct choice of orthogonal gauge groups needs to be made \cite{Cabrera:2017ucb}.

\item When the orthosymplectic Coulomb branch monopole formula is applied to a quiver, the conformal dimension of all monopole operators must be positive for the Hilbert series to be well formed.

\end{enumerate}


Leaving the discussion of conformal dimension to section \ref{subsec:BCDCoulomb}, it is remarkable that a procedure exists for a partial resolution of these complexities, and indeed forms the basis for Coulomb branch constructions for the \textit{unrefined} Hilbert series of nilpotent orbits of special orthogonal groups in \cite{Cabrera:2017ucb}. The method draws on the Barbasch-Vogan map\footnote{A particularly clear description of this map is given in equation (5) of \cite{Achar:2002p}. A summary of dual maps between partitions and their appearance in the literature can be found in \cite[sec.~4]{Cabrera:2017njm}.} $d_{BV}(\rho)$ \cite{barbasch_vogan_1985}, which provides a bijection between the partitions of real vector representations associated with $B$ series special nilpotent orbits and those of pseudo-real vector representations associated with $C$ series special nilpotent orbits. By making use of Higgs equivalences, to select canonical linear quivers of type ${\cal L}_{BC}$,  ${\cal L}_{CD}$ or ${\cal L}_{DC}$, which can be done for all \textit{special} nilpotent orbits, the $d_{BV}(\rho)$ map can be extended to identify candidates for Coulomb branch constructions of Hilbert series of $BCD$ Slodowy slices, in each case starting from a homomorphism $\rho$.

The specific transformations from the partitions $\rho^T$ to the sequences $\sigma$ are summarised in table \ref {tab:BCD1a}.
\begin{table}[htp]
\begin{center}
\begin{tabular}{|c|c|c|c|}
\hline
$\text {Group}$&${\cal \bar O}_{\rho}$& Transformation &${\cal S}_{{\cal N}, \rho}$\\
\hline
$ A $&$   {Higgs \left[{\cal L}_A (\rho ^T)\right]}     $&$ \rho  = {\left( {{\rho ^T}} \right)^T}   $&$        {Coulomb \left[ {{{\cal L}_A} ( \rho)} \right]}        $\\
$ B $&$     {Higgs \left[{\cal L}_B (\rho ^T)\right]}    $&$ \sigma  \equiv {\left. {{{\left( {{{\left( {{{\left( { {\rho ^T}} \right)}_{N \to N - 1}}} \right)}_C}} \right)}^T}} \right|_{CD}}     $&$        {Coulomb \left[ {{{\cal L}_{CD}} (\sigma)} \right]}        $\\
$ C $&$    {Higgs \left[{\cal L}_C (\rho ^T)\right]}      $&$\sigma  \equiv {\left. {{{\left( {{{\left( {{{\left( { {\rho ^T}} \right)}_{N \to N + 1}}} \right)}_B}} \right)}^T}} \right|_{BC}}  $&$         {Coulomb \left[ {{{\cal L}_{BC}} (\sigma)} \right]}       $\\
$ D $&$    {Higgs \left[{\cal L}_D (\rho ^T)\right]}     $&$ \sigma  = {\left. {{{\left( {{{\left( {{\rho ^T}} \right)}_D}} \right)}^T}} \right|_{DC}}  $&$ {Coulomb\left[ {{{\cal L}_{DC}} ( \sigma)} \right]} $\\
\hline
\end{tabular}
\end{center}
\caption{Coulomb Branch Quiver Candidates for Slodowy Slices}
\label{tab:BCD1a}
\end{table}
Within these; $\rho^T$ indicates the transpose of a partition; $\rho_{N \to N \pm 1}$ indicates incrementing(decrementing) the first(last) term of a partition by 1; $\rho_B$, $\rho_C$, or $\rho_D$ indicates \textit{collapsing} a partition to a lower partition that is a valid $B$, $C$, or $D$ partition \cite{Collingwood:1993fk}; $|_{BC}$ or $|_{CD}$ indicates shifting $D$ or $B$ nodes in a linear quiver to a `Higgs equivalent' quiver that consists purely of $B-C$ or of $C-D$ pairs of nodes. The transformations can be written more concisely as   $\sigma  = {\left. {d_{BV}{{\left( \rho  \right)}^T}} \right|_{canonical}}$.
The resulting linear quivers, ${\cal L}_{CD }\left(\sigma \right)$, ${\cal L}_{BC }\left(\sigma \right)$ and ${\cal L}_{DC }\left(\sigma \right)$, whose Coulomb branches are candidates for Slodowy slices of $BCD$ groups up to rank 4, are included in figures \ref{fig:BCD3a} through \ref{fig:BCD5b}. 

The quivers ${\cal L}_{DC} ((1^N))$ and ${\cal B}_D ({\mathbf N_f}(1^N))$ for the Higgs and Coulomb branch constructions of the Slodowy slice to the trivial nilpotent orbit are the same. These tables also include identified quivers of type ${\cal D}_G ({\mathbf N_f}([{d_{BV}}( \rho ) ] ))$, whose Higgs branch Hilbert series match those of ${{\cal B}_{B/C/D}} ({\mathbf N_f(\rho)})$.

\FloatBarrier

\paragraph{Type IIB String Theory Brane Systems}

Note that all the resulting quivers, presented in figures \ref{fig:BCD3a} through \ref{fig:BCD5b} represent $3d \ \mathcal N=4$ gauge theories that admit an embedding in Type IIB string theory. They correspond to the effective gauge theory living on the world-volume of D3-branes suspended along one spatial direction between NS5-branes and D5-branes. This is achieved by taking the construction of \cite{Hanany:1996ie} and introducing O3-planes \cite{Feng:2000eq}. This type of system was further explored in \cite{Gaiotto:2008ak} where the Coulomb branches and Higgs branches were described in terms of nilpotent orbits and Slodowy slices, and the label $T_\sigma^\rho (G)$ was introduced to denote the SCFT at the superconformal fixed point. These brane systems and $3d$ quivers were also studied in \cite{Cabrera:2017njm}, finding the physical realization of transverse slices between closures of nilpotent orbits that are adjacent in their corresponding Hasse diagrams. This phenomenon has been named the \textit{Kraft-Procesi transition}.

\subsection{Higgs Branch Constructions}
\label{subsec:BCDHiggs}

In the case of the balanced unitary quivers ${\cal D}_D ({\mathbf N_f})$, based on $D$ series Dynkin diagrams, the calculation of Higgs branch Hilbert series proceeds similarly to the $A$ algebras. This leads to a Higgs branch formula that is comparable to \ref{eq:Aquivers5}, modified to include the connection of three pairs of bifundamental fields to the central node. The dimension formula \ref{eq:Aquivers6} remains unchanged.

In the case of orthosymplectic quivers of type ${{\cal B}_{B/C/D}} ({\mathbf N_f})$, modifications to the $A$ series Higgs branch formula are required. The $O/USp$ alternating chains are taken to comprise bifundamental (half) hypermultiplet fields that transform in vector representations $[1,0,\ldots,0]_{B/D} \otimes [1,0,\ldots,0]_{C}$. Also, it is necessary to average the integrations over the disconnected $SO$ and $O^-$ components of the $O$ gauge groups; this requires precise choices both of the character for the vector representation of $O^-$ and of the HKQ associated with the integration over $O^-$, as explained in \cite{Hanany:2016gbz}.\footnote{The effect of non-connected $O$ group components is also discussed in \cite{Bourget:2017tmt}.}

In other respects, the calculation of the Higgs branch of a balanced $BCD$ quiver follows a similar Weyl integration to the $A$ series. The general Higgs branch formula for $BCD$ series Slodowy slices is:

\begin{equation} 
\label{eq:BCDquivers1a}
\begin{aligned}
g_{HS}^{Higgs[{\cal B}_{B/C/D}({{\bf{N}}_{\bf{f}}}(\rho ))]} &=\\
\frac{1}{{{2^{\# O}}}}\sum\limits_{O \pm } {~}& \oint\limits_{G_1 \left( {{N_1}} \right) \otimes  \ldots G_k \left( {{N_{k}}} \right)}{d\mu}{~}\prod\limits_{n = 1}^{k} {\frac{{PE\left[ {{{[vector]}_{G_n({N_n})}} \otimes {{[vector]}_{G_{f_n}({N_{{f_n}}})}},t} \right]}}{{HKQ\left[ {G_n({N_n}),t} \right]}}}  \\
 \times & \prod\limits_{n = 1}^{k - 1} {PE\left[ {{{[vector]}_{G_n({N_n})}} \otimes {{[vector]}_{G_{n+1}({N_{n + 1}})}},t} \right]}.
 \end{aligned}
\end{equation}
In \ref{eq:BCDquivers1a}, $G_n$ alternates between $O(N)$ and $USp(N)$, ${d\mu}$ is the Haar measure for the ${G_1\left( {{N_1}} \right) \otimes  \ldots G_k \left( {{N_{k}}} \right)}$ product group, $HKQ\left[ {G_n ({N_n}),t} \right]$ is the HyperKahler quotient for a gauge node, and the summation indicates that the calculation is averaged over the non-connected $SO$ and $O^-$ components of $O$ gauge groups \cite{Hanany:2016gbz}. 

The character $[vector]_{O(2r)^ {-}}=[vector]_{O(2)^ {-}} \oplus [vector]_{USp(2r-2)} $, where $[vector]_{O(2)^ {-}}$ is (the trace of) a diagonal matrix with eigenvalues $\{1,-1\}$. The HKQ is given by $HKQ \left[ {G_n ({N_n}),t} \right] = PE \left[ [adjoint ]_{G_n},t^2 \right]$, where for the $O(2r)^-$ component of an $O(2r)$ group, $[adjoint]_{O(2r)^ {-}} \equiv {\Lambda ^2} \left[ [vector]_{O(2r)^{-}} \right]$.\footnote{For further detail, see \cite{Hanany:2016gbz}.} 

The structure of the Higgs branch formula can be used to identify the dimensions of the Hilbert series. In essence, each bifundamental field contributes HS generators according to its dimensions (being the product of the dimensions of the $O$ and $USp$ vectors), and each gauge group offsets the generators by HS relations numbering twice the dimension of the gauge group (once for the Weyl integration and once for the HKQ). This gives a rule for the dimensions of a Slodowy slice calculated from a balanced ${{\cal B}_{B/C/D}} ({\mathbf N_f}(\rho))$ quiver:

\begin{equation} 
\label{eq:BCDquivers2}
\begin{aligned}
\left| g_{HS}^{Higgs[{\cal B}({{\bf{N}}_{\bf{f}}}(\rho ))]} \right| & = {{\bf{N}}_f}( \rho  ) \cdot {\bf{N}}( \rho  ) - \frac{1}{2}{\bf{N}}( \rho  ) \cdot {\bf{A}} \cdot {\bf{N}}( \rho  ) + {\bf{N}}( \rho  ) \cdot {\bf{K}},\\
 \end{aligned}
\end{equation}
where $K_n = \left\{ \begin{array}{l} + 1{~}\text{ if}{~}{{ G_n = B/D}}\\- 1{~}\text{ if}{~}{{G_n = C}}\end{array} \right.$ and \ref{eq:BCDquivers1} is used to calculate ${\bf{N}}( \rho)$.


\subsection{Coulomb Branch Constructions}
\label{subsec:BCDCoulomb}

While the $O/USp$ version of the monopole formula \ref{eq:monOSp} derives from \ref{eq:mon0} by following similar general principles to the unitary monopole formula \ref{eq:mon1}, there are several aspects and subtleties that require discussion:
\begin{equation} 
\label{eq:monOSp}
\begin{aligned}
g_{HS}^{{\text{Coulomb}}}( {{\bf f},t^2} ) \equiv \sum\limits_{o,s} {} {}P_{o/s}^{G}( t^2 ){~}{{\bf f}^{o/s}}{~}{t^{2 \Delta (o,s)}}.
 \end{aligned}
\end{equation}
\begin{enumerate}

\item Monopole lattice. The lattice of monopole charges depends on the symmetry group. For $SU$ and $USp$ groups, points in the monopole charge lattice correspond to sets of ordered integers and are in bijective correspondence with highest weight Dynkin labels. However, the monopole charge lattices of orthogonal groups only span the vector sub-lattices and exclude weight space states whose spinor Dynkin labels sum to an odd number. Labelling monopole charges as ${\bf q} \equiv (q_1,\ldots,q_r)$ for unitary nodes, ${\bf s} \equiv (s_1,\ldots, s_r)$ for symplectic nodes and ${\bf o} \equiv (o_1,\ldots,o_r)$ for orthogonal nodes, the relationships between monopole and integer weight space lattices can be summarised as in table \ref{tab:BCD1b}.

\begin{table}[htp]
\footnotesize
\begin{center}
\begin{tabular}{|c|c|c|c|}
\hline
Group &  Monopole Lattice & Basis Transformations & Dynkin Labels \\
$      $&$        $&$    $&$     [n_1, \ldots, n_r]      $\\
\hline
$   {{U(r)}}   $&
$   {\infty  > {q_1} \ge  \ldots {q_i} \ge  \ldots {q_r} >  - \infty }    $&
$   {{q_i} \equiv \sum\nolimits_{j = i}^r {{n_j}} }    $&
${\left\{ \begin{array}{l} \infty  > {n_{i < r}} \ge 0\\ \infty  > {n_r} >  - \infty  \end{array} \right.}   $\\
\hline
$   {{A_r}}   $&
$   {\infty  > {q_1} \ge  \ldots {q_i} \ge  \ldots {q_r} \ge 0}    $&
$    {{q_i} \equiv \sum\nolimits_{j = i}^r {{n_j}} }     $&
$   {\infty  > {n_i} \ge 0}   $\\
\hline
$   {{B_r}}   $&
$    {\infty  > {o_1} \ge  \ldots {o_i} \ge  \ldots {o_r} \ge 0}      $&
$    {{o_i} \equiv \sum\nolimits_{j = i}^{r - 1} {{n_j} + {n_r}/2} }     $&
$ {\left\{ \begin{array}{l} \infty  > {n_i} \ge 0\\ {n_r} = 2k \end{array} \right.}   $\\
\hline
$   {{C_r}}   $&
$    {\infty  > {s_1} \ge  \ldots {s_i} \ge  \ldots {s_r} \ge 0}      $&
$   {{s_i} \equiv \sum\nolimits_{j = i}^r {{n_j}} }      $&
$  {\infty  > {n_i} \ge 0}    $\\
\hline
$   {{D_r}}   $&
$   {\infty  > {o_1} \ge  \ldots {o_i} \ge  \ldots \left| {{o_r}} \right| \ge 0 }    $&
$    {\left\{ \begin{array}{l} {o_{i < r}} \equiv \sum\nolimits_{j = i}^{r - 2} {{n_j} + \left( {{n_{r - 1}} + {n_r}} \right)/2} \\ {o_r} = \left( { - {n_{r - 1}} + {n_r}} \right)/2 \end{array} \right.} $&
$  {\left\{ \begin{array}{l} \infty  > {n_i} \ge 0\\ {n_{r + 1}} + {n_r} = 2k \end{array} \right.} $\\
\hline
\end{tabular}
\end{center}
\caption{Monopole and Dynkin Label Lattices}
\label{tab:BCD1b}
\end{table} 

\item Characters. The definition of conformal dimension draws on the characters of the bifundamental scalar fields in the hyper multiplets and of the adjoint scalars in the vector multiplets: the weights of the fugacities in the characters become coefficients of the monopole charges $q$ in $\Delta(q)$. These characters take a relatively simple form in the monopole lattice basis, compared with the weight space integer (Dynkin label) basis, as shown in tables \ref{tab:BCD1c} and \ref{tab:BCD1d}. CSA fugacities are taken as $\{x_1,\ldots,x_r\}$ in the weight space integer (Dynkin label) basis, or $\{y_1,\ldots,y_r\}$ in the monopole basis.

\begin{table}[htp]
\begin{center}
\begin{tabular}{|c|c|c|}
\hline
Group &$    \begin{array}{c}  \text{Monopole Basis} \\ \text{Vector/Fundamental} \end{array} $&$    \begin{array}{c} \text{Weight Space Basis}\\  \text{Vector/Fundamental}\end{array} $ \\
\hline
$   {{U(r)}}   $&
$   {\sum\nolimits_{i = 1}^r {{y_i}} } $&
$   {{x_1} + \sum\nolimits_{i = 2}^r {{{{x}}_i}{{/}}{{{x}}_{i - 1}}} }  $\\
\hline
$   {{A_r}}   $&
$   {\sum\nolimits_{i = 1}^r {{y_i}}  + \prod\nolimits_{i = 1}^r {1/{y_i}} }   $&
$   {{x_1} + \sum\nolimits_{i = 2}^r {{{{x}}_i}{{/}}{{{x}}_{i - 1}} + 1/{{{x}}_r}} } $\\
\hline
$   {{B_r}}   $&
$   {1 + \sum\nolimits_{i = 1}^r {{y_i} + \sum\nolimits_{i = 1}^r {1/{y_i}} } }   $&
$    \begin{array}{c} 1 + 1/{x_1} + {x_1}\\ + \sum\nolimits_{i = 2}^{r - 1} {\left( {{{x}_{i - 1}}{/}{{x}_i} + {{x}_i}{/}{{x}_{i - 1}}} \right)}  + {{x}_{r - 1}}/{x_r^2} +{ x_r^2}/{{x}_{r - 1}}\end{array}  $\\
\hline
$   {{C_r}}   $&
$ {\sum\nolimits_{i = 1}^r {{y_i} + \sum\nolimits_{i = 1}^r {1/{y_i}} } }     $&
$  {1/{x_1} + {x_1} + \sum\nolimits_{i = 2}^r {\left( {{{x}_{i - 1}}{/}{{x}_i} + {{x}_i}{/}{{x}_{i - 1}}}\right)}}$\\
\hline
$   {{D_r}}   $&
$   {\sum\nolimits_{i = 1}^r {{y_i} + \sum\nolimits_{i = 1}^r {1/{y_i}} } } $&
$  \begin{array}{c}
1/{x_1} + {x_1} + \sum\nolimits_{i = 2}^{r - 2} {\left( {{{{x}}_{i - 1}}{{/}}{{{x}}_i} + {{{x}}_i}{{/}}{{{x}}_{i - 1}}} \right)} \\
{{ +  }}{{{x}}_{r - 2}}{{/(}}{{{x}}_{r - 1}}{{{x}}_r}{{)  +  }}{{{x}}_{r - 1}}{{/}}{{{x}}_r}{{  + }}{{{x}}_r}{{/}}{{{x}}_{r - 1}}{{  +  (}}{{{x}}_{r - 1}}{{{x}}_r}{{)/}}{{{x}}_{r - 2}}
\end{array}   $\\
\hline
\end{tabular}
\end{center}
\caption{Vector/Fundamental Characters}
\label{tab:BCD1c}
\end{table}

\begin{table}[htp]
\begin{center}
\begin{tabular}{|c|c|c|}
\hline
Group & Monopole Basis  \\
\hline
$   {{U(r)}}   $&
$ {r + \sum\nolimits_{i \ne j}^{} {{y_i}/{y_j}} }  $\\
\hline
$   {{A_r}}   $&
$ {r + \prod\nolimits_{i = 1}^r {1/{y_i}\left( {\sum\nolimits_{j = 1}^r {1/{y_j}} } \right) + \prod\nolimits_{i = 1}^r {{y_i}\left( {\sum\nolimits_{j = 1}^r {{y_i}} } \right) + \sum\nolimits_{i \ne j}^{} {{y_i}/{y_j}} } } }  $\\
\hline
$   {{B_r}}   $&
$  {r + \sum\nolimits_{i = 1}^r {\left( {{y_i} + 1/{y_i}} \right) + \sum\nolimits_{i < j}^{} {\left( {{y_i}{y_j} + {y_i}/{y_j} + {y_j}/{y_i} + 1/( {{y_i}{y_j}} )} \right)} } }   $\\
\hline
$   {{C_r}}   $&
$ {r + \sum\nolimits_{i = 1}^r {\left( {y_i^2 + 1/y_i^2} \right) + \sum\nolimits_{i < j}^{} {\left( {{y_i}{y_j} + {y_i}/{y_j} + {y_j}/{y_i} + 1/( {{y_i}{y_j}} )} \right)} } }   $\\
\hline
$   {{D_r}}   $&
$  {r + \sum\nolimits_{i < j}^{} {\left( {{y_i}{y_j} + {y_i}/{y_j} + {y_j}/{y_i} + 1/( {{y_i}{y_j}} )} \right)} }   $\\
\hline
\end{tabular}
\end{center}
\caption{Adjoint Characters}
\label{tab:BCD1d}
\end{table} 

\FloatBarrier

\item Conformal dimension. The contributions to conformal dimension of the $O/USp$ bifundamental fields linking gauge or flavour nodes, and of the $O/USp$ gauge nodes, follow from \ref{eq:mon0} in a similar manner to the unitary case \ref{eq:mon3}, starting from the relevant characters: the coefficients $\{0, \pm 1, \pm 2 \}$ of the monopole charges $\{{\bf o},{\bf s}\}$ in the conformal dimension formula match the weights (exponents) of the ${\bf y}$ fugacities in the characters of the respective bifundamental or adjoint representations in the monopole basis. Tables \ref{tab:BCD1e} and \ref{tab:BCD1f} show the resulting contributions from the various types of gauge node and bifundamental field.

\begin{table}[htp]
\begin{center}
\begin{tabular}{|c|c|c|}
\hline
Gauge Group & $\Delta(\text{Node})$  \\
\hline
$   {{U(r)}}   $&
$ { - \sum\nolimits_{1 \le i < j \le r} {| {{q_i} - {q_j}} |} }$\\
\hline
$   {{B_r}}   $&
$  { - \sum\nolimits_{i = 1}^r {| {{o_i}} |}  - \sum\nolimits_{1 \le i < j \le r} {| {{o_i} \pm {o_j}} |} }  $\\
\hline
$   {{C_r}}   $&
$  { - 2\sum\nolimits_{i = 1}^r {| {{s_i}} |}  - \sum\nolimits_{1 \le i < j \le r} {| {{s_i} \pm {s_j}} |} } $\\
\hline
$   {{D_r}}   $&
$  { - \sum\nolimits_{1 \le i < j \le r} {| {{o_i} \pm {o_j}}|} } $\\
\hline
\end{tabular}
\end{center}
\caption{Gauge Node Conformal Dimensions}
\label{tab:BCD1e}
\end{table} 

\begin{table}[htp]
\begin{center}
\begin{tabular}{|c|c|c|}
\hline
Gauge Groups & $\Delta(\text{Bifundamental})$  \\
\hline
$ {{U({{r_1}})} - {U({{r_2}})}}   $&
$ {\frac{1}{2}\sum\limits_{i = 1}^{{r_1}} {\sum\limits_{j = 1}^{{r_2}} {| {q_{1,i}} - {q_{2,j}}|} } }$\\
\hline
$  {{B_{{r_1}}} - {C_{{r_2}}}}   $&
$ {\frac{1}{2}\sum\limits_{j = 1}^{{r_2}} {| {{s_j}} |}  + \frac{1}{2}\sum\limits_{i = 1}^{{r_1}} {\sum\limits_{j = 1}^{{r_2}} {| {{o_i} \pm {s_j}} |} } } $\\
\hline
$ {D_{{r_1}}} - {C_{{r_2}}}  $&
$ {\frac{1}{2}\sum\limits_{i = 1}^{{r_1}} {\sum\limits_{j = 1}^{{r_2}} {| {{o_i} \pm {s_j}} |} } }$\\
\hline
\end{tabular}
\end{center}
\caption{Bifundamental Conformal Dimensions}
\label{tab:BCD1f}
\end{table}

\FloatBarrier

\item Symmetry factors. The residual symmetries for a flux (whether $\bf{o}$, $\bf{s}$, or $\bf{q}$) over a gauge node can be fixed from the sub-group of the $O/USp/U$ gauge group identified by the Dynkin diagram formed by those monopole charges that equal their successors (or, equivalently, correspond to zero Dynkin labels). Note that the symmetry factors may belong to a sub-group from a different series to the gauge node.

\item $O$ vs $SO$ gauge nodes. Both the characters of vector irreps and symmetry factors depend on whether a $D$ series gauge node is taken as $SO$ or as $O$. As noted in \cite{Cremonesi:2014uva}, the Casimirs of an $O(2n)$ symmetry group are the same as those of $SO(2n+1)$, due to the absence of a Pfaffian in $O(2n)$ (since the determinant of representation matrices can be of either sign). The Coulomb branch calculations for Slodowy slices herein are based entirely on $SO$ gauge nodes. This is a choice consistent with the results in \cite{Cabrera:2017ucb}. When these results are translated to the brane configurations, the action of the Lusztig's Canonical Quotient ${\cal \bar A}(\mathcal O)$ related to each quiver can be seen in terms of  \textit{collapse transitions} \cite{Cabrera:2017njm} performed in the branes. Each time a collapse transition moves two half D5-branes away from each other all magnetic lattices of the orthogonal gauge nodes in between are acted upon by a diagonal $\mathbb{Z}_2$ action. The brane configurations \cite{Feng:2000eq,Gaiotto:2008ak,Cabrera:2017njm} for linear quivers ${{\cal L}_{B/C/D}} (\sigma)$ do not present this effect, and therefore all gauge nodes are $SO$.  

\item Fugacities. In the unitary monopole formula, $\bf{z}$ in \ref{eq:mon1} can be treated as a fugacity for the simple roots of the group for which the quiver is a balanced Dynkin diagram. As discussed in \cite{Cremonesi:2014kwa}, such a treatment cannot be extended to the $O/USp$ monopole formula due to the non-unitary gauge groups involved. Thus, while it can be helpful to include fugacities ${\bf f} \equiv ({f_1,\ldots,f_r})$ during the calculation of Coulomb branches, their interpretation is unclear. Such issues do not affect the validity of the \textit{unrefined} Hilbert series ultimately obtained by setting $\forall f_i: f_i \to 1$.

\end{enumerate}

In order for a Coulomb branch Hilbert series not to lead to divergences when the fugacities $\bf f$ are set to unity, it is necessary that no sub-lattice of the monopole lattice (other than the origin) should have a conformal dimension of zero (to ensure that the fugacities $\bf f$ only appear as generators when coupled with $t^k$, where $k > 0$). A necessary (albeit not always sufficient) condition on $O/USp$ quivers can be formulated by examining unit shifts away from the origin of the monopole lattice. This is similar to the ``good or ugly,  but not bad" balance condition on unitary quivers \cite{Gaiotto:2008ak}.

\begin{table}[htp]
\begin{center}
\begin{tabular}{|c|c|c|}
\hline
Gauge Group Chain & ${{\Delta _r}\left( {1,0 \ldots 0} \right)}$  \\
\hline
$ {{C_{{r_1}}} - {D_r} - {C_{{r_2}}}}  $&
$ {{r_1} + {r_2} - 2r + 2}  $\\
\hline
$ {{C_{{r_1}}} - {B_r} - {C_{{r_2}}}}  $&
$  {{r_1} + {r_2} - 2r + 1} $\\
\hline
$ {{B_{{r_1}}} - {C_r} - {B_{{r_2}}}}  $&
$  {{r_1} + {r_2} - 2r+1 } $\\
\hline
$ {{B_{{r_1}}} - {C_r} - {D_{{r_2}}}}   $&
$  {{r_1} + {r_2} - 2r +1/2 } $\\
\hline
$ {{D_{{r_1}}} - {C_r} - {D_{{r_2}}}}  $&
$  {{r_1} + {r_2} - 2r } $\\
\hline
\end{tabular}
\end{center}
\caption{Quiver Chain Unit Conformal Dimensions}
\label{tab:BCD1g}
\end{table}

In table \ref{tab:BCD1g} we examine the unit conformal dimensions that result, based on tables \ref{tab:BCD1e} and \ref{tab:BCD1f}, from setting a single monopole charge ($o_1$, or $s_1$) on a central gauge node in a chain of three nodes to unity, depending on the ranks of the nodes involved. We can use this table to check that no gauge node in a quiver is necessarily ``bad". For example, the central gauge node in the chain $D_2 - C_1 - D_1$ is assigned a unit conformal dimension of $1$ and is a ``good" node. Quivers with zero conformal dimension are identified as such in figures \ref{fig:BCD3a} through \ref{fig:BCD5b}. Their Hilbert series clearly do not match those of the Higgs branch constructions for Slodowy slices, and are not tabulated here.

Providing (i) a nilpotent orbit ${\cal O}_{\rho}$ is special (so that the Barbasch-Vogan map can be uniquely applied), and (ii) that the quiver ${{\cal L}_{BC/CD/DC}} (\sigma(\rho))$ does not suffer from zero conformal dimension, the $O/USp$ monopole formula \ref{eq:monOSp} can be used to calculate \textit{unrefined} Hilbert series for Slodowy slices; these match those calculated on the Higgs branch of ${{\cal B}_{B/C/D}} ({\mathbf N_f}(\rho))$ using \ref{eq:Aquivers5}.

\subsection{Hilbert Series}
\label{subsec:BCDHilbert}

The Hilbert series of the Slodowy slices of algebras $B_1$ to $B_4$, $C_1$ to $C_4$ and $D_2$ to $D_4$, calculated as above, are summarised in tables \ref{tab:BCD2},  \ref{tab:BCD3},  \ref{tab:BCD4},  \ref{tab:BCD5} and \ref{tab:BCD6}. The refined Hilbert series are based on the Higgs branches of the balanced quivers ${{\cal B}_{B/C/D}} ({\mathbf N_f}(\rho))$.

Whenever the flavour symmetry groups are from the $B$ or the $D$ series, a choice has to be made between the characters of $SO(N)$ or $O(N)^-$. In the tables, $B/D$ flavour nodes have been taken as $SO$ type, with the exception of $B_0$ where the $O(1)$ fugacity $k_i= \pm 1$ has been used (with indices dropped where no ambiguity arises)\footnote{Note that if one wishes to read the generators of the chiral ring from the quiver as described in Section \ref{sec:OfQBCD}, then all fugacities $k_i$ need to be set to $1$.}.

The Hilbert series are presented in terms of their generators, or $PL[HS]$, using character notation $[n_1,\ldots, n_r]_G$ to label irreps. Symmetrisation of these generators using the $PE$ recovers the refined Hilbert series. The underlying adjoint maps  \ref{eq:SS9} can readily be recovered from the generators by inverting \ref{eq:SS11}. The HS can be unrefined by replacing irreps of the global symmetry groups by their dimensions.

\begin{sidewaystable}[htp]
\begin{center}
\begin{tabular}{|c|c|c|c|c|}
\hline
${\begin{array}{c} \text{Nilpotent}\\\text{Orbit} \end{array}}$
&${\begin{array}{c} \text{Dimension}\\ {| {\cal S}_{\cal N,\rho}|} \end{array}}$
&${\begin{array}{c}  \text{Symmetry} \\ F\end{array}}$
&$ \text{Generators of HS} \equiv \text{PL[HS]}$
&$ \text{Unrefined HS} $\\
\hline
$[0]$&$ 2  $&$B_1   $&$  [2] t^2-t^4   $&$\frac {  (1 - t^4)}{(1 - t^2)^3}    $\\
$[2]$&$ 0  $&$  \emptyset   $&$  0   $&$  1    $\\
\hline
$[00]$&$ 8  $&$  B_2   $&$  [0,2] t^2-t^4-t^8   $ & $\frac{(1 - t^4) (1 - t^8)}{(1 - t^2)^{10}}$\\
$[01]$&$ 4  $&$  C_1 \otimes B_0 $&$[2] t^2+[1] k t^3-t^8   $ & $ \frac{ (1 - t^8)}{(1 - t^2)^3 (1 - t^3)^2 }   $\\
$[20]$&$ 2  $&$  D_1 \otimes B_0 $&$ t^2+(1) k t^4- t^8   $ & $ \frac{ (1 - t^8)}{(1 - t^2) (1 - t^4)^2} $\\
$[22]$&$ 0  $&$   \emptyset  $&$ 0   $ & $ 1    $\\
\hline
$[000]$&$ 18  $&$B_3$ & $[0,1,0] t^2-t^4-t^8-t^{12} $&$ \frac{(1 - t^4) (1 - t^8) (1 - t^{12})}{(1 - t^2)^{21}}   $\\
$[010]$&$ 10 $&$B_1 \otimes C_1$  &  $[2]_B t^2+[2]_C t^2+[2]_B [1]_C t^3 -t^8-t^{12}   $&$ \frac {(1 - t^8) (1 - t^{12})}{(1 - t^2)^6 (1 - t^3)^6}   $\\
$[200]$&$ 8  $&$ D_2 \otimes B_0$&$ [2,0] t^2 +  [0,2] t^2+[1,1] k t^4-t^8-t^{12} $&$\frac {(1 - t^8) (1 - t^{12})}{(1 - t^2)^6 (1 - t^4)^4}  $\\
$[101]$&$ 6  $&$  C_1 \otimes B_0 $&      $[2] t^2+[1] k t^3+t^4+[1] k t^5-t^8-t^{12}  $&$ \frac {(1 - t^8) (1 - t^{12})}{(1 - t^2)^3 (1 - t^3)^2 (1 - t^4) (1 - t^5)^2} $\\
$[020]$&$ 4  $&$ D_1 \otimes B_0$ &      $t^2+(1) k t^4+(2) t^4 +t^6-t^8-t^{12}  $&$ \frac {(1 - t^8) (1 - t^{12})}{(1 - t^2) (1 - t^4)^4 (1 - t^6)} $\\
$[220]$&$ 2  $&$  D_1 \otimes B_0  $&       $t^2+(1) k t^6-t^{12}  $&$ \frac {(1 - t^{12})}{(1 - t^2) (1 - t^6)^2} $\\
$[222]$&$ 0  $&$    \emptyset  $&$ 0   $&$ 1    $\\
\hline
\end{tabular}
\end{center}
\text{N.B. $(n)$ denotes the character of the $D_1 \equiv SO(2)$ reducible representation $q^n+q^{-n}$ of $U(1)$.}\\
\text{$k$ denotes the character $\pm 1$ if $B_0 \to O(1)$ or $1$ if $B_0 \to SO(1)$.}
\caption{Hilbert Series for Slodowy Slices of $B_1$, $B_2$, and $B_3$}
\label{tab:BCD2}
\end{sidewaystable}

\begin{sidewaystable}[htp]
\footnotesize
\begin{center}
\begin{tabular}{|c|c|c|c|c|}
\hline
${\begin{array}{c} \text{Nilpotent}\\\text{Orbit} \end{array}}$
&${\begin{array}{c} \text{Dimension}\\ {| {\cal S}_{\cal N,\rho}|} \end{array}}$
&${\begin{array}{c}  \text{Symmetry} \\ F\end{array}}$
&$ \text{Generators of HS} \equiv \text{PL[HS]}$
&$ \text{Unrefined HS} $\\
\hline
$[0000]$&$ 32  $&$  B_4   $&$  [0,1,0,0] t^2-t^4-t^8-t^{12}-t^{16}   $&$\frac {(1 - t^4) (1 - t^8) (1 - t^{12}) (1 - t^{16})}{(1 - t^2)^{36}} $\\

$[0100]$&$ 20  $&$  B_2 \otimes C_1                $&$[0,2] t^2 + [2] t^2+[1,0] [1] t^3 -t^8-t^{12}-t^{16}   $&$\frac{(1 - t^8) (1 - t^{12}) (1 - t^{16})}{(1 - t^2)^{13} (1 - t^3)^{10}} $\\

$[2000]$&$ 18  $&$D_3 \otimes B_0                  $&$[0,1,1] t^2+ [1,0,0] k t^4-t^{8}-t^{12}-t^{16} $&$\frac{(1- t^8) (1-t^{12}) (1-t^{16})}{(1-t^2)^{15} (1 - t^4)^6}  $\\

$[0001]$&$ 16  $&$   C_2 \otimes B_0                 $&$[2,0] t^2+[1,0] k t^3+[0,1] t^4-t^8-t^{12}-t^{16}$&$\frac {(1 - t^8) (1 - t^{12}) (1 - t^{16})}{(1 - t^2)^{10} (1 - t^3)^4 (1-t^4)^5} $\\

$[1010]$&$ 12  $&$   C_2 \otimes D_1 \otimes B_0                $&$[2] t^2+[1] (1) t^3+[1] k t^3 +(1) k t^4+ t^4+[1] k t^5 +t^2 -t^8-t^{12}-t^{16} $&$\frac {(1 - t^8) (1 - t^{12}) (1 - t^{16})}{(1 - t^2)^4 (1 - t^3)^6 (1 - 
     t^4)^3 (1 - t^5)^2} $\\
     
$[0200]$&$ 10  $&$  B_1 \otimes D_1 $&$[2]t^2+(2)t^4+[2](1)t^4+t^2+t^6-t^8-t^{12}-t^{16} $&$ \frac{(1 - t^8) (1 - t^{12}) (1 - t^{16})}{(1 - t^2)^4 (1 - t^4)^8 (1 - t^6)}  $\\

$[0020]$&$ 8  $&$      B_1  $&$[2]t^2+[4]t^4+[2]t^6-t^8-t^{12}-t^{16} $&$ \frac{(1 - t^8) (1 - t^{12}) (1 - t^{16})}{(1 - t^2)^3 (1 - t^4)^5 (1 - t^6)^3}  $\\

$[2200]$&$ 8  $&$      D_2 \otimes B_0 $&$[2,0]t^2+[0,2]t^2+[1,1] k t^6 -t^{12}-t^{16}$&$ \frac{(1 - t^{12}) (1 - t^{16})}{(1 - t^2)^6 (1 - t^6)^4}  $\\

$[0201]$&$ 6  $&$    C_1 \otimes B_0  $&$[2]t^2+[1] k t^5+[2] t^6 -t^{12}-t^{16}   $&$ \frac{(1 - t^{12}) (1 - t^{16})}{(1 - t^2)^3 (1 - t^5)^2 (1 - t^6)^3}  $\\

$[2101]$&$ 6  $&$  C_1 \otimes B_0   $&$ [2] t^2+[1] k t^5+[1] k t^7+t^4-t^{12}-t^{16}   $&$ \frac{(1 - t^{12}) (1 - t^{16})}{(1 - t^2)^3 (1 - t^4) (1 - t^5)^2 (1 - t^7)^2}  $\\

$[2020]$&$ 4  $&$ B_0 \otimes B_0 \otimes B_0  $&$(k_1 k_3+k_3 k_5) t^4 +(k_1 k_5+k_3 k_5) t^6+ k_3 k_5 t ^8+t^4 - t^{12}-t^{16}  $&$ \frac{(1 - t^{12}) (1 - t^{16})}{(1 - t^4)^3 (1 - t^6)^2 (1 - t^8)}  $\\

$[2220]$&$ 2  $&$    D_1 \otimes B_0               $&$(1) k t^8+t^2-t^{16}   $&$\frac{1-t^{16}}{(1-t^2) (1-t^8)^2}  $\\

$[2222]$&$ 0  $&$  \emptyset $           &$ 0   $&$ 1    $\\
\hline
\end{tabular}
\end{center}
\text{N.B. $(n)$ denotes the character of the $D_1 \equiv SO(2)$ reducible representation $q^n+q^{-n}$ of $U(1)$.}\\
\text{$k$ denotes the character $\pm 1$ if $B_0 \to O(1)$ or $1$ if $B_0 \to SO(1)$.}
\caption{Hilbert Series for Slodowy Slices of $B_4$.}
\label{tab:BCD3}
\end{sidewaystable}

\begin{sidewaystable}[htp]
\begin{center}
\begin{tabular}{|c|c|c|c|c|}
\hline
${\begin{array}{c} \text{Nilpotent}\\\text{Orbit} \end{array}}$
&${\begin{array}{c} \text{Dimension}\\ {| {\cal S}_{\cal N,\rho}|} \end{array}}$
&${\begin{array}{c}  \text{Symmetry} \\ F\end{array}}$
&$ \text{Generators of HS} \equiv \text{PL[HS]}$
&$ \text{Unrefined HS} $\\
\hline
$[0]$&$ 2  $&$C_1   $&$  [2] t^2-t^4   $&$\frac {  (1 - t^4)}{(1 - t^2)^3}    $\\
$[2]$&$ 0  $&$  \emptyset   $&$  0   $&$  1    $\\
\hline
$[00]$&$ 8  $&$  C_2   $&$  [2,0] t^2-t^4-t^8   $ & $\frac{(1 - t^4) (1 - t^8)}{(1 - t^2)^{10}}$\\
$[10]$&$ 4  $&$  C_1 \otimes B_0 $&$[2] t^2+[1] k t^3-t^8   $ & $ \frac{ (1 - t^8)}{(1 - t^2)^3 (1 - t^3)^2 }   $\\
$[02]$&$ 2  $&$  D_1 $&$ t^2+(1) t^4- t^8   $ & $ \frac{ (1 - t^8)}{(1 - t^2) (1 - t^4)^2} $\\
$[22]$&$ 0  $&$   \emptyset  $&$ 0   $ & $ 1    $\\
\hline
$[000]$&$ 18  $&$C_3$ & $[2,0,0]t^2-t^4-t^8-t^{12} $&$\frac{(1-t^4) (1-t^8) (1-t^{12})}{(1-t^2)^{21}}  $\\
$[100]$&$ 12 $&$C_2 \otimes B_0$  &  $[2,0] t^2+[1,0] k t^3-t^8-t^{12}  $&$\frac{(1-t^8) (1-t^{12})}{(1-t^2)^{10} (1-t^3)^4}  $\\
$[010]$&$ 8  $&$ C_1 \otimes D_1$&$ [2]t^2+[1](1)t^3+(2)t^4+t^2-t^8-t^{12} $&$\frac{(1-t^8) (1-t^{12})}{(1-t^2)^4 (1-t^3)^4 (1-t^4)^2} $\\
$[002]$&$ 6  $&$  B_1 $&      $[2]t^2+[4]t^4-t^8-t^{12} $&$\frac{(1-t^8) (1-t^{12})}{(1-t^2)^3 (1-t^4)^5} $\\
$[020]$&$ 4  $&$ C_1 $ &      $[2]t^2+[2]t^6-t^8-t^{12} $&$\frac{(1-t^8) (1-t^{12})}{(1-t^2)^3 (1-t^6)^3}$\\
$[210]$&$ 4  $&$  C_1 \otimes B_0  $&       $[2]t^2+[1] k t^5-t^{12}  $&$\frac{1-t^{12}}{(1-t^2)^3 (1-t^5)^2} $\\

$[202]$&$ 2  $&$  B_0 \otimes B_0  $&       $t^4+k_2 k_4 t^4+k_2 k_4 t^6-t^{12}  $&$\frac{1-t^{12}}{(1-t^4)^2 (1-t^6)} $\\
$[222]$&$ 0  $&$    \emptyset  $&$ 0   $&$ 1    $\\
\hline
\end{tabular}
\end{center}
\text{N.B. $(n)$ denotes the character of the $D_1 \equiv SO(2)$ reducible representation $q^n+q^{-n}$ of $U(1)$.}\\
\text{$k$ denotes the character $\pm 1$ if $B_0 \to O(1)$ or $1$ if $B_0 \to SO(1)$.}
\caption{Hilbert Series for Slodowy Slices of $C_1$, $C_2$, and $C_3$}
\label{tab:BCD4}
\end{sidewaystable}

\begin{sidewaystable}[htp]
\footnotesize
\begin{center}
\begin{tabular}{|c|c|c|c|c|}
\hline
${\begin{array}{c} \text{Nilpotent}\\\text{Orbit} \end{array}}$
&${\begin{array}{c} \text{Dimension}\\ {| {\cal S}_{\cal N,\rho}|} \end{array}}$
&${\begin{array}{c}  \text{Symmetry} \\ F\end{array}}$
&$ \text{Generators of HS} \equiv \text{PL[HS]}$
&$ \text{Unrefined HS} $\\
\hline
$[0000]$&$ 32  $&$  C_4   $&$  [2,0,0,0] t^2-t^4-t^8-t^{12}-t^{16}   $&$\frac {(1 - t^4) (1 - t^8) (1 - t^{12}) (1 - t^{16})}{(1 - t^2)^{36}} $\\
$[1000]$&$ 24  $&$  C_3 \otimes B_0 $&${[2,0,0]}t^2+[1,0,0] k t^3-t^8-t^{12}-t^{16}   $&$\frac{(1-t^8) (1-t^{12}) (1-t^{16})}{(1-t^2)^{21} (1-t^3)^6}$\\
$[0100]$&$ 18  $&$C_2 \otimes D_1 $&${[2,0]}t^2+[1,0](1)t^3+(2)t^4+t^2-t^8-t^{12}-t^{16}$&$\frac{(1-t^8) (1-t^{12}) (1-t^{16})}{(1-t^2)^{11} (1-t^3)^8 (1-t^4)^2}  $\\
$[0010]$&$ 14 $&$   C_1 \otimes B_1 $&${[2]_B}t^2+[2]_C t^2+[2]_B [1]_C t^3+[4]_B t^4 -t^8-t^{12}-t^{16}$&$\frac{(1-t^8) (1-t^{12}) (1-t^{16})}{(1-t^2)^6 (1-t^3)^6 (1-t^4)^5} $\\
$[0002]$&$ 12  $&$   D_2 $&${[2,0]}t^2+[0,2]t^2+[2,2]t^4-t^8-t^{12}-t^{16} $&$\frac{(1-t^8) (1-t^{12}) (1-t^{16})}{(1-t^2)^6 (1-t^4)^9} $\\ 
$[2100]$&$ 12  $&$  C_2 \otimes B_0 $&$[2,0] t^2+[1,0] k t^5 -t^{12} - t^{16}$&$ \frac{(1-t^{12}) (1-t^{16})}{(1-t^2)^{10} (1-t^5)^4}  $\\
$[0200]$&$ 10  $&$     C_1 \otimes C_1  $&$[2]t^2+[2]t^2+[1][1]t^4+[2]t^6-t^8-t^{12}-t^{16} $&$\frac{(1-t^8) (1-t^{12}) (1-t^{16})}{(1-t^2)^6 (1-t^4)^4 (1-t^6)^3} $\\
$[0110]$&$ 8  $&$      C_1 \otimes B_0 $&$[2]t^2+[1] k t^3+[1] k t^5+ [2]t^6+t^4-t^8-t^{12}-t^{16}$&$\frac{(1-t^8) (1-t^{12}) (1-t^{16})}{(1-t^2)^3 (1-t^3)^2 (1-t^4) (1-t^5)^2 (1-t^6)^3} $\\
$[2010]$&$ 8  $&$    C_1 \otimes B_0 \otimes B_0  $&$[2] t^2+[1] k_2 t^3+[1] k_4 t^5+k_2 k_4 t^4+k_2 k_4t^6+t^4-t^{12}-t^{16} $&$\frac{(1-t^{12}) (1-t^{16})}{(1-t^2)^3 (1-t^3)^2 (1-t^4)^2 (1-t^5)^2 (1-t^6)} $\\
$[2002]$&$ 6  $&$  D_1 \otimes B_0   $&$ (1)k t^4+(2) t^4+(1)k t^6+t^2+t^4-t^{12}-t^{16}  $&$ \frac{(1-t^{12}) (1-t^{16})}{(1-t^2) (1-t^4)^5 (1-t^6)^2} $\\
$[0202]$&$ 4  $&$D_1  $&$(2)t^4+(2)t^8+t^6-t^{12}-t^{16} $&$\frac{(1-t^{12}) (1-t^{16})}{(1-t^2) (1-t^4)^2 (1-t^6) (1-t^8)^2} $\\
$[2210]$&$ 4  $&$    C_1 \otimes B_0  $ & $[2] t^2+[1] k t^7-t^{16}  $&$\frac{1-t^{16}}{(1-t^2)^3 (1-t^7)^2}  $\\

$[2202]$&$ 2  $&$   B_0   \otimes B_0               $&$k_2 k_6 t^6+k_2 k_6 t^8+t^4-t^{16} $&$\frac{1-t^{16}}{(1-t^4) (1-t^6) (1-t^8)}  $\\
$[2222]$&$ 0  $&$  \emptyset $           &$ 0   $&$ 1    $\\
\hline
\end{tabular}
\end{center}
\text{N.B. $(n)$ denotes the character of the $D_1 \equiv SO(2)$ reducible representation $q^n+q^{-n}$ of $U(1)$.}\\
\text{$k$ denotes the character $\pm 1$ if $B_0 \to O(1)$ or $1$ if $B_0 \to SO(1)$.}
\caption{Hilbert Series for Slodowy Slices of $C_4$.}
\label{tab:BCD5}
\end{sidewaystable}

\begin{sidewaystable}[htp]
\footnotesize
\begin{center}
\begin{tabular}{|c|c|c|c|c|}
\hline
${\begin{array}{c} \text{Nilpotent}\\\text{Orbit} \end{array}}$
&${\begin{array}{c} \text{Dimension}\\ {| {\cal S}_{\cal N,\rho}|} \end{array}}$
&${\begin{array}{c}  \text{Symmetry} \\ F\end{array}}$
&$ \text{Generators of HS} \equiv \text{PL[HS]}$
&$ \text{Unrefined HS} $\\
\hline
$[00]$&$ 4  $&$  D_2   $&$[2,0] t^2+[0,2] t^2-2 t^4 $ & $\frac{(1-t^4)^2}{(1-t^2)^6}$\\
$[20]$&$ 2  $&$  C_1  \cong A_1  $&$[2]t^2-t^4   $ & $\frac{1-t^4}{(1-t^2)^3}$\\
$[02]$&$ 2  $&$  C_1 \cong A_1 $&$[2]t^2-t^4   $ & $\frac{1-t^4}{(1-t^2)^3}$\\
$[22]$&$ 0  $&$   \emptyset  $&$ 0   $ & $ 1    $\\
\hline
$[000]$&$ 12  $&$D_3$ & $[0,1,1]t^2-t^4-t^6-t^8$&$\frac{(1-t^4) (1-t^6) (1-t^8)}{(1-t^2)^{15}} $\\
$[011]$&$ 6 $&$C_1 \otimes D_1$  &  $[2]t^2+[1](1) t^3 +t^2-t^6-t^8  $&$\frac{(1-t^6) (1-t^8)}{(1-t^2)^4 (1-t^3)^4}  $\\
$[200]$&$ 4  $&$ B_1 \otimes B_0$&$[2] t^2+[2] k t^4-k t^6-t^8 $&$\frac{(1-t^6) (1-t^8)}{(1-t^2)^3 (1-t^4)^3}$\\
$[022]$&$ 2  $&$  D_1 $&      $(2) t^4+ t^2 - t^8 $&$\frac{1-t^8}{(1-t^2) (1-t^4)^2} $\\
$[222]$&$ 0  $&$ \emptyset $ &   $ 0$ &$1$\\
\hline
$[0000]$&$ 24  $&$D_4$ & $[0,1,0,0] t^2-t^4-2 t^8-t^{12}$ & $\frac{(1-t^4) (1-t^8)^2 (1-t^{12})}{(1-t^2)^{28}} $\\
$[0100]$&$ 14 $&$D_2 \otimes C_1$  &  $[2,0] t^2+ [0,2] t^2+[2] t^2 +[1,1][1] t^3- 2 t^8-t^{12} $&$\frac{(1-t^8)^2 (1-t^{12})}{(1-t^2)^9 (1-t^3)^8}  $\\
$[0002]$&$ 12  $&$ C_2$&$[2,0]t^2+[0,1]t^4 -2t^8-t^{12} $&$\frac{(1-t^8)^2 (1-t^{12})}{(1-t^2)^{10} (1-t^4)^5}$\\
$[0020]$&$ 12  $&$ C_2$&$[2,0]t^2+[0,1]t^4 -2t^8-t^{12} $&$\frac{(1-t^8)^2 (1-t^{12})}{(1-t^2)^{10} (1-t^4)^5}$\\
$[2000]$&$ 12  $&$ B_2 \otimes B_0 $ & $ [0,2] t^2+[1,0] k t^4- k t^8-t^8-t^{12}$ &$\frac{(1-t^8)^2 (1-t^{12})}{(1-t^2)^{10} (1-t^4)^5}$\\
$[1011]$&$ 8  $&$  C_1 \otimes B_0 \otimes B_0 $ &   $ [2]t^2+[1](k_1+k_3) t^3+ [1] k_1 t^5+ t^4+k_1 k_3 t^4-k_1 k_3 t^8 -t^8-t^{12}$ &$\frac{(1-t^8)^2 (1-t^{12})}{(1-t^2)^3 (1-t^3)^4 (1-t^4)^2 (1-t^5)^2}$\\
$[0200]$&$ 6  $&$ D_1 \otimes D_1 $ &   $2 t^2+(1)(1)t^4+(2)t^4+t^6-2 t^8-t^{12}$ &$\frac{(1-t^8)^2 (1-t^{12})}{(1-t^2)^2 (1-t^4)^6 (1-t^6)}$\\

$[0202]$&$ 4  $&$ C_1 $ &   $ [2]t^2+[2]t^6-t^8-t^{12}$ &$\frac{(1-t^8) (1-t^{12})}{(1-t^2)^3 (1-t^6)^3}$\\
$[0220]$&$ 4  $&$ C_1 $ &   $ [2]t^2+[2]t^6-t^8-t^{12}$ &$\frac{(1-t^8) (1-t^{12})}{(1-t^2)^3 (1-t^6)^3}$\\
$[2200]$&$ 4  $&$B_1 \otimes B_0 $ &   $ [2]t^2+[2] k t^6-k t^8-k t^{12}$ &$\frac{(1-t^8) (1-t^{12})}{(1-t^2)^3 (1-t^6)^3}$\\
$[2022]$&$ 2  $&$ B_0 \otimes B_0  $ &   $ k_3 k_5 t^4+t^4+k_3 k_5 t^6-t^{12}$ &$\frac{1-t^{12}}{(1-t^4)^2 (1-t^6)}$\\
$[2222]$&$ 2  $&$ \emptyset $ &   $ 0$ &$1$\\
\hline
\end{tabular}
\end{center}
\text{N.B. $(n)$ denotes the character of the $D_1 \equiv SO(2)$ reducible representation $q^n+q^{-n}$ of $U(1)$.}\\
\text{$k$ denotes the character $\pm 1$ if $B_0 \to O(1)$ or $1$ if $B_0 \to SO(1)$.}
\caption{Hilbert Series for Slodowy Slices of $D_2$, $D_3$, and $D_4$}
\label{tab:BCD6}
\end{sidewaystable}

Many observations can be made about these Hilbert series.

\begin{enumerate}

\item As expected, (i) the Slodowy slice to the trivial nilpotent orbit $\mathcal{S}_{\mathcal{N},(1^N)}$ has the same Hilbert series as the nilpotent cone, (ii) the slice to the sub-regular orbit has the Hilbert series of a Kleinian singularity of type $\hat A_{2r-1}$ for the $B$ series, $\hat D_{r+1}$ for the $C$ series, and $\hat D_{r}$ for the $D$ series, and (iii) the slice to the maximal nilpotent orbit is trivial.

\item The Slodowy slices $\mathcal{S}_{\mathcal{N},\rho}$ are all complete intersections, giving a good answer to the question posed in \cite{Hanany:2011db}.

\item The adjoint maps can contain singlet generators at even powers of $t$ up to (twice) the degree of the highest Casimir of $G$; these generators may be cancelled by one or more Casimir relations.

\item The global symmetry groups of the Slodowy slice generators include mixed $BCD$ Lie groups (or $A$ series isomorphisms), as well as finite groups of type $B_0$, and descend in rank as the dimension of the Slodowy slice reduces. Different Slodowy slices may share the same symmetry group, while having inequivalent embeddings into $G$.

\item The sub-regular Slodowy slices of non-simply laced algebras match those of specific simply laced algebras, in accordance with their Kleinian singularities, as listed in table \ref{table:SS1}. In the case of Slodowy slices of $C_n$ nilpotent orbits with vector partitions of type $(2n-k,k)$, it was identified in \cite{henderson_licata_2014} that these isomorphisms with $D_{n+1}$ extend further down the Hasse diagram: ${\cal S}_{{\cal N},C (2n-k,k)} \equiv {\cal S}_{{\cal N},D (2n-k+1,k+1)}$. This occurs  due to matching chains of Kraft-Procesi transitions \cite{Kraft:1982fk} within such slices.

\item We have not attempted an exhaustive analysis of $Z_2$ factors associated with the choice of $SO$ vs $O$ flavour groups and the ensuing subtleties. 

For example, the slices ${\cal S}_{{\cal N},B[20]}$ and ${\cal S}_{{\cal N},C[02]}$ have the global symmetries $D_1 \otimes B_0$ and $D_1$, respectively, with the $B$ series Slodowy slice having an extra $B_0$ fugacity ($k = \pm 1$), notwithstanding the isomorphism between the $B$ and $C$ Lie algebras.

Similarly, in the case of $D_{4}$, the spinor pair slices, respectively ${\cal S}_{{\cal N},D[0020]}$/${\cal S}_{{\cal N},D[0002]}$ or ${\cal S}_{{\cal N},D[0220]}$/${\cal S}_{{\cal N},D[0202]}$, only carry a $C_{1 \text{ or } 2}$ series symmetry, while the corresponding vector slices of the same dimension, ${\cal S}_{{\cal N},D[2000]}$ or ${\cal S}_{{\cal N},D[2200]}$, carry a  $B_0   \otimes B_{1 \text{ or } 2}$ symmetry.

\end{enumerate}

Whilst Higgs branch constructions based on the balanced quivers of type ${{\cal B}_{B/C/D}} ({\mathbf N_f}(\rho))$ are available for all Slodowy slices, Coulomb branch constructions based on ${{\cal L}_{BC/CD/DC}}$ quivers or Higgs branch constructions based on the quivers of type ${\cal D}_{G} ({\mathbf N_f})$ are not generally available:

\begin{enumerate}

\item In the cases calculated, the slice to a sub-regular nilpotent orbit always has a Coulomb branch construction.

\item Many $BCD$ Slodowy slices do not have Coulomb branch constructions as ${{\cal L}_{BC/CD/DC}}$ quivers, either because their underlying nilpotent orbits are not special, or due to zero conformal dimension problems under the $O/USp$ monopole formula. While the issue of zero conformal dimension ($\Delta = 0$) is less prevalent for low dimension Slodowy slices, the problem is inherent in maximal $B_r - C_r - B_{r-1}$ sub-chains, and so affects many $C$ series Slodowy slices; certain other quivers are also problematic. 

\item Other than $A$ series isomorphisms, the quivers of type ${\cal D}_{G} ({\mathbf N_f})$ only provide Higgs branch constructions for $D$ series Slodowy slices of low dimension. The nilpotent orbits underlying these Slodowy slices are dual, under the Barbasch-Vogan map, to (minimal or near-to-minimal) nilpotent orbits of Characteristic height 2, for which Coulomb branch constructions using the unitary monopole formula are known \cite{Hanany:2017ooe}, plus some others, such as ${\cal S}_{{\cal N},D[0200]}$. These Dynkin diagram quivers have $S(U \otimes \ldots U )$ flavour nodes and their refined Hilbert series may not replicate all the possible combinations of orthogonal group characters.

\end{enumerate}

These matters are discussed further in the concluding section.

\FloatBarrier


\subsection{Matrix Generators for Orthosymplectic Quivers}
\label{subsec:BCDgenerators}

In the case of $BCD$ series, prescriptions are similarly available for obtaining the generators of the chiral ring corresponding to a Slodowy slice directly from the partition data or from the Higgs branch quiver.

\subsubsection{Vector Decomposition}
\label{sec:BCDVD}

From \ref{eq:BCDquivers3} and the alternating nature of the quiver, it follows that the character of the vector representation of $G$ decomposes into vector representations of an $O/USp$ product group, tensored with the $SU(2)$ embedding:

\begin{equation} 
\label{eq:BCDgens1}
\begin{aligned}
\rho :\chi _{vector}^{O(N)} \to \mathop  \oplus \limits_{\scriptstyle [n] \atop
\scriptstyle bosonic} [n]_{\rho}{~}\chi _{vector}^{O \left(N_{{f_{n + 1}}}\right)}\mathop  \oplus \limits_{\scriptstyle [n] \atop
\scriptstyle fermionic} [n]_{\rho}{~}\chi _{vector}^{USp \left(N_{{f_{n + 1}}}\right)},\\
\rho :\chi _{vector}^{USp(N)} \to \mathop  \oplus \limits_{\scriptstyle[n]\atop
\scriptstyle bosonic} [n]_{\rho}{~}\chi _{vector}^{USp \left(N_{{f_{n + 1}}} \right)}\mathop  \oplus \limits_{\scriptstyle [n] \atop
\scriptstyle fermionic} [n]_{\rho}{~}\chi _{vector}^{O \left( N_{{f_{n + 1}}} \right)},\\
\end{aligned}
\end{equation}
where ${[n]_{\rho}}$ are bosonic (odd dimension) or fermionic (even dimension) irreps of the $SU(2)$ associated with the nilpotent orbit embedding $\rho$. The requirement that the partition $\rho$ obeys the $BCD$ selection rules ensures that the $USp$ irreps are all of even dimension. Once this decomposition has been identified, the mapping of the adjoint of $G$ into matrix generators \ref{eq:SS7} follows, either by symmetrising the $USp$ vector, or by antisymmetrising the $O$ vector. This can be checked against the adjoint partition $\rho :\chi _{adjoint}^G$. Note that a choice can be made whether to use the $SO$ form of orthogonal group characters or the $O^{-}$ form.

\subsubsection{Generators from Quiver Paths}
\label{sec:OfQBCD}

For orthosymplectic quivers, the method in section \ref{sec:OfQA} can be applied, with a few changes. An operator ${\mathcal P}_{ij}(a)$ formed from a path in the quiver is defined identically. However, for orthosymplectic quivers, $\mathcal P_{ij}(a) = \mathcal P_{ji}(a)^T$, and a path yields only one generator when $i \neq j$. Other differences follow from the irreducible representations of the operators $\mathcal P_{ij}(a)$ and the gauge group invariants. There are two cases:

\begin{enumerate}
	\item $i\neq j$. The operator transforms in the defining representation of the initial flavour group and the defining representation of the final flavour group. For example, if the flavour node at position $i$ is $O(7)$ and the flavour node at position $j$ is $USp(4)$, $\mathcal P_{ij}(a)$ transforms in the irrep of dimension $7\times 4$.
	\item $i=j$. The operator has two indices that transform under the flavour group at position $i$. They are symmetrized if the gauge node at the mid point of the path is of $O$-type, or antisymmetrized if the gauge node is of $USp$-type.
\end{enumerate}
The set of operators $\mathcal P_{ij}(a)$ gives us all the generators of the chiral ring. The relations are inherited from those of the \textit{nilpotent cone} $\mathcal N$, and for $\mathcal S_{\mathcal N,\rho}$ are always the Casimir invariants of $G$. 

Now, an $O(N_{f_i})$ flavour node (of rank $>0$) always contributes (at least) a path $\mathcal P_{ii}(1)$ of length 2 that starts at $O(N_{f_i})$, goes to the gauge node $USp(N_i)$ and comes back to $O(N_{f_i})$. Since the gauge node in the middle of the path is $USp$, the operator transforms in the second antisymmetrization $\Lambda^2[fund.]_O=[adjoint]_O$. Similarly, a $USp(N_{f_i})$ flavour node always contributes (at least) a path $\mathcal P_{ii}(1)$ of length 2  that starts at $USp(N_{f_i})$, goes to the gauge node $O(N_i)$ and comes back to $USp(N_{f_i})$. Since the gauge node in the middle of the path is $O$, the operator transforms in the second symmetrization $Sym^2[fund.]_{USp}=[adjoint]_{USp}$. Consequently, the adjoint of every flavour group appears as a generator at path length 2.

\paragraph {Example} Consider the balanced quiver based on the partition $(2^2,1^4)$, whose Higgs branch is the the Slodowy slice $\CS_{{\cal N},(4,2)}$ to the nilpotent orbit $D[0100]$:
\be
	{{\cal B}_{D}{({\bf N_f}(2^2,1^4)})} ={~} \ \node{\wver{}{\,\, O(4)}}{USp(4)}-\node{\wver{}{\,\,USp(2)}}{O(6)}-\node{}{USp(4)}-\node{}{O(4)}-\node{}{USp(2)}-\node{}{O(2)}.
\ee
%
%
The decomposition of $G$ to $SU(2) \otimes F$ is:

\be \label{eq:exbcd1}
	SO(8) \rightarrow SU(2)_{\rho} \otimes O(4) \otimes USp(2).
\ee
The Hilbert series of the chiral ring of operators in the Higgs branch has generators ${\mathcal P}_{ij}(a)$ given by the quiver paths in table \ref{tab:BCD7}. 
\begin{table}[htp]
\begin{center}
\begin{tabular}{|c|c|c|}
\hline
$   {\cal P}_{ij}(a)         $&     Quiver Path        &       Generator       \\
\hline
$ {\cal P} _{11}(1)        $& $    \node{\fdu{}{\,\, O(4)}}{USp(4)}\ \ \node{}{O(6)}\node{}{USp(4)}\node{}{O(4)}\node{}{USp(2)}\node{}{O(2)}       $ &  $   \Lambda^2([1,1]t) = [2,0]t^2+[0,2]t^2   $  \\
\hline
$ {\cal P} _{2,2}(1) $&$     \node{}{USp(4)}\ \ \node{\fdu{}{\,\,USp(2)}}{O(6)}\node{}{USp(4)}\node{}{O(4)}\node{}{USp(2)}\node{}{O(2)}  $ &  $     Sym^2([1]t) = [2]t^2    $   \\
\hline
$  {\cal P} _{2,2}(2)        $& $   \node{}{USp(4)}\lrarr\node{\fdu{}{\,\,USp(2)}}{O(6)}\node{}{USp(4)}\node{}{O(4)}\node{}{USp(2)}\node{}{O(2)}    $  &    $   \Lambda^2([1]t^2) = [0]t^4    $ \\
\hline
$  {\cal P} _{1,2}(1)        $&$   \node{\fd{}{\,\, O(4)}}{USp(4)}\rarr\node{\fu{}{\,\,USp(2)}}{O(6)}\node{}{USp(4)}\node{}{O(4)}\node{}{USp(2)}\node{}{O(2)} $  &  $    [1,1][1] t^3    $ \\
\hline
\end{tabular}
\end{center}
\caption{Generators for Slodowy Slice to $D[0100]$.}
\label{tab:BCD7}
\end{table}
%
%
For $D_4$ the Casimirs give relations, $-t^4 - 2t^8 - t^{12}$, therefore, the PL[HS] read directly from the quiver is:

\be 
	PL[g_{HS}^{Higgs[{{\cal B}_{D} ({ \bf N_f} (2^2,1^4))}]}] =  [2,0]t^2+[0,2]t^2 + [2]t^2 +  [1,1][1] t^3- 2t^8 - t^{12}.
\ee
\FloatBarrier
\subsubsection{Matrices and Relations}
\label{BCDrelations}
Finally, in tables \ref{tab:BCD8910} to \ref{tab:BCD12} we provide a set of algebraic varieties described by matrices such that their HS have been computed to be identical to those of the corresponding Slodowy slices $\CS_{\mathcal N,\rho}$ of $B_1$ to $B_3$ nilpotent orbits. The analysis can in principle be continued to higher rank.

\begin{table}[h]
	\centering
	\footnotesize
	\begin{tabular}{|c|c|c|c|c|}
	\hline
	Orbit & Partition & Dim. & $\begin{array}{c}\text{Generators;}\\ \text{Degree}\end{array}$ & Relations \\ \hline
	B[0]&$(1^3)$&2& $\begin{array}{rc} M_{3\times 3}; & 2 \\  \end{array}$&$\begin{array}{rl} tr(M^2) &=0  \end{array}$ \\  \hline
	B[2] & $(3)$ & 0 & - & - \\ 
	\hline
	\hline
	B[00] & $(1^5)$ & 8 & $\begin{array}{rc} M_{5\times 5}; & 2 \end{array}$ & $\begin{array}{rl} tr(M^2) &= 0 \\ tr(M^4)&=0  \end{array}$  \\ \hline
	B[01] & $(2^2,1)$ & 4 & $\begin{array}{rc} N_{2\times 2}; & 2 \\  A_{2\times 1} ; & 3 \end{array}$ & $\begin{array}{rl} tr((N\Omega)^4) &=A^T\Omega N\Omega A \end{array}$  \\ \hline
	B[20] & $(3,1^2)$ & 2 & $\begin{array}{rc} M_{2\times 2}; & 2 \\  A_{2\times 1} ; & 4 \end{array}$ & $\begin{array}{rl} tr(M^4) &= A^TA \end{array}$  \\ \hline
	B[22] & $(5)$ & 0 & - & - \\ \hline
	\hline
	B[000] & $(1^7)$ & 18 & $\begin{array}{rc} M_{7\times 7}; & 2 \end{array}$ & $\begin{array}{rl}  tr(M^2) &= 0 \\ tr(M^4)&=0 \\ tr(M^6)&=0  \end{array}$  \\ \hline
	{B[010]} & $(2^2,1^3)$ & 10 & $\begin{array}{rc} M_{3\times 3}; & 2 \\ N_{2\times 2}; & 2 \\  A_{3\times 2} ; & 3 \end{array}$ & $\begin{array}{rl} tr(M^4) + tr((N\Omega) ^{4}) &=tr(A\Omega A^TM) +tr(A^TA\Omega N\Omega)\\ tr(M^6) + tr((N\Omega)^6) &=  tr((A\Omega A^T)^2) \end{array}$  \\ \hline
	B[200] & $(3,1^4)$ & 8 & $\begin{array}{rc} M_{4\times 4}; & 2 \\  A_{4\times 1} ; & 4 \end{array}$ & $\begin{array}{rl} tr(M^4) &=A^T A \\ tr(M^6) &= A ^T M^2 A \end{array}$  \\ \hline
	
	B[101] & $(3,2^2)$ & 6 & $\begin{array}{rc} N_{2\times 2}; & 2\\  A_{2\times 1} ; & 3 \\ M_{2\times 2}; & 4 \\  B_{2\times 1} ; & 5 \end{array}$ & $\begin{array}{rl} tr((N\Omega)^4 +(M\Omega)^2)&=  B^T\Omega A\\ tr((N\Omega)^6 +(M\Omega)^3)&= B^T \Omega M\Omega A  \end{array}$  \\ \hline
	
	B[020] & $(3^2,1)$ & 4 & $\begin{array}{rc} M_{2\times 2}; & 2\\  A_{2\times 1} ; & 4 \\ N_{2\times 2}; & 4 \\  O_{2\times 2} ; & 6 \end{array}$ & $\begin{array}{rl} tr(N)&=0\\ tr(M^4+ N^2)  &= A^T  A  \\ tr(M^6+N^3+O^2) &=A^T N A  \end{array}$  \\ \hline
	B[220] & $(5,1^2)$ & 2 & $\begin{array}{rc} M_{2\times 2}; & 2 \\  A_{2\times 1} ; & 6 \end{array}$ & $\begin{array}{rl} tr(M^6) &  = A^T A  \end{array}$  \\ \hline
	B[222] & $(7)$ & 0 & - & - \\ \hline
	\end{tabular}
	\caption[$B_1$, $B_2$ and $B_3$ Slodowy Slice Varieties]
	{$B_1$, $B_2$ and $B_3$ varieties, generated by complex matrices $M$, $N$, $O$, $A$ and $B$ and their relations, which have Hilbert series matching Slodowy slices $\CS_{\mathcal N,\rho}$. The matrices $M=-M^T$ and $O=-O^T$ are antisymmetric, $N=N^T$ is symmetric and $\Omega$ represents a square matrix that is antisymmetric and invariant under the action of $USp(2n)$.}
	\label{tab:BCD8910}
\end{table}

\begin{table}[h]
	\centering
		\small
	\begin{tabular}{|c|c|c|c|c|}
	\hline
	Orbit & Partition & Dim. & $\begin{array}{c}\text{Generators;}\\ \text{Degree}\end{array}$ & Relations \\ \hline
	C[0]&$(1^2)$&2& $\begin{array}{rc} N_{2\times 2}; & 2 \\  \end{array}$&$\begin{array}{rl} tr(N^2) &=0  \end{array}$ \\  \hline
	C[2] & $(2)$ & 0 & - & - \\ \hline
	\hline
	C[00] & $(1^4)$ & 8 & $\begin{array}{rc} N_{4\times 4}; & 2 \end{array}$ & $\begin{array}{rl} tr((N\Omega)^2) &= 0 \\ tr((N\Omega)^4)&=0  \end{array}$  \\ \hline
	
	C[10] & $(2,1^2)$ & 4 & $\begin{array}{rc} N_{2\times 2}; & 2 \\  A_{2\times 1} ; & 3 \end{array}$ & $\begin{array}{rl} tr((N\Omega)^4) &=A^T\Omega N\Omega A \end{array}$  \\ \hline
	
	C[02] & $(2^2)$ & 2 & $\begin{array}{rc} M_{2\times 2}; & 2 \\  N_{2\times 2} ; & 4 \end{array}$ & $\begin{array}{rl}tr(N)&=0 \\ tr(M^4) &= tr(N^2) \end{array}$  \\ \hline
	C[22] & $(4)$ & 0 & - & - \\ \hline
	\hline
	
	C[000] & $(1^6)$ & 18 & $\begin{array}{rc} N_{6\times 6}; & 2 \end{array}$ & $\begin{array}{rl} tr((N\Omega)^2) &= 0 \\ tr((N\Omega)^4)&=0 \\ tr((N\Omega)^6)&=0  \end{array}$  \\ \hline
	
	C[100] & $(2,1^4)$ & 12 & $\begin{array}{rc} N_{4\times 4}; & 2  \\  A_{4\times 1} ; & 3 \end{array}$ & $\begin{array}{rl} tr((N\Omega)^4)  &= A^T \Omega N \Omega A\\  tr((N\Omega)^6) & = A^T \Omega (N {\Omega})^3 A \end{array}$  \\ 
	\hline
	C[010] & $(2^2,1^2)$ & 8 & $\begin{array}{rc} N_{2\times 2}; & 2 \\ M_{2\times 2};& 2\\  A_{2\times 2};& 3 \\P_{2\times2} ; & 4 \end{array}$ & $\begin{array}{rl} tr(P)&=0\\tr((N\Omega)^4+M^4+P^2) &=tr(A^T\Omega N\Omega A) \\ tr((N\Omega)^6+M^6+P^3) &= tr(A ^T\Omega ( N\Omega) ^2\Omega A) \end{array}$  \\ \hline
	
	C[002] & $(2^3)$ & 6 & $\begin{array}{rc} M_{3\times 3}; & 2\\   N_{3\times 3}; & 4 \\  \end{array}$ & $\begin{array}{rl} tr(N)&=0 \\tr(M^4)&=tr(N^2)\\tr(M^6)&=tr(N^3) \end{array}$  \\ \hline
	
	C[020] & $(3^2)$ & 4 & $\begin{array}{rc} N_{2\times 2}; & 2\\ M_{2\times 2}; & 4 \\  P_{2\times 2} ; & 6 \end{array}$ & $\begin{array}{rl} tr(M \Omega)&=0\\ tr(P \Omega)&=0 \\tr((N\Omega)^4+ (M\Omega )^2)  &= 0  \\ tr((N\Omega )^6+(M\Omega )^3+(P\Omega )^2) &=0 \end{array}$  \\ \hline
	
	C[210] & $(4,1^2)$ & 4 & $\begin{array}{rc} N_{2\times 2}; & 2 \\  A_{2\times 1} ; & 5 \end{array}$ & $\begin{array}{rl} tr((N\Omega)^6) &= A^T\Omega N \Omega A  \end{array}$  \\ \hline
	C[202] & $(4,2)$ & 2 & $\begin{array}{rc} N_{1\times 1}; & 4 \\  A_{1\times 1} ; & 4\\  P_{1\times 1} ; & 6 \end{array}$ & $\begin{array}{rl} tr(N A^2) &=  tr(P^2)  \end{array}$  \\
	 \hline
	C[222] & $(6)$ & 0 & - & - \\ 
	\hline
	\end{tabular}
	\caption[$C_1$, $C_2$ and $C_3$ Slodowy Slice Varieties]
	{$C_1$, $C_2$ and $C_3$ varieties, generated by complex matrices $M$, $N$, $O$, $P$, $A$ and $B$ and their relations, which have Hilbert series matching Slodowy slices $\CS_{\mathcal N,\rho}$. The matrices $M=-M^T$ and $O=-O^T$ are antisymmetric, $N=N^T$ and $P=P^T$ are symmetric and $\Omega$ represents a square matrix that is antisymmetric and invariant under the action of $USp(2n)$.}
	\label{tab:BCD11}
\end{table}

\begin{table}[h]
	\centering
		\small
	\begin{tabular}{|c|c|c|c|c|}
	\hline
	Orbit & Partition & Dim. & $\begin{array}{c}\text{Generators;}\\ \text{Degree}\end{array}$ & Relations \\ \hline
	D[00] & $(1^4)$ & 4 & $\begin{array}{rc} M_{4\times 4}; & 2 \end{array}$ & $\begin{array}{rl} tr(M^2) &= 0 \\ pf(M)&=0  \end{array}$  \\ \hline
	D[20] & $(2^2)$ & 2 & $\begin{array}{rc} N_{2\times 2}; & 2 \\  \end{array}$ & $\begin{array}{rl} tr((N\Omega)^2) &=0 \end{array}$  \\ \hline
	D[02] & $(2^2)$ & 2 & $\begin{array}{rc} N_{2\times 2}; & 2 \\  \end{array}$ & $\begin{array}{rl} tr((N\Omega)^2) &=0 \end{array}$   \\ \hline
	D[22] & $(4)$ & 0 & - & - \\ \hline
	\hline
	D[000] & $(1^6)$ & 12 & $\begin{array}{rc} M_{6\times 6}; & 2 \end{array}$ & $\begin{array}{rl}  tr(M^2) &= 0 \\ tr(M^4)&=0 \\ pf(M)&=0  \end{array}$  \\ \hline
	D[011] & $(2^2,1^2)$ & 6 & $\begin{array}{rc} M_{2\times 2}; & 2 \\ N_{2\times 2}; & 2 \\  A_{2\times 2} ; & 3 \end{array}$ & $\begin{array}{rl} tr(A\Omega A^T\Omega) &= 0\\ tr(M^4) + tr((N\Omega)^4) &=  tr(A\Omega A^T M +A\Omega N \Omega A^T) \end{array}$  \\ \hline
	D[200] & $(3,1^3)$ & 4 & $\begin{array}{rc} M_{3\times 3}; & 2 \\  A_{3\times 1} ; & 4 \end{array}$ & $\begin{array}{rl} \epsilon^{ijk}M_{ij}A_k &= 0 \\ tr(M^4) &= A ^T A \end{array}$  \\ \hline
	D[022] & $(3^2)$ & 2 & $\begin{array}{rc} M_{2\times 2}; & 2 \\ N_{2\times 2}; & 4 \end{array}$ & $\begin{array}{rl} tr(N)&=0\\ tr(M^4)&=tr(N^2)\end{array}$  \\ \hline
	D[222] & $(5,1)$ & 0 & - & - \\ \hline
	\end{tabular}
	\caption[$D_2$ and $D_3$ Slodowy Slice Varieties]
	{$D_2$ and $D_3$ varieties, generated by complex matrices $M$, $N$, and $A$ and their relations, which have Hilbert series matching Slodowy slices $\CS_{\mathcal N,\rho}$. The matrix $M=-M^T$ is antisymmetric, $N=N^T$ is symmetric and $\Omega$  represents a square matrix that is antisymmetric and invariant under the action of $USp(2n)$. $pf()$ denotes the Pfaffian.}
	\label{tab:BCD12}
\end{table}

\FloatBarrier

\section{Discussion and Conclusions}
\label{sec:Conclusions}

\paragraph{Higgs Branch}

We have presented constructions for quivers whose Higgs branches yield Hilbert series corresponding to the Slodowy slices of the nilpotent orbits of  $A_1$ to $A_5$ plus $BCD$ algebras up to rank 4. There are essentially two families of quivers, the balanced unitary type $\{{{\cal B}_{A}} = {\cal D}_A, {\cal D}_D  \}$ and the canonically balanced orthosymplectic type $\{ {{\cal B}_{B/C/D}} \}$. The balanced unitary quivers have gauge nodes in the pattern of the parent algebra Dynkin diagram and yield constructions for Slodowy slices of simply laced algebras, including all $A$ series slices and $D$ series slices of low dimension. The orthosymplectic quivers yield constructions of all $BCD$ Slodowy slices.

The global symmetry $F$ of a Slodowy slice descends from that of the parent group $G$ (in the case of the slice to the trivial nilpotent orbit), via subgroups of $G$, down to ${\mathbb Z}_2$ symmetries (for the slices of some near maximal nilpotent orbits). The grading of the Hilbert series is such that (i) the sets of Slodowy slices and nilpotent orbits intersect at the nilpotent cone and at the origin and (ii) the sub-regular slices match the known singularities \cite{brieskorn1970singular,slodowy_1980, Cabrera:2017njm}. In between, we have shown how the Slodowy slice symmetry groups and mappings of $G$ representations to $SU(2) \otimes F$ follow, via the Higgs branch formula, from the $SU(2)$ homomorphisms into $G$ of the associated nilpotent orbits.

We anticipate that these results generalise to Classical groups of arbitrary rank.

\paragraph{Coulomb Branch}
As is known, in the case of the $A$ series, the existence of a bijection between partitions and their transposes (the Luztig-Spaltenstein map) leads to a complete set of Coulomb branch constructions for Slodowy slices; these yield the same set of Hilbert series as the Higgs branch constructions. The Coulomb branch constructions are based on applying the unitary monopole formula to linear quivers ${{\cal L}_{A}}$, which are not generally balanced.

In the case of the $BCD$ series, however, other than for accidental isomorphisms with the $A$ series, this study has clarified that (i) the existence of suitable linear orthosymplectic quivers $\{ {{\cal L}_{BC}}, {{\cal L}_{CD}}, {{\cal L}_{DC}} \}$ is limited to the Slodowy slices of special nilpotent orbits, (ii) within these, the applicability of the Coulomb branch orthosymplectic monopole formula is restricted to those quivers that have positive conformal dimension, and (iii) the resulting Hilbert series are only available in unrefined form.


\paragraph{Slodowy Slice Formula}
The refined Hilbert series of a Slodowy slice can also be obtained directly from the mapping of the adjoint representation of $G$ into $SU(2) \otimes F$, using \ref{eq:SS11}. This mapping follows from the decomposition of the fundamental/vector of $G \rightarrow SU(2) \otimes F$ under \ref{eq:Agens1} or \ref{eq:BCDgens1}.

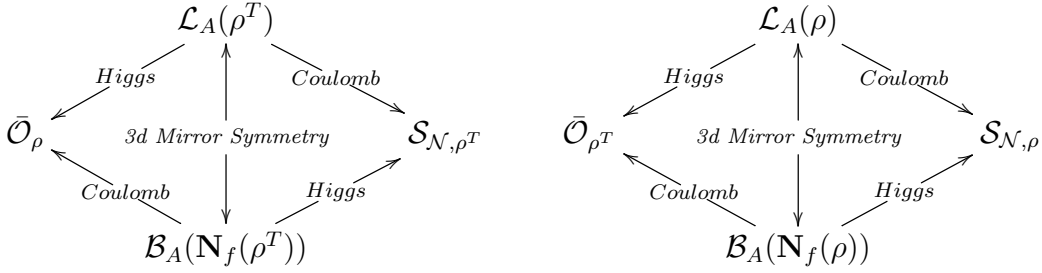
\begin{figure}[htp]
\centering
\begin{displaymath}
    \xymatrix{
  &   {{\cal L}_A(\rho ^T)} \ar[dl]|{Higgs} \ar[dr]|{Coulomb} & & &  {{\cal L}_A(\rho)}   \ar[dl]|{Higgs} \ar[dr]|{Coulomb} & \\
  {\bar {\cal O}}_\rho  &\text{\scriptsize \it 3d~Mirror Symmetry} \ar[d] \ar[u] &  {{\cal S}}_{{{\cal N}},{\rho^T} }  & {\bar {\cal O}}_{\rho^T} & \text{\scriptsize \it 3d~Mirror Symmetry} \ar[d] \ar[u] & {{\cal S}}_{{{\cal N}},{\rho} }  \\
   &  {{\cal B}_A( {{\mathbf N_f}( \rho ^T)})}  \ar[ul]|{Coulomb} \ar[ur]|{Higgs} & & & {{\cal B}_A( {{\mathbf N_f}( \rho)})}   \ar[ul]|{Coulomb} \ar[ur]|{Higgs}  & }
\end{displaymath}
\caption[A Series 3d Mirror Symmetry]{A Series 3d Mirror Symmetry. All constructions give refined Hilbert series for a partition $\rho$ and its dual $\rho^T$ under the Lusztig-Spaltenstein map.}
\label{fig:msA}
\end{figure}

\paragraph{Dualities and $3d$ Mirror Symmetry}
The $A$ series findings verify the known $3d$ mirror symmetry relations \ref{eq:Aquivers4} and \ref{eq:Aquivers4aa}. Under these, linear or balanced quivers based on partitions $\rho$ can be used either for Higgs branch or Coulomb branch constructions; one combination yields a Slodowy slice and the other combination yields a (generally different) dual nilpotent orbit under the Lusztig-Spaltenstein map $\rho^T$, as illustrated in figure \ref{fig:msA}.

The analysis of $BCD$ series quivers shows, however, that such a picture of dualities \cite{Gaiotto:2008ak} does not extend to the $BCD$ series, other than in a limited way, due to the various restrictions on Coulomb branch constructions, discussed above. The refined (i.e. faithful) HS relationships for nilpotent orbits of the $BCD$ series can be summarised:

\begin{equation} 
\label{eq:conc2}
\begin{aligned}
{{\cal S}}_{{{\cal N}},\rho } & = Higgs \left[ {{\cal B}_{B/C/D}( {{\mathbf N_f}( \rho)})} \right],\\
{\bar {\cal O}}_\rho & = Higgs\left[ {{\cal L}_{B/C/D}(\rho ^T)} \right],
 \end{aligned}
\end{equation}
and, for $D$ series Dynkin type quivers of Characteristic height 2:

\begin{equation} 
\label{eq:conc3}
\begin{aligned}
{ \cal \bar O}_\rho = Coulomb\left[ {{{\cal D}_D}([\rho] )} \right],\\
S_{ {\cal N} , {\rbv} } = Higgs\left[ {{{\cal D}_D}([\rho ])} \right],
 \end{aligned}
\end{equation}
where $d_{BV}(\rho)$ is the dual partition to $\rho$ under the $D$ series Barbasch-Vogan map.

\begin{figure}[htp]
\centering
\scriptsize
\begin{displaymath}
    \xymatrix{
  &   {{\cal L}_{BC/CD/DC}(\rho ^T)} {\ar[dl]|{\color{black} {\mathop {Higgs}\limits_{(O)}}}}    \ar@{.>}[dr]|{\mathop {Coulomb}\limits_{(SO)} } & & &  {{\cal L}_{BC/CD/DC}({\rbv}^T)}   \ar@{-->}[dl]|{{\mathop {Higgs}\limits_{(O)}}} \ar@{.>}[dr]|{\mathop {Coulomb}\limits_{(SO)} } & \\
  {\bar {\cal O}}_\rho  & \text{\scriptsize \it 3d~Mirror Symmetry} \ar[d] \ar[u]   &  {{\cal S}}_{{{\cal N}},{\rbv} }  & {\bar {\cal O}}_{\rbv} & \text{\scriptsize \it 3d~Mirror Symmetry} \ar[d] \ar[u]  & {{\cal S}}_{{{\cal N}},{\rho} }  \\
   &  {{\cal B}_{B/C/D}( {{\mathbf N_f}( \rbv)})}  \ar@{.>}[ul]|{\mathop {Coulomb}\limits_{(O/SO)} } \ar@{-->}[ur]|{{\mathop {Higgs}\limits_{(O)}}} & & & {{\cal B}_{B/C/D}( {{\mathbf N_f}( \rho)})}   \ar@{.>}[ul]|{\mathop {Coulomb}\limits_{(O/SO)} } \ar[ur]|{{\mathop {Higgs}\limits_{(O)}}}  & }
\end{displaymath}
\caption[BCD Series 3d Mirror Symmetry]{BCD Series 3d Mirror Symmetry. Solid arrows indicate Higgs branches which give \textit{refined} Hilbert series for a partition $\rho$. Dashed arrows indicate Higgs branches which give refined Hilbert series for the Barbasch-Vogan dual partition $\rbv$ of a \textit{special} nilpotent orbit. Dotted arrows indicate Coulomb branches which give \textit{unrefined} Hilbert series for those \textit{special} nilpotent orbits whose quivers have positive conformal dimension.}
\label{fig:msBCD}
\end{figure}
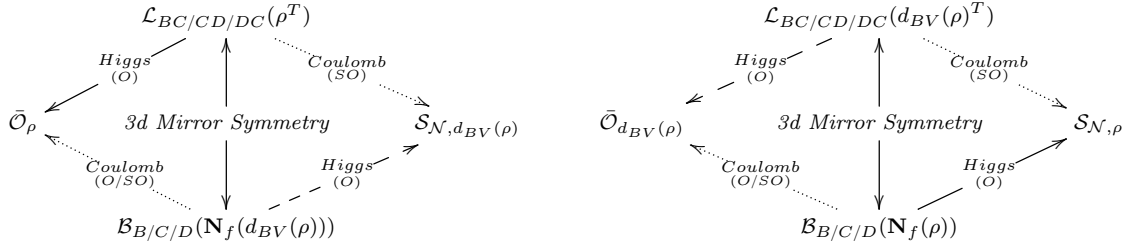

If we restrict ourselves (i) to \textit{special} nilpotent orbits, (ii) to quivers with positive conformal dimension, and (iii) to \textit{unrefined} Hilbert series, then we can summarise the more limited $3d$ mirror symmetry for the $BCD$ series as in figure \ref{fig:msBCD}.

Note that even for these cases there is a further obstruction: the difference between $SO$ and $O$ nodes in the quiver \cite{Cremonesi:2014uva,Cabrera:2017ucb}. For the A series, $3d$ mirror symmetry involves a pair of quivers for which the Coulomb branch and Higgs branch are swapped. In the BCD series however, once the gauge algebra of the quiver is specified there is still the question of whether the gauge groups are orthogonal or special orthogonal. As shown in figure \ref{fig:msBCD} a different choice needs to be made depending on the branch of the quiver.
This is not quite the same as $3d$ mirror symmetry. 

On the other hand, there is a pair of SCFTs, $T_\sigma^\rho(G)$ and $T^\sigma_\rho(G^\vee)$ \cite{Gaiotto:2008ak,Benini:2010uu,Chacaltana:2012zy}, which are predicted to have precisely the two different gauge algebras depicted in one of the diagrams of figure \ref{fig:msBCD}: if $T_\sigma^\rho(G)$ corresponds to quiver ${\cal L}_{BC/CD/DC}(\rho ^T)$, then $T^\sigma_\rho(G^\vee)$ has the quiver ${\cal B}_{B/C/D}( {{\mathbf N_f}( \rbv)})$, along with the Higgs and Coulomb branches depicted in the same diagram. However, the present results, together with \cite{Cremonesi:2014uva,Hanany:2016gbz,Cabrera:2017ucb}, show that this cannot be the case, since there are factors of $\mathbb{Z}_2$ in the gauge group of the quiver for $T_\sigma ^\rho (G)$ that differ depending on the branch being computed. This is a very intriguing point that needs to be addressed in future studies, especially since it has consequences for the way effective gauge theories can be employed to understand the dynamics of Dp-branes in the presence of Op-planes.

Thus, it is the Higgs branch that provides the means to conduct a refined analysis of the HS of $BCD$ series nilpotent orbits and Slodowy slices. These represent only a subset of the $BCD$ series moduli spaces, $S_{\rho_1,\rho_2}\equiv \bar{\mathcal{O}}_{\rho_1}\cap S_{\rho_2}$, which include nilpotent orbits $S_{\rho,trivial}$ and Slodowy slices $S_{{\cal N},\rho}$ as limiting cases.\footnote{Such $BCD$ series moduli spaces $S_{\rho_1,\rho_2}$ generalise naturally to any pair of nilpotent orbits (unlike $T_\sigma^\rho(O/USp)$ theories, which are restricted to special orbits).}. The indications are that Higgs branch methods should provide a fruitful means of analysing such spaces.

\paragraph{Further Work}

Besides a study of quivers for $S_{\rho_1,\rho_2}$ moduli spaces, it would be interesting to extend this analysis to the Slodowy slices of Exceptional groups. While Higgs branch quiver constructions are not available for nilpotent orbits of Exceptional groups, a limited number of Coulomb branch quiver constructions are known. For Slodowy slices, where the situation is somewhat reversed by dualities, some Higgs branch constructions should be available, based, for example, on Dynkin diagrams of the E series.

With respect to the Coulomb branch, it would be interesting to understand (i) whether some non-linear fugacity map can be developed for the orthosymplectic monopole formula in order to obtain \textit{refined} Hilbert series, and (ii) whether a modified monopole formula can be found that avoids the zero conformal dimension problem associated with many orthosymplectic quivers. A recent advance has been made on this front in \cite{Assel:2018exy}, where Coulomb branches of \textit{bad} quivers with a single $C_r$ gauge node have been computed. A case that also appears in our study is the quiver $[D_{2r}]-(C_r)$, where the expected Slodowy slices are formed in quite a surprising way\footnote{\cite{Assel:2018exy} computes that there are two most singular points in this Coulomb branch, related by a $\mathbb{Z}_2$ action. Crucially, at each point, an SCFT denoted $T_{USp(2r),2r}$ has a Coulomb branch identical to the expected Slodowy slice (identified in \cite{Assel:2018exy} as the Higgs branch of the corresponding ${\cal D}_{G} ({\mathbf N_f})$ quiver). }. It remains a challenge to develop such techniques to obtain Coulomb branch calculations for the Slodowy slices of the other quivers with $\Delta = 0$ in our tables.

More generally, the family of transverse spaces and symmetry breaking associated with Slodowy slices provides a rich basis set of quivers that can be extended or used as building blocks to understand the relationships between a wide array of quiver theories and their Higgs and/or Coulomb branches.



\paragraph{Acknowldgements}

We would like to thank Stefano Cremonesi and Benjamin Assel for helpful conversations during the development of this project.
S.C. is supported by an EPSRC DTP studentship EP/M507878/1. A.H. is supported by STFC Consolidated Grant ST/J0003533/1, and EPSRC Programme Grant EP/K034456/1.

\appendix

\section{Notation and Terminology}
\label{apx:1}

We refer to Slodowy slices and nilpotent orbits either by their Lie algebras $\mathfrak g$, or by the Lie groups $G$ in which they transform. While such references are relatively interchangeable for $USp$ groups, with Lie algebras of $C$ type, it can be important to distinguish between $O$ and $SO$ forms of orthogonal groups, which may share the same $B$ or $D$ type Lie algebra, but whose representations have different characters. We have sought to highlight those areas where this distinction is important in the text.

We freely use the terminology and concepts of the Plethystics Program, including the Plethystic Exponential (``PE"), its inverse, the Plethystic Logarithm (``PL"), the Fermionic Plethystic Exponential (``PEF") and, its inverse, the Fermionic Plethystic Logarithm(``PFL"). For our purposes:
\begin{equation} 
\label{eq:intro1}
\begin{aligned}
PE\left[ {\sum\limits_{i = 1}^d {{A_i}} ,t} \right] & \equiv \prod\limits_{i = 1}^d {\frac{1}{{\left( {1 - {A_i}t} \right)}}},\\
PE\left[ { - \sum\limits_{i = 1}^d {{A_i}} ,t} \right] & \equiv \prod\limits_{i = 1}^d {\left( {1 - {A_i}t} \right)},\\
PE\left[ {\sum\limits_{i = 1}^d {{A_i}} , - t} \right] & \equiv \prod\limits_{i = 1}^d {\frac{1}{{\left( {1 + {A_i}t} \right)}}} ,\\
 PE\left[ { - \sum\limits_{i = 1}^d {{A_i}} , - t} \right] & \equiv PEF\left[ {\sum\limits_{i = 1}^d {{A_i}} ,t} \right] & \equiv \prod\limits_{i = 1}^d {\left( {1 + {A_i}t} \right)},\\
 \end{aligned}
\end{equation}
where $A_i$ are monomials in weight or root coordinates or fugacities. The reader is referred to \cite{Benvenuti:2006qr} or \cite{Hanany:2014dia} for further detail.

We present the characters of a group $G$ either in the generic form ${\cal X}^{G}(x_i)$, or as $[irrep]_{G}$, or using Dynkin labels as ${[ {{n_1}, \ldots , {n_r}}]_{G}}$, where $r$ is the rank of $G$. We often represent singlet irreps implicitly via their character $1$. We typically label unimodular Cartan subalgebra (``CSA") coordinates for weights within characters by $x \equiv (x_1 \ldots x_r)$ and simple root coordinates by $z \equiv (z_1 \ldots z_r)$, dropping subscripts if no ambiguities arise. The Cartan matrix $A_{ij}$ mediates the canonical relationship between simple root and CSA coordinates as ${z_i} = \prod\limits_j {x_j^{{A_{ij}}}}$ and ${x_i} = \prod\limits_j {z_j^{{A^{ - 1}}_{ij}}}$.

We label field (or R-charge) counting variables with $t$, adding subscripts if necessary. Under the conventions in this paper, the fugacity $t$ corresponds to an R-charge of 1/2 and $t^2$ corresponds to an R-charge of 1.
We may refer to \textit{series}, such as $1 + f + {f^2} + \ldots $, by their \textit{generating functions} $1/\left( {1 - f} \right)$. Different types of generating function are indicated in table \ref{tab1}; amongst these, the refined HS faithfully encode the group theoretic information about a moduli space.
\begin{table}[htp]
\small
\centering
\begin{tabular}{|c|c|c|}
\hline
$\text{Generating~Function}$&$ \text{Notation}$&$ \text{Definition} $\\
\hline
$\text{Refined~HS~(Weight~coordinates)}$&$ {{g^{G}_{HS}}( {{x},{t}} )}$&$\sum\limits_{n = 0}^\infty {{a_n}({x})}{t^n} $\\
$\text{Refined~HS~(Simple root~coordinates)}$&$ {{g^{G}_{HS}}( {{z},{t}} )}$&$\sum\limits_{n = 0}^\infty {{a_n}({z})}{t^n} $\\
$\text{Unrefined~HS}$&$ {{g^{G}_{HS}}\left( t \right)}$&$ \sum\limits_{n = 0}^\infty {{a_n}} {t^n} \equiv \sum\limits_{n = 0}^\infty {{a_n}({1})}{t^n} $\\
\hline
\end{tabular}
\caption{Types of Generating Function}
\label{tab1}
\end{table}
%


\clearpage

\bibliographystyle{JHEP}

\bibliography{RJKBibLib}


\end{document}